\documentclass{article}

\usepackage{arxiv}

\usepackage[utf8]{inputenc} % allow utf-8 input
\usepackage[T1]{fontenc}    % use 8-bit T1 fonts
\usepackage{hyperref}       % hyperlinks
\usepackage{url}            % simple URL typesetting
\usepackage{booktabs}       % professional-quality tables
\usepackage{amsfonts}       % blackboard math symbols
\usepackage{nicefrac}       % compact symbols for 1/2, etc.
\usepackage{microtype}      % microtypography
\usepackage{lipsum}
\usepackage{graphicx}
\usepackage{subfigure}
\usepackage{amssymb}
\usepackage{amsmath}
\usepackage[ruled,vlined]{algorithm2e}
\graphicspath{ {./images/} }

\title{Full waveform inversion by model extension: theory, design and optimization}

\author{
    Guillaume Barnier \\
    Geophysics Department\\
    Stanford University\\
    Stanford, CA 94305 \\
    \texttt{gbarnier@sep.stanford.edu} \\
\And
    Ettore Biondi \\
    Seismology Laboratory\\
    California Institute of Technology\\
    Pasadena, CA 91125\\
    \texttt{ettore88@sep.stanford.edu} \\
\And
    Robert G. Clapp \\
    Geophysics Department\\
    Stanford University\\
    Stanford, CA 94305 \\
    \texttt{bob@sep.stanford.edu} \\
\And
 Biondo Biondi \\
  Geophysics Department\\
  Stanford University\\
  Stanford, CA 94305 \\
  \texttt{biondo@sep.stanford.edu} \\  
}

\begin{document}
\maketitle

%% Abstract
\begin{abstract}
%%%%%%%%%%%%%%%%%%%%%%%%%%%%%%%%%%%%%%%%%%%%%%%%%%%%%%%%%%%%%%%%%%%%%%%%%%%%%%%%%%%%%%
%%%%%%%%%%%%%%%%%%%%%%%%%%%%%%%%%%%% Abstract %%%%%%%%%%%%%%%%%%%%%%%%%%%%%%%%%%%%%%%%
%%%%%%%%%%%%%%%%%%%%%%%%%%%%%%%%%%%%%%%%%%%%%%%%%%%%%%%%%%%%%%%%%%%%%%%%%%%%%%%%%%%%%%
We describe a new method, full waveform inversion by model extension (FWIME) that recovers accurate acoustic subsurface velocity models from seismic data, when conventional methods fail. We leverage the advantageous convergence properties of wave-equation migration velocity analysis (WEMVA) with the accuracy and high-resolution nature of acoustic full waveform inversion (FWI) by combining them into a robust mathematically-consistent workflow with minimal need for user inputs. The novelty of FWIME resides in the design of a new cost function and a novel optimization strategy to combine the two techniques, making our approach more efficient and powerful than applying them sequentially. We observe that FWIME mitigates the need for accurate initial models and low-frequency long-offset data, which can be challenging to acquire. Our new objective function contains two components. First, we modify the forward mapping of the FWI problem by adding a data-correcting term computed with an extended demigration operator, whose goal is to ensure phase matching between predicted and observed data, even when the initial model is inaccurate. The second component, which is a modified WEMVA cost function, allows us to progressively remove the contributions of the data-correcting term throughout the inversion process. The coupling between the two components is handled by the variable projection method, which reduces the number of adjustable hyper-parameters, thereby making our solution simple to use. We devise a model-space multi-scale optimization scheme by re-parametrizing the velocity model on spline grids to control the resolution of the model updates. We generate three cycle-skipped 2D synthetic datasets, each containing only one type of wave (transmitted, reflected, refracted), and we analyze how FWIME successfully recovers accurate solutions with the same procedure for all three cases. In a second paper, we apply FWIME to challenging realistic examples where we simultaneously invert all wave modes.

\end{abstract}

%% Introduction
\section{Introduction}
%%%%%%%%%%%%%%%%%%%%%%%%%%%%%%%%%%%%%%%%%%%%%%%%%%%%%%%%%%%%%%%%%%%%%%%%%%%%%%%%%%%%%%
%%%%%%%%%%%%%%%%%%%%%%%%%%%%%%%%% Introduction %%%%%%%%%%%%%%%%%%%%%%%%%%%%%%%%%%%%%%%
%%%%%%%%%%%%%%%%%%%%%%%%%%%%%%%%%%%%%%%%%%%%%%%%%%%%%%%%%%%%%%%%%%%%%%%%%%%%%%%%%%%%%%
% Standard imaging sequential process
The standard seismic imaging process is based on a separation of model scales and is typically conducted in three sequential steps \cite[]{claerbout1985imaging,yilmaz2001seismic,biondi2014simultaneous}. First, conventional tomographic techniques such as WEMVA are employed to retrieve low-resolution velocity models of the subsurface \cite[]{woodward1992wave,symes1994inversion,clapp2000geologically,clapp2004incorporating,sava2004wave,yang2013wavefield,zhang2015velocity,estebanThesis}. Such algorithms have heuristically been found to be more immune to initial velocity model choices \cite[]{stolk2002smooth}. The second step is usually the most challenging one and consists in recovering higher-resolution Earth models by applying an advanced iterative method referred to as acoustic full waveform inversion (FWI), first formulated by \cite{lailly1983seismic,tarantola1984inversion}. In the past decade, successful applications of FWI on large-scale 3D field data have demonstrated the method’s ability at producing accurate and useful solutions \cite[]{sirgue2010thematic,baeten2013use,shen2018full}. However, its success relies on already having an accurate initial model in relation with the frequency content of the recorded data, and failing to satisfy this requirement may lead FWI to recover non-physical solutions, which correspond to spurious local minima present in the loss function that is being minimized \cite[]{virieux2009overview}. This problem can potentially be mitigated with the presence of coherent low-frequency long-offset data, and gradually increasing the bandwidth of the inverted signal may drive the solution to the correct minimum \cite[]{bunks1995multiscale,brenders2007waveform,fichtner2010full}. Unfortunately, such type of data can be challenging and costly to acquire \cite[]{dellinger2016wolfspar,bate2021ultra}, and may therefore not be available, especially in companies' legacy datasets \cite[]{warner2016adaptive}. 

In the last step, higher-resolution maps depicting the various interfaces between rock layers are obtained by applying a seismic migration algorithm, such as reverse-time migration (RTM) \cite[]{baysal1983reverse}, although high-frequency FWI might become an attractive alternative \cite[]{lazaratos2011improving,routh2017impact}. The reliability of such maps is essential for energy companies to ensure safety and efficiency during exploration, drilling and production. Unfortunately, the quality of migrated images heavily depends on the accuracy of the velocity macro-model provided by the FWI step. 

% Proposed methods when the standard way fails
When conventional tomographic techniques are unable to produce accurate enough initial models for FWI, an information gap between the low and high wavenumbers of the subsurface model is present within the seismic velocity estimation sequence \cite[]{claerbout1985imaging,barnier2020full}. To bridge this gap, multiple methods have been proposed where the conventional FWI problem is modified by either directly extending the unknown model search space \cite[]{symes2008migration,fleury2012bi,biondi2014simultaneous,huang2015born,huang2017full,guo2017velocity,barnier2020full,metivier2021receiver}, by relaxing certain constraints related to the physics of the problem \cite[]{van2013mitigating,van2014new,warner2014adaptive,warner2016adaptive,guasch2019adaptive,aghamiry2019improving,li2021extended}, or by measuring the data misfit with a different norm \cite[]{metivier2016measuring,metivier2018optimal}. From an optimization standpoint, the common goal behind all of these approaches is to design alternate and more convex objective functions that share the same global minimum as FWI.  

To our knowledge, \cite{symes2008migration} was the first author to introduce the idea of combining the robustness of migration velocity analysis (MVA) with the accuracy of FWI into one cost function by using a concept referred to as extended modeling. Following this idea, many authors have proposed different methods to efficiently implement this concept. \cite{biondi2014simultaneous} developed a technique referred to as tomographic full waveform inversion and showed very promising results. However, the proposed algorithm, based on a nested-scheme approach, requires the user to tune many hyper-parameters inherent to their inverse problem formulation. \cite{huang2015born} showed an interesting solution using the variable projection method \cite[]{golub1973differentiation,rickett2013variable}, but the workflow was still based on an explicit model-scale separation between low- and high-wavenumber components, preventing it from fully bridging the information gap previously described.

% FWIME
We build upon the work of \cite{barnier2018full,barnier2020full}, and we present a novel algorithm that produces accurate acoustic velocity models by successfully pairing both WEMVA and FWI techniques into one robust, mathematically consistent, and user-friendly workflow. The main novelty of this approach, namely full waveform inversion by model extension (FWIME), comes from our cost function design and the optimization strategy we devise, which results in a more efficient and powerful technique than simply applying WEMVA and FWI separately and/or sequentially. In this paper, we do not mathematically prove that our new objective function is more suited for gradient-descent optimization (i.e., free of local minima), but we provide strong numerical evidence to support this claim. 

Understanding how this new algorithm operates is important for future applications by non-expert users. Our new objective function contains two components. In the first component, we modify the conventional FWI forward mapping operator (i.e., the acoustic isotropic constant-density finite-difference modeling operator) by adding a data-correcting term generated by the linear mapping of an extended model perturbation into the data space such that the observed data is fully fitted at any given precision. This linear mapping, referred to as extended modeling, is the most important tool of our method: it allows us to linearly predict any event in the data space regardless of the accuracy of the initial velocity model, thereby creating a non-physical forward mapping that always ensures wiggle matching between predicted and observed data.

Being able to match the observed data with this additional term does not mathematically guarantee any improvement in the convexity of our new cost function, such as a reduction of the number of local minima. One necessary condition is that for the expected target model, the contribution of this additional term should eventually vanish. Therefore, we add a second component that allows us to progressively remove the contributions of the data-correcting term throughout the inversion process by eliminating all the energy present within the extended model perturbation. This additional component, referred to as the FWIME annihilator, possesses similar beneficial convergence properties as the conventional WEMVA objective function. Moreover, we make use of the variable projection method, which provides three advantages. First, it gives us better control on the phase alignment between predicted and observed data. It also allows us to avoid separating our unknown velocity model parameters into a background (low-wavenumber component) and a reflectivity (high-wavenumber perturbation). Finally, it handles the coupling between the two components of the objective function in a mathematically consistent fashion, thereby reducing the number of optimization hyper-parameters to two. 

We devise a new optimization process which is a crucial ingredient of our method. We adapt the general model-space multi-scale method presented in \cite{barnier2019waveform} to our FWIME scheme. We parametrize our unknown velocity model on B-spline grids, and we simultaneously invert the full data-bandwidth while gradually refining the grid spacing with iterations \cite[]{zhao2015multiscale,tavakoli2019matrix}. The inverted model for a given grid is then used as the initial guess for the following inversion performed with a finer grid. The grid refinement rate allows us to better control and slowly increase the wavenumber content (i.e., the spatial resolution) of the model updates, which constitutes the multi-scale aspect of our technique. In addition, this approach alleviates the need for tedious data filtering and event selection typically required for conventional FWI \cite[]{shen2014early,brittan2019fwi} because all wave modes are inverted at once with the same procedure.

% Paper outline
We begin by providing a theoretical and mathematical framework of our algorithm design. We analyze the role of each term/component of the FWIME objective function and we illustrate their properties with 2D numerical examples. We then describe our new model-space multi-scale optimization approach. Finally, we successfully invert cycle-skipped data on three 2D synthetic tests where conventional FWI converges to unsatisfactory local minima. Each test is designed to generate simple datasets containing only one wave mode (transmission, reflection or refraction) in order to demonstrate that FWIME can invert any type of seismic data using the same framework and without the need for intensive hyper-parameter tuning.

%%%%%%%%%%%% Loss function design %%%%%%%%%%%
\section{FWIME formulation}
%%%%%%%%%%%%%%%%%%%%%%%%%%%%%%%%%%%%%%%%%%%%%%%%%%%%%%%%%%%%%%%%%%%%%%%%%%%%%%%%%%%%%%
%%%%%%%%%%%%%%%%%%%%%%%%%%%% FWIME formulation %%%%%%%%%%%%%%%%%%%%%%%%%%%%%%%%%%%%%%%
%%%%%%%%%%%%%%%%%%%%%%%%%%%%%%%%%%%%%%%%%%%%%%%%%%%%%%%%%%%%%%%%%%%%%%%%%%%%%%%%%%%%%%
We briefly review the theory of FWI and WEMVA, which are the two main family of algorithms on which FWIME is based. Then, we formally define our proposed objective function and we provide some insight on the design of FWIME. 
\subsection{Full waveform inversion (FWI)}
In a conventional FWI workflow, the following objective function is minimized

\begin{eqnarray}
    \label{eqn:fwi.obj}
    \Phi_{\rm FWI}(\mathbf{m}) &=& \frac{1}{2} \left\| \mathbf{f}(\mathbf{m}) - \mathbf{d}^{obs} \right\|^2_2,
\end{eqnarray}

where $\mathbf{m} \in \mathbb{R}^{N_m}$ is the unknown seismic velocity model (discretized and parametrized on a finite-difference grid), $\mathbf{f}: \; \mathbb{R}^{N_m} \mapsto \mathbb{R}^{N_d}$ is the discretized acoustic isotropic constant-density forward modeling operator conducted for a collection of source/receiver pairs, and $\mathbf{d}^{obs} \in \mathbb{R}^{N_d}$ is the seismic data recorded at a set of receivers' locations. $N_m= N_z \times N_x \times N_y$ is the dimension of the model space, where $N_z$, $N_x$, and $N_y$ are the number of grid points in each spatial direction. $N_d = N_t \times N_{tr}$ is the dimension of the data space, where $N_t$ is the number of time samples per trace, and $N_{tr}$ is the number of traces. In the following, we assume the existence and uniqueness of a true model $\mathbf{m}_t$ such that $\mathbf{f}(\mathbf{m}_t) = \mathbf{d}^{obs}$, which implies that equation~\ref{eqn:fwi.obj} possesses a unique global minimum. Due to the large problem scales encountered in 3D field applications, the minimization of equation~\ref{eqn:fwi.obj} is commonly performed using gradient-based optimization methods. Since $\mathbf{f}$ is nonlinear with respect to $\mathbf{m}$, $\Phi_{\rm FWI}$ usually bears multiple local minima. 

When the initial model $\mathbf{m}_0$ is close enough to the true model $\mathbf{m}_t$, it may belong to a basin of attraction of objective function defined in equation~\ref{eqn:fwi.obj} whose minimum coincides with the global minimum $\mathbf{m}_t$, in which case a gradient-based optimization scheme is sufficient to recover the global minimizer of $\Phi_{\rm FWI}$. Unfortunately, when the initial model is too inaccurate for providing a decent wave propagation simulation (and thus does not lie within the global basin of attraction), a gradient-based optimization scheme may converge to a local minimum corresponding to a non-physical and uninformative representation of the Earth's subsurface. In the data space, some of the events in the predicted data may initially be shifted by more than half of one cycle from their counterpart in the recorded data, which is referred to as ``cycle-skipping". Their cross-correlation contribution to the model update will be zero or in the wrong direction \cite[]{virieux2009overview,yao2019tackling}. An even worse situation occurs when some predicted data attempt to interpret non-corresponding recorded data leading to an apparent misfit reduction. Such wrong association may lead the iterative process into a meaningless local minimum. Throughout this paper, we carefully distinguish the concept of ``cycle-skipping" (a data-space phenomenon) from ``converging to a local minimum" (a model-space phenomenon). 

\subsection{Wave-equation migration velocity analysis (WEMVA)}
WEMVA belongs to a family of techniques aimed at estimating the optimal background velocity model that improves the quality, coherency and focusing of migrated images computed with wave-equation based modeling operators \cite[]{biondi1999wave}. The goal of the optimization process is to minimize objective functions of the following form

\begin{eqnarray}
    \label{eqn:wemva.obj}
    \Phi_{\rm WEMVA}(\mathbf{m}) &=& \frac{1}{2} \left\| \mathbf{E} \left (\tilde{\mathbf{I}}(\mathbf{m}) \right) \right\|^2_2,
\end{eqnarray}

where $\mathbf{m} \in \mathbb{R}^{N_m}$ is the unknown seismic velocity model, and $\tilde{\mathbf{I}} \in \mathbb{R}^{N_{\tilde{p}}}$ is an extended migrated image defined by 

\begin{eqnarray}
    \label{eqn:image.extended}
    \tilde{\mathbf{I}}(\mathbf{m}) &=& \tilde{\mathbf{B}}^*(\mathbf{m}) \mathbf{d}^{ref},
\end{eqnarray}
where $\mathbf{d}^{ref} \in \mathbb{R}^{N_d} $ is a subset of the observed data assumed to be primary reflected events. Operator $\tilde{\mathbf{B}}: \mathbb{R}^{N_{\tilde{p}}} \mapsto \mathbb{R}^{N_d}$ denotes the extended Born modeling operator, and $*$ symbolizes the adjoint process. In this paper, we use the $\sim$ symbol to refer to all extended operations and operators. Extensions typically include horizontal subsurface offsets $\mathbf{h}$, time lags $\tau$, subsurface reflection angles $\gamma$, or seismic sources $\mathbf{s}$ \cite[]{biondi1999wave,biondi2004angle,sava2004wave,biondi2014simultaneous,perrone2014linearized,zhang2015velocity}. When specifically dealing with time lags and subsurface offsets, we define the ``physical space" (or ``physical plane" for 2D applications) as the set of points whose coordinates on the extended axis/axes are zero ($\tau=0$ s and $\mathbf{h}=\mathbf{0}$). $N_{\tilde{p}} = N_m \times N_{ext}$ is the dimension of the extended space, where $N_{ext}$ refers to the extension size (i.e., the number of points on the extended axis/axes). $\mathbf{E}: \mathbb{R}^{N_{\tilde{p}}} \mapsto \mathbb{R}^{N_{\tilde{p}}}$ is an operator that typically measures and enhances the defocusing of $\tilde{\mathbf{I}}$, and computes an image residual used to iteratively update the background velocity model $\mathbf{m}$. Thus, minimizing equation~\ref{eqn:wemva.obj} corresponds to finding an optimal model such that the image defocusing is reduced. Various approaches have been developed to design efficient enhancing operators $\mathbf{E}$ by (1) evaluating the curvature of the subsurface-angle-domain common image gathers (ADCIGs) \cite[]{zhang2015velocity}, by (2) employing the differential semblance optimization (DSO) operator that penalizes the first derivative along the angle axis of ADCIGs, or by (3) measuring the lack of focusing in subsurface-offset extended images \cite[]{symes1991velocity,symes1994inversion,shen2008automatic}. While WEMVA methods tend to be less sensitive to the accuracy of the initial model, their output usually lacks vertical resolution because only the transmission effects of the velocity are used for surface acquisition geometries \cite[]{almomin2016tomographic}.

\subsection{FWIME objective function}
We provide a formal mathematical definition of the FWIME cost function. We begin by the following initial formulation,

\begin{eqnarray}
    \label{eqn:fwime.obj.novp}
    \Psi_{\epsilon}(\mathbf{m},\mathbf{\tilde{p}}) &=& \frac{1}{2} \left\| \mathbf{f}(\mathbf{m}) + \tilde{\mathbf{B}} (\mathbf{m}) \mathbf{\tilde{p}} - \mathbf{d}^{obs} \right\|^2_2 + \frac{\epsilon^2}{2} \left\| \mathbf{D} \tilde{\mathbf{p}} \right\|^2_2,
\end{eqnarray}

where $\mathbf{m} \in \mathbb{R}^{N_m}$ is the unknown discretized seismic velocity model, $\tilde{\mathbf{B}}$ denotes the extended Born modeling operator (whose adjoint is employed and defined in equation~\ref{eqn:image.extended}), and $\tilde{\mathbf{p}} \in \mathbb{R}^{N_{\tilde{p}}}$ is an extended perturbation in either time lags $\tau$ or horizontal subsurface offsets $\mathbf{h}$. $N_{\tilde{p}} = N_m \times N_{ext}$ is the dimension of the extended space. In the FWIME framework, the unknown velocity model $\mathbf{m}$ is never extended. The diagonal matrix $\mathbf{D}$ is a modified invertible modified form of the DSO operator that enhances the energy of the extended perturbation $\tilde{\mathbf{p}}$ \cite[]{symes1991velocity,symes1994inversion}. $\epsilon$ is the trade-off parameter between the two components of $\Psi_{\epsilon}$. Its value is set at the initial step and kept fixed throughout the optimization process. In equation~\ref{eqn:fwime.obj.novp}, $\tilde{\mathbf{B}}$ is linear with respect to $\tilde{\mathbf{p}}$ but nonlinear with respect to the velocity model $\mathbf{m}$, while $\mathbf{D}$ is linear with respect to $\tilde{\mathbf{p}}$. Therefore, $\Psi_{\epsilon}$ is quadratic with respect to $\tilde{\mathbf{p}}$ (for a fixed $\mathbf{m}$). By employing the variable projection method to minimize equation~\ref{eqn:fwime.obj.novp}, we can explicitly express $\tilde{\mathbf{p}}$ as a function of $\mathbf{m}$ and remove it from the dependencies of $\Psi_{\epsilon}$ in equation~\ref{eqn:fwime.obj.novp}. The FWIME objective function may be formally defined:

\begin{eqnarray}
    \boxed{
    \begin{array}{rcl}        
    \label{eqn:fwime.obj}
    \Phi_{\epsilon}(\mathbf{m}) = \dfrac{1}{2} \left\| \mathbf{f}(\mathbf{m}) + \tilde{\mathbf{B}} (\mathbf{m}) \mathbf{\tilde{p}}_{\epsilon}^{opt}(\mathbf{m}) - \mathbf{d}^{obs} \right\|^2_2 + \dfrac{\epsilon^2}{2} \left\| \mathbf{D}{\mathbf{\tilde p}}_{\epsilon}^{opt}(\mathbf{m}) \right\|^2_2.
    \end{array}
    }
\end{eqnarray}

Equation~\ref{eqn:fwime.obj} introduces two modifications compared to equation~\ref{eqn:fwime.obj.novp}. The objective function now solely depends on the velocity model $\mathbf{m}$, and 
the extended perturbation $\tilde{\mathbf{p}}$ has been replaced by $\mathbf{\tilde{p}}_{\epsilon}^{opt} (\mathbf{m})$, which is defined as the minimizer of the following quadratic objective function (for fixed values of $\mathbf{m}$ and $\epsilon$),

\begin{eqnarray}
    \label{eqn:vp.obj}
    \Phi_{\epsilon,\mathbf{m}}(\mathbf{\mathbf{\tilde{p}}}) &=& \frac{1}{2} \left\| \tilde{\mathbf{B}}(\mathbf{m})\mathbf{\tilde{p}}  - \left ( \mathbf{d}^{obs} - \mathbf{f}(\mathbf{m}) \right ) \right\|^2_2 + \frac{\epsilon^2}{2} \left\| \mathbf{D} \tilde{\mathbf{p}} \right\|^2_2.
\end{eqnarray}

The Hessian of $\Phi_{\epsilon,\mathbf{m}}$ is given by the following expression,
\begin{eqnarray}
    \label{eqn:hessian.vp.obj}
    \mathbf{H}_{\Phi_{\mathbf{m},\epsilon}} &=& \tilde{\mathbf{B}}^*(\mathbf{m}) \tilde{\mathbf{B}}(\mathbf{m}) + \epsilon^2 \mathbf{D}^* \mathbf{D}.
\end{eqnarray}

For $\epsilon >0$, $\mathbf{H}_{\Phi_{\mathbf{m},\epsilon}}$ is a positive definite operator as the sum of a positive semi-definite operator $\tilde{\mathbf{B}}^*(\mathbf{m}) \tilde{\mathbf{B}}(\mathbf{m})$ and a positive definite operator, $\epsilon^2 \mathbf{D}^* \mathbf{D}$. Hence, $\Phi_{\epsilon,\mathbf{m}}$ has a unique minimizer (and minimum), referred to as the optimal extended perturbation and denoted by $\tilde{\mathbf{p}}_{\epsilon}^{opt}$, which depends nonlinearly on the velocity model $\mathbf{m}$. Its analytical expression is given by the formal solution of the normal equations

\begin{eqnarray}
    \label{eqn:pert.opt}
    \tilde{\mathbf{p}}_{\epsilon}^{opt}(\mathbf{m}) &=& \tilde{\mathbf{B}}_{\epsilon, \mathbf{D}}^{\dagger}(\mathbf{m}) \left ( \mathbf{d}^{obs} - \mathbf{f}(\mathbf{m}) \right ),
\end{eqnarray}

where $\tilde{\mathbf{B}}_{\epsilon, \mathbf{D}}^{\dagger}$ is the pseudo-inverse of $\tilde{\mathbf{B}}$ in equation~\ref{eqn:vp.obj}:

\begin{eqnarray}
    \label{eqn:vp.pseudo.inv}
    \tilde{\mathbf{B}}_{\epsilon, \mathbf{D}}^{\dagger} (\mathbf{m})&=& \big ( \tilde{\mathbf{B}}^*(\mathbf{m}) \; \tilde{\mathbf{B}}(\mathbf{m}) + \epsilon^2 {\mathbf{D}^*} {\mathbf{D}} \big )^{-1} \tilde{\mathbf{B}}^*(\mathbf{m}). 
\end{eqnarray}

In practice, the minimization of the quadratic inner problem defined in equation~\ref{eqn:vp.obj} is performed iteratively using a linear conjugate-gradient algorithm \cite[]{aster2018parameter}, and is referred to as the variable projection step in FWIME. In this paper, the minimization of the nonlinear problem formulated in equation~\ref{eqn:fwime.obj} is performed with L-BFGS \cite[]{nocedal1980updating}. 

The advantage of recasting the inverse problem expressed in equation~\ref{eqn:fwime.obj.novp} into the problem expressed in equation~\ref{eqn:fwime.obj} with a linear sub-problem using the variable projection method (equation~\ref{eqn:vp.obj}) can be further analyzed from an optimization point of view. Equation~\ref{eqn:fwime.obj.novp} is defined on an increased search space with the use of the additional extended variable $\mathbf{\tilde{p}}$. Since $\Psi_{\epsilon}$ is quadratic with respect to $\mathbf{\tilde{p}}$ for a fixed $\mathbf{m}$, the global minimum of $\Psi_{\epsilon}$ will be reached for the pair $(\mathbf{m}^{opt},\mathbf{\tilde{p}}^{opt})$ where $\mathbf{\tilde{p}}^{opt}$ minimizes the quadratic cost function $\mathbf{\tilde{p}} \mapsto \Psi_{\epsilon}(\mathbf{m}^{opt}, \mathbf{\tilde{p}})$, which corresponds to equation~\ref{eqn:vp.obj}. The main advantage of our compact formulation (equation~\ref{eqn:fwime.obj}) is that we now formally invert for a single physical non-extended parameter $\mathbf{m}$ (representing the unknown seismic velocity) while still benefiting from an extended search space embedded within $\tilde{\mathbf{p}}_{\epsilon}^{opt}$. Consequently, we do need to implement convoluted alternate-direction or gradient-scaling optimization techniques that are usually indispensable for multi-parameter inversions \cite[]{operto2013guided,biondi2014simultaneous,le2019anisotropic}. 

\subsection{$\text{FWIME} \approx \text{FWI}+\text{WEMVA}$}
In this section, we provide a high-level description of the main intuition that led to the design of our objective function. In equation~\ref{eqn:fwime.obj}, we introduce a data-correcting term $\mathbf{g}_{\epsilon}(\mathbf{m})=\tilde{\mathbf{B}} (\mathbf{m}) \mathbf{\tilde{p}}_{\epsilon}^{opt}(\mathbf{m})$ in the data-fitting term of the FWI objective function (equation~\ref{eqn:fwi.obj}), whose goal is to ensure that $\mathbf{f}(\mathbf{m})+\mathbf{g}_{\epsilon}(\mathbf{m})$ fully fits the entire dataset $\mathbf{d}^{obs}$ at any given precision, even for inaccurate initial velocity models $\mathbf{m}$. As we show in the next section, the use of an extended perturbation and extended Born operator is absolutely necessary to ensure that the data can be accurately fit. The conventional FWI loss function (equation~\ref{eqn:fwi.obj}) is modified as follows,

\begin{eqnarray}
    \label{eqn:fwi.g.obj}
    \Phi(\mathbf{m}) &=& \frac{1}{2} \left\| \mathbf{f}(\mathbf{m}) + \mathbf{g}_{\epsilon}(\mathbf{m}) - \mathbf{d}^{obs} \right\|^2_2.
\end{eqnarray}

The right side of equation~\ref{eqn:fwi.g.obj} is referred to as the data-fitting component of FWIME. Assuming that such $\mathbf{g}_{\epsilon}(\mathbf{m})$ exists and that it can be computed, we now have an enhanced non-physical modeling operator $\mathbf{f}+\mathbf{g}_{\epsilon}$ that can match the observed data as accurately as needed. With this additional term, the observed data are not ``cycle-skipped" anymore but there is no guarantee that the objective function is convex nor free of local minima. In fact, if defined as such, the objective function defined in equation~\ref{eqn:fwi.g.obj} is constant and null for all models. We use the necessary condition which requires that for the true model, our data prediction should eventually be generated by $\mathbf{f}$ only, without the need for a correcting term. We add an annihilating component $A(\mathbf{m}) = \left\| \mathbf{D}{\mathbf{\tilde p}}_{\epsilon}^{opt}(\mathbf{m}) \right\|^2_2$ to the objective function that will gradually reduce the contribution of $\mathbf{g}_{\epsilon}$ during the minimization of $\Phi$, while still ensuring data-matching. Therefore, the FWIME objective function can be symbolically expressed in the following form

\begin{eqnarray}
\label{eqn:annihilator.obj}
    \Phi(\mathbf{m})_{\epsilon} &=& \frac{1}{2} \left\| \mathbf{f}(\mathbf{m}) + \mathbf{g}_{\epsilon}(\mathbf{m}) - \mathbf{d}^{obs} \right\|^2_2 + \frac{\epsilon^2}{2} A(\mathbf{m}).
\end{eqnarray}

During the optimization process, the annihilating component gradually forces the $L_2$-norm of $\tilde{\mathbf{p}}_{\epsilon}^{opt}(\mathbf{m})$ to vanish, thereby reducing the contribution of the data-correction term $\tilde{\mathbf{B}} (\mathbf{m}) \mathbf{\tilde{p}}_{\epsilon}^{opt}(\mathbf{m})$. Eventually, if $\mathbf{f}+\mathbf{g}_{\epsilon} \approx \mathbf{d}^{obs}$ and $\mathbf{g}_{\epsilon}$ has vanished, the FWIME inverted model coincides with the global minimum of conventional FWI. In this paper, we do not mathematically prove that our strategy is more suited for gradient-descent optimization methods, but we provide plenty of numerical examples that support this statement. 

An intuitive way to understand the advantage of this new formulation is that the data-correcting term $\mathbf{g}_{\epsilon}(\mathbf{m})$ can be adjusted to reduce and control the relative weight of the data-fitting component with respect to the annihilator in the cost function. In addition, the annihilating component gives the freedom to create judicious operators based on WEMVA techniques, which have been heuristically observed to guide the inversion to the global minimum in a more robust manner \cite[]{symes1991velocity,symes1994inversion,stolk2002smooth}. The rate at which the annihilator reduces the contribution of the data-correcting term is controlled by the value of $\epsilon$ and is key to ensure smooth convergence \cite[]{fu2017discrepancy}. Finally, equation~\ref{eqn:annihilator.obj} can be interpreted as the sum of two terms: a data-fitting component which is a modified FWI problem where the forward modeling combines physical wave propagation with an unphysical additional term, and an annihilating component that possesses similar features as a WEMVA objective function:

\begin{eqnarray}
    \boxed {
        \begin{array}{rcl}   
            \label{eqn:data-fit.and.annihilator.obj}
            \text{FWIME} \approx \text{FWI} + \text{WEMVA}.
        \end{array}
    }        
\end{eqnarray}

%%%%%%%% Dissection of loss function %%%%%%%%
\section{Dissection of the FWIME objective function}
%%%%%%%%%%%%%%%%%%%%%%%%%%%%%%%%%%%%%%%%%%%%%%%%%%%%%%%%%%%%%%%%%%%%%%%%%%%%%%%%%%%%%%
%%%%%%%%%%%%%%%%%%%%%%%%%%%% Objective function dissection %%%%%%%%%%%%%%%%%%%%%%%%%%%
%%%%%%%%%%%%%%%%%%%%%%%%%%%%%%%%%%%%%%%%%%%%%%%%%%%%%%%%%%%%%%%%%%%%%%%%%%%%%%%%%%%%%%
In this section, we provide a thorough analysis of the four constitutive blocks of FWIME: (1) the data-correcting term, (2) the extended optimal perturbation, (3) the annihilator, and (4) the trade-off parameter. We illustrate how they are effectively computed on a 2D numerical example based on the Marmousi2 model \cite[]{martin2006marmousi2}. 

%%%%%%%%%%%%%%%%%%%%%%%%%%%%%%% Data-correcting term %%%%%%%%%%%%%%%%%%%%%%%%%%%%%%%%%
\subsection{The data-correcting term: a powerful tool}
\label{data_correctin_term_subsection}
When the current velocity model estimate $\mathbf{m}$ is inaccurate, some events present in the recorded data $\mathbf{d}^{obs}$ may be incorrectly modeled (or even missed) by the conventional forward mapping $\mathbf{f}(\mathbf{m})$. The goal of the data-correcting term is to predict these missing events contained in the residual $\mathbf{d}^{obs}-\mathbf{f}(\mathbf{m})$ by combining the concept of extended Born modeling with the variable projection method. Mathematically, this task can be expressed by finding an extended perturbation $\tilde{\mathbf{p}}$ such that 

\begin{eqnarray}
    \label{eqn:data.matching}
\tilde{\mathbf{B}} (\mathbf{m}) \tilde{\mathbf{p}} &\approx& \mathbf{d}^{obs} - \mathbf{f}(\mathbf{m}),
\end{eqnarray}

which can be recast into an optimization problem where the goal is to find the minimizer $\tilde{\mathbf{p}}_0^{opt}$ of the following cost function,

\begin{eqnarray}
    \label{eqn:data.matching.least.squares}
    \Phi_{0,\mathbf{m}}(\mathbf{\mathbf{\tilde{p}}}) &=& \frac{1}{2} \left\| \tilde{\mathbf{B}}(\mathbf{m})\mathbf{\tilde{p}}  - \big ( \mathbf{d}^{obs} - \mathbf{f}(\mathbf{m}) \big ) \right\|^2_2.
\end{eqnarray}

Equation~\ref{eqn:data.matching.least.squares} is indeed a specific case of the FWIME variable projection step (equation~\ref{eqn:vp.obj}) for $\epsilon = 0$. It has been numerically observed that for an appropriate extended Born modeling operator and extension type, the minimizer $\tilde{\mathbf{p}}_0^{opt}$ of equation~\ref{eqn:data.matching.least.squares} exists and satisfies equation~\ref{eqn:data.matching} up to numerical precision, regardless of the accuracy of $\mathbf{m}$ \cite[]{symes2008migration,barnier2020full}. That is, $\Phi_{0,\mathbf{m}}(\tilde{\mathbf{p}}_0^{opt})=0$. This result means that we can always find/compute a data-correcting that can accurately predict any type of events from the observed data that the physical wave-equation modeling operator $\mathbf{f}(\mathbf{m})$ failed to capture (if we set $\epsilon = 0$). 

In practice, setting $\epsilon = 0$ is not useful because the FWIME objective function (equation~\ref{eqn:fwime.obj}) would be constant and null for all models $\mathbf{m}$. In fact, the $\epsilon$-value is initially adjusted to control the level of data-fitting. Low $\epsilon$-values will lead the data-correcting term to better satisfy equation~\ref{eqn:data.matching}, and vice-versa. While extended modeling is a powerful tool, its main downside remains the computational cost: the estimation of $\tilde{\mathbf{p}}_{\epsilon}^{opt}$ (i.e., the minimization of equations~\ref{eqn:vp.obj} or \ref{eqn:data.matching.least.squares}) is equivalent to conducting an extended linearized waveform inversion \cite[]{leader2015separation}. Moreover, the more inaccurate $\mathbf{m}$ is, the larger the extended space is required to be for the data-correcting term to satisfy equation~\ref{eqn:data.matching}.

\begin{figure}[t]
    \centering
    \label{fig:data_correcting_model_true}\includegraphics[width=0.45\linewidth]{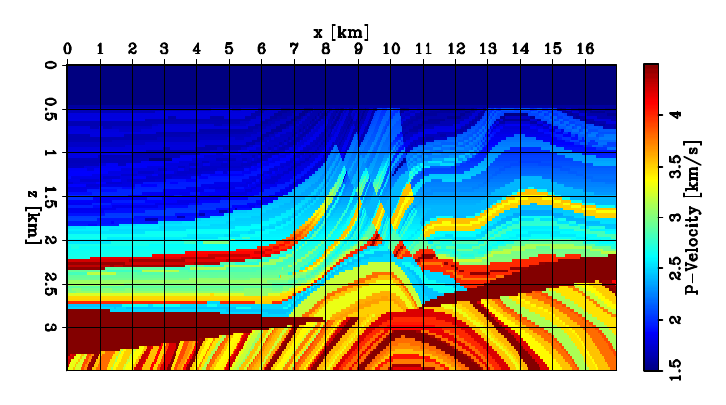}
    \label{fig:data_correcting_model_init}\includegraphics[width=0.45\linewidth]{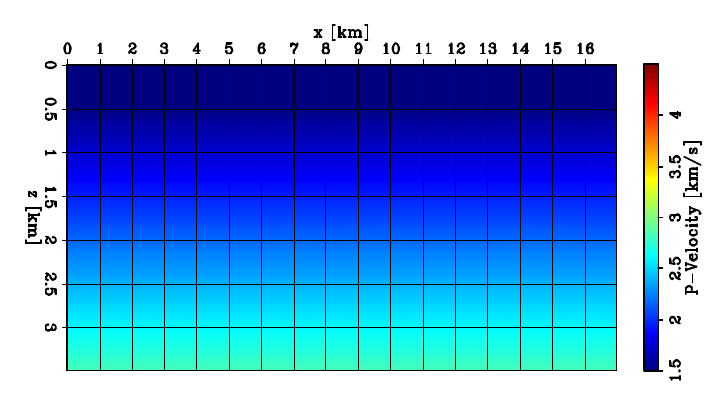}\\
    \caption{2D panels of velocity models. (a) Marmousi2 model. (b) Initial model $\mathbf{m}_0$. }
    \label{fig:data_correcting_model}
\end{figure}

\begin{figure}[t]
    \centering
    \label{fig:data_correcting_model_1d_5km}\includegraphics[width=0.3\linewidth]{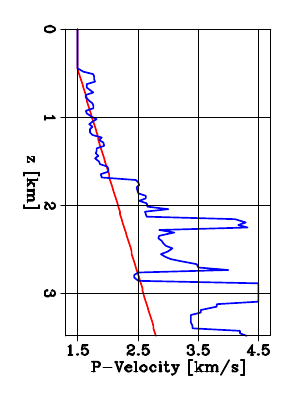} \hspace{20mm}
    \label{fig:data_correcting_model_1d_10km}\includegraphics[width=0.3\linewidth]{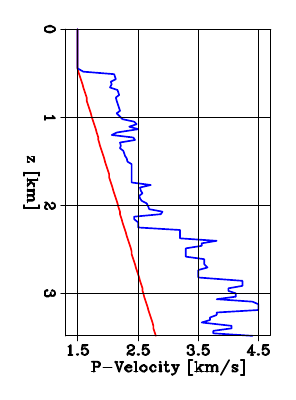}
    \caption{Depth velocity profiles extracted at (a) $x = 5$ km and (b) $x = 10$ km. The blue and red curves represent the true and initial models, respectively.}
    \label{fig:data_correcting_model_1d}
\end{figure}

We compute and show the existence of the data-correcting term on a numerical example based on the Marmousi2 model (Figure~\ref{fig:data_correcting_model_true}) when $\mathbf{m}$ is extremely inaccurate. We generate noise-free pressure data with a two-way acoustic finite-difference propagator using a grid-spacing of 30 m in both directions. At the surface, we place 140 sources every 120 m, and 567 receivers every 30 m. Data are modeled with a wavelet containing energy restricted to the 4-13 Hz frequency range, and are recorded for 7 s. The initial model $\mathbf{m}_0$ (Figure~\ref{fig:data_correcting_model_init}) is laterally invariant and linearly increasing with depth (with a 500 m water layer on top). Figure~\ref{fig:data_correcting_model_1d} shows two velocity profiles extracted at $x = 5$ km and $x = 10$ km, respectively. 

% 2D velocity model 

% 1D velocity model 

\begin{figure}[tbhp]
    \centering
    \label{fig:data_correcting_dat_true}\includegraphics[width=0.3\linewidth]{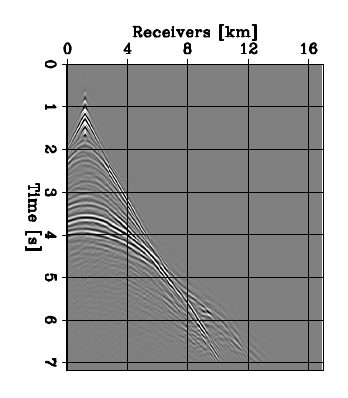}
    \label{fig:data_correcting_dat_init}\includegraphics[width=0.3\linewidth]{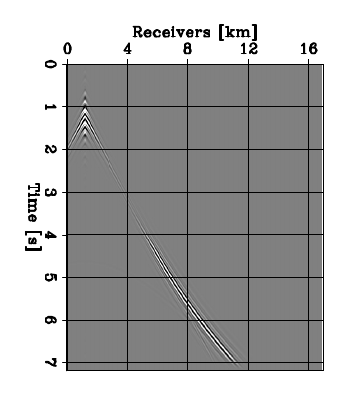}
    \label{fig:data_correcting_dat_init_diff}\includegraphics[width=0.3\linewidth]{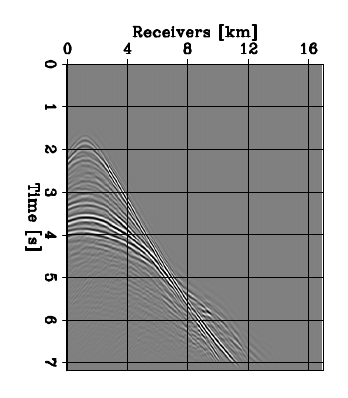}    
    \caption{Shot gathers generated by a source located at $x = 1.2$ km. (a) Observed data, $\mathbf{d}^{obs}$. (b) Predicted data with the initial model, $\mathbf{f}(\mathbf{m}_0)$. (c) Initial data residual, $\mathbf{d}^{obs}-\mathbf{f}(\mathbf{m}_0)$. All panels are displayed with the same grayscale.}
    \label{fig:data_correcting_data_obs}
\end{figure}

Figure~\ref{fig:data_correcting_dat_true} shows a representative shot gather of the observed data, $\mathbf{d}^{obs}$, generated by a source placed at $x = 1.2$ km. The recorded data contain both reflected and refracted energy. Figure~\ref{fig:data_correcting_dat_init} displays the analogous shot gather computed with the initial model, $\mathbf{f}(\mathbf{m}_0)$, which fails to predict most of the events in the observed data, and Figure~\ref{fig:data_correcting_dat_init_diff} is the initial data residual, $\mathbf{d}^{obs}-\mathbf{f}(\mathbf{m}_0)$. 

% Initial data

\begin{figure}[t]
    \centering
    \includegraphics[width=0.5\linewidth]{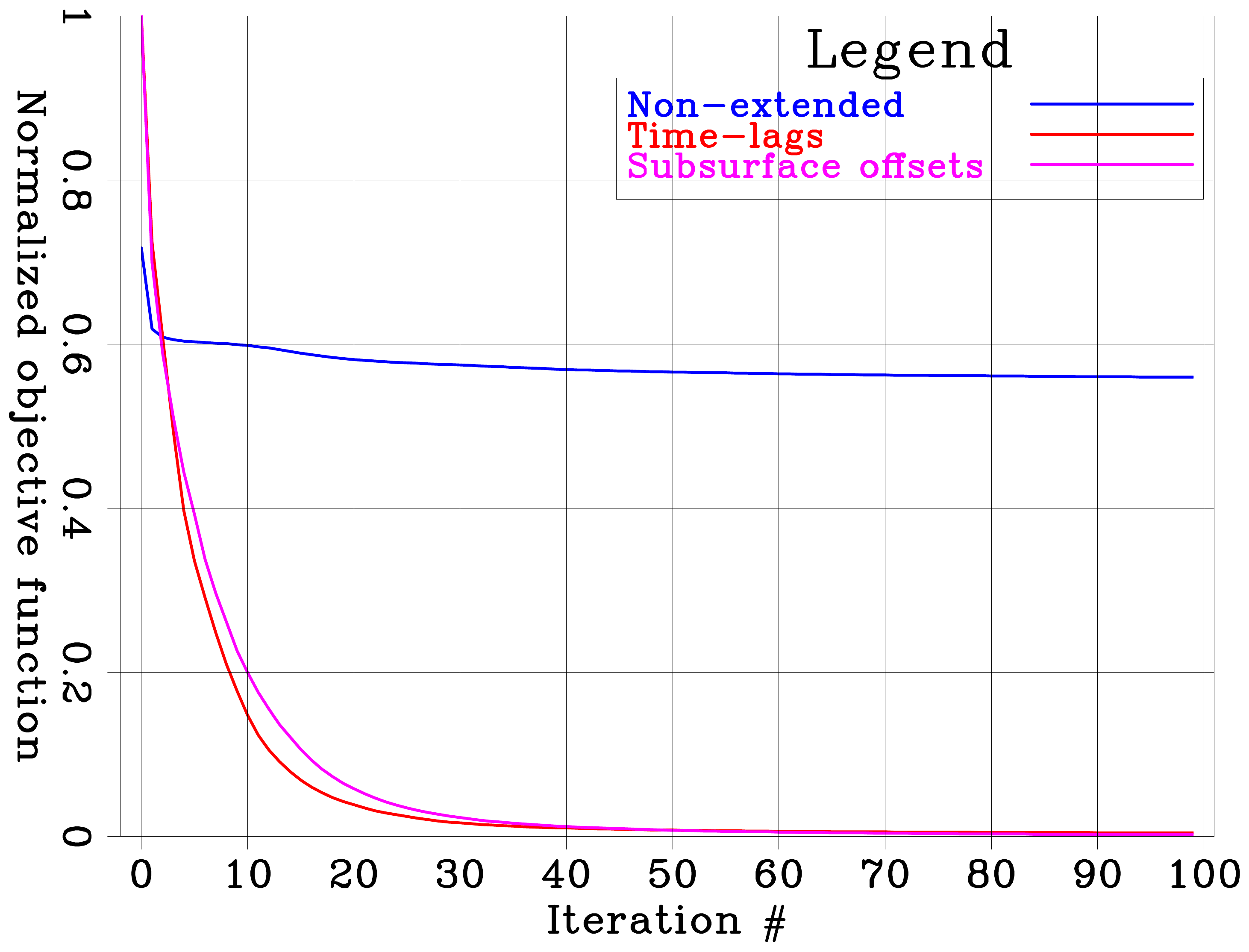}
    \caption{Normalized objective functions corresponding to the minimization of equation~\ref{eqn:vp.obj} with three different Born operators. Non-extended Born (blue curve), time-lag extension (red curve), and horizontal subsurface offset extension (pink curve).}
    \label{fig:vp_obj}
\end{figure}

We set $\epsilon = 0$ and we minimize equation~\ref{eqn:vp.obj} (variable projection) with 100 iterations of linear conjugate gradient using three different forms of $\tilde{\mathbf{B}}$: (1) non-extended (conventional Born), (2) extended with time lags $\tau$, and (3) extended with horizontal subsurface offsets $h_x$. For both time lags and subsurface offsets, we use 141 points of extension which allows $\tau$ to range from -1.12 s to 1.12 s, and $h_x$ to range from -2.1 km to 2.1 km. In this example, a large number of extended points are needed to account for the (unrealistic) inaccuracy of the initial velocity model $\mathbf{m}_0$. The convergence curves for these optimizations are shown in Figure~\ref{fig:vp_obj}. Both time-lag (red curve) and horizontal subsurface offset (pink curve) extensions manage to reduce the objective function value by more than $99.5\%$, and a more accurate matching can be obtained by conducting more iterations. Not surprisingly, the conventional (non-extended) Born operator is unable to match the data misfit (blue curve). 

% Vp objective functions

\begin{figure}[t]
    \centering
    \label{fig:data_correcting_vp_init}\includegraphics[width=0.3\linewidth]{Fig/sep20-dataInitDiff.pdf}
    \label{fig:data_correcting_vp_pred}\includegraphics[width=0.3\linewidth]{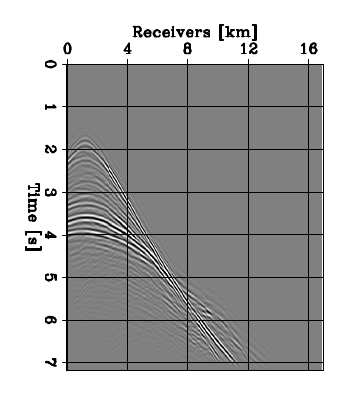}
    \label{fig:data_correcting_vp_error}\includegraphics[width=0.3\linewidth]{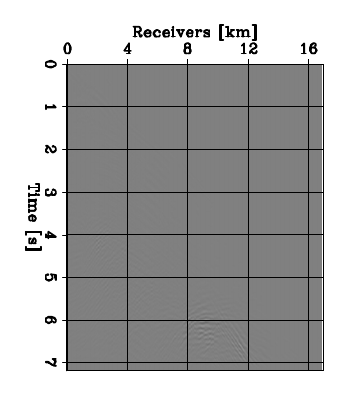} 
    \caption{Shot gathers generated by a source located at $x = 1.2$ km. (a) Initial data misfit, $\mathbf{d}^{obs} - \mathbf{f}(\mathbf{m}_0)$. (b) FWIME data-correcting term computed using a horizontal subsurface-offset extension, $\tilde{\mathbf{B}}(\mathbf{m}_0)\tilde{\mathbf{p}}_{0}^{opt}(\mathbf{m}_0)$. (c) Difference between the data-correcting term and the initial data-residual, $\tilde{\mathbf{B}}(\mathbf{m}_0)\tilde{\mathbf{p}}_{0}^{opt}(\mathbf{m}_0) - \big ( \mathbf{d}^{obs} - \mathbf{f}(\mathbf{m}_0) \big )$. All panels are displayed with the same grayscale.}
    \label{fig:data_correcting_vp}
\end{figure}

\begin{figure}[t]
    \centering
    \label{fig:data_correcting_vp_init_noExt}\includegraphics[width=0.3\linewidth]{Fig/sep20-dataInitDiff.pdf}
    \label{fig:data_correcting_vp_pred_noExt}\includegraphics[width=0.3\linewidth]{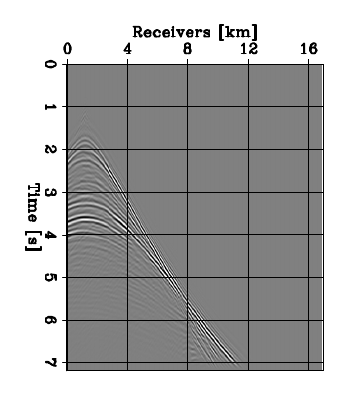}
    \label{fig:data_correcting_vp_error_noExt}\includegraphics[width=0.3\linewidth]{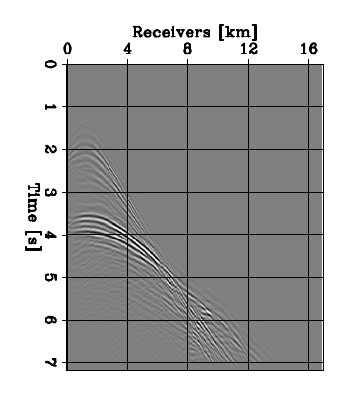}  
    \caption{Shot gathers generated by a source located at $x = 1.2$ km. (a) Initial data misfit, $\mathbf{d}^{obs} - \mathbf{f}(\mathbf{m}_0)$. (b) FWIME data-correcting term computed using a non-extended Born operator, $\mathbf{B}(\mathbf{m}_0)\mathbf{p}_{0}^{opt}(\mathbf{m}_0)$. (c) Difference between the data-correcting term and the initial data-residual, $\mathbf{B}(\mathbf{m}_0) \mathbf{p}_{0}^{opt}(\mathbf{m}_0) - \big ( \mathbf{d}^{obs} - \mathbf{f}(\mathbf{m}_0) \big )$. All panels are displayed with the same grayscale.}
    \label{fig:data_correcting_vp_noExt}
\end{figure}

Most importantly, the effectiveness of extended modeling can be appreciated in the data space. Figure~\ref{fig:data_correcting_vp_init} shows a shot gather extracted from the initial data residual, $\mathbf{d}^{obs}-\mathbf{f}(\mathbf{m}_0)$. Figure~\ref{fig:data_correcting_vp_pred} is the FWIME data-correcting term, $\tilde{\mathbf{B}}(\mathbf{m}_0)\tilde{\mathbf{p}}_{0}^{opt}(\mathbf{m}_0)$, computed after minimization of equation~\ref{eqn:vp.obj} with a horizontal subsurface-offset extension. This result indicates that with the use of an extended linear Born modeling operator, the data-correcting term is able to capture all the events (with the correct amplitudes) that were missed by the inaccurate nonlinear prediction shown in Figure~\ref{fig:data_correcting_vp_init}, even when the velocity model is totally incorrect. Finally, Figure~\ref{fig:data_correcting_vp_error} is the difference between the data-correcting term and the initial data residual,  $\tilde{\mathbf{B}}(\mathbf{m}_0)\tilde{\mathbf{p}}_{0}^{opt}(\mathbf{m}_0) - \big ( \mathbf{d}^{obs} - \mathbf{f}(\mathbf{m}_0) \big )$, which is numerically close to zero. Figure~\ref{fig:data_correcting_vp_noExt} shows analogous panels for the non-extended optimization. As expected, the data-correcting term stemming from the non-extended Born operator fails to match most of the deeper reflections and the refracted energy at larger offsets (Figure~\ref{fig:data_correcting_vp_error_noExt}). 

% Data-correcting term with extension

% Data-correcting term without extension

%%%%%%%%%%%%%%%%%%%%%%% The optimal extended perturbation %%%%%%%%%%%%%%%%%%%%%%%%%%%%%
\subsection{The optimal extended perturbation}
\label{popt}
In FWIME, the optimal extended perturbation $\tilde{\mathbf{p}}_{\epsilon}^{opt}$ is the output of a linear mapping of the conventional FWI data residual into an extended (non-physical) model space (according to equation~\ref{eqn:pert.opt}). Even though computing $\tilde{\mathbf{p}}_{\epsilon}^{opt}$ is mechanically and computationally equivalent to conducting an extended least-squares reverse-time migration (ELSRTM), its purpose is fundamentally different. ELSRTM is driven by physics, and only a subset of the recorded data, the primary reflected events $\mathbf{d}^{ref}$, is inverted. The aim is to find a coherent, focused, and geologically consistent extended image $\mathbf{\tilde{I}}^{opt}$ such that its demigration produces a set of reflections that match the ones present in the recorded data \cite[]{leader2015separation}. This process is done by minimizing the following quadratic function,

\begin{eqnarray}
    \label{eqn:elsrtm}
    \varphi_{\mathbf{m}}(\mathbf{\tilde{I}}) &=& \frac{1}{2} \left\| \tilde{\mathbf{B}}(\mathbf{m})\mathbf{\tilde{I}}  - \mathbf{d}^{ref} \right\|^2_2,
\end{eqnarray}

where $\mathbf{m} \in \mathbb{R}^{N_m}$ is the (fixed) background velocity model, and $\mathbf{\tilde{I}} \in \mathbb{R}^{N_{\tilde p}}$ is an extended image expected to only contain short-wavelength components (e.g., seismic reflectors mapping interfaces between rocks layers). Low-wavenumber features may arise with the use of two-way wave-equation propagators, but are usually considered noise and removed with various techniques \cite[]{youn2001depth,fletcher2005suppressing,guitton2007least,fei2010blending,liu2011effective}. The optimization problem defined in equation~\ref{eqn:elsrtm} can also be regularized to incorporate a priori geological subsurface information and mitigate the effects of uneven illumination patterns in the migrated image \cite[]{prucha2004imaging}.

In FWIME, $\tilde{\mathbf{p}}_{\epsilon}^{opt}$ is computed with the same mathematical mechanism as in equation~\ref{eqn:elsrtm}, and may therefore share similar features with a conventional extended image \cite[]{sava2006time}. However, there are two main differences. During the FWIME variable projection step, the full data mismatch $\mathbf{d}^{obs} - \mathbf{f}(\mathbf{m})$ is inverted (equation~\ref{eqn:vp.obj}). This term may include all types of waves such as transmissions, refractions, and (but not only) reflections, as shown in the previous numerical example in Figure~\ref{fig:data_correcting_dat_init_diff}. The mapping of this data mismatch from the data space into $\tilde{\mathbf{p}}_{\epsilon}^{opt}$ will potentially introduce low- and high-wavenumber events with certain characteristics in the extended space that can provide quantitative information about the errors in the current velocity model estimate $\mathbf{m}$. Thus no filtering nor restriction on the wavenumber content of $\mathbf{p}_{\epsilon}^{opt}$ should be applied at any stage of the FWIME workflow (we want to keep all the information within $\mathbf{p}_{\epsilon}^{opt}$). Consequently, $\mathbf{p}_{\epsilon}^{opt}$ possesses the same the dimensions as an extended image, but serves as a metric in the extended-model space to assess the error between the physical prediction $\mathbf{f}(\mathbf{m})$ and the observed data $\mathbf{d}^{obs}$. Complex overlapping events present in the data space can be mapped and more easily untangled into the extended space of $\mathbf{p}_{\epsilon}^{opt}$.

Moreover, in a noise-free environment and assuming the FWIME optimization scheme can converge to the unique global minimum $\mathbf{m}_t$, the data-prediction error $\mathbf{d}^{obs}-\mathbf{f}(\mathbf{m}_t)$ should eventually vanish, which implies that $\tilde{\mathbf{p}}_{\epsilon}^{opt}(\mathbf{m}_t)$ must also vanish (according to equation~\ref{eqn:pert.opt}). Hence, unlike the output of conventional ELSRTM or previously proposed velocity-estimation methods \cite[]{biondi2004angle,sava2004wave,biondi2014simultaneous}, the ultimate goal of $\tilde{\mathbf{p}}_{\epsilon}^{opt}$ is not to become a focused or enhanced image of the subsurface, and no physical meaning is assigned to this variable. It is simply a tool used to model the events in the recorded data that were missed by the physical prediction $\mathbf{f}(\mathbf{m})$.

\begin{figure}[t]
    \centering
    \label{fig:offset_xcig5}\includegraphics[width=0.23\linewidth]{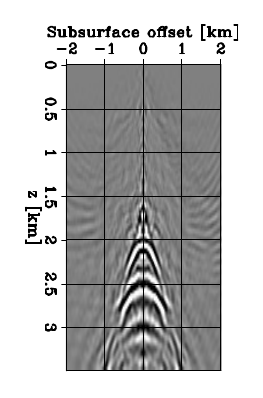}
    \label{fig:offset_xcig8}\includegraphics[width=0.23\linewidth]{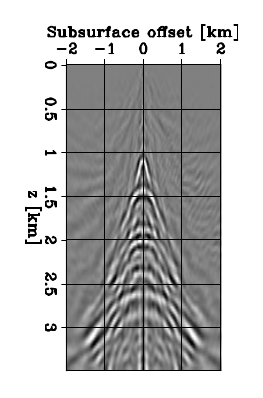}
    \label{fig:offset_xcig10}\includegraphics[width=0.23\linewidth]{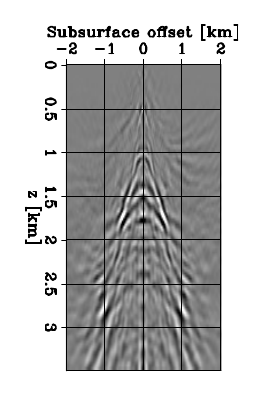}
    \label{fig:offset_xcig13}\includegraphics[width=0.23\linewidth]{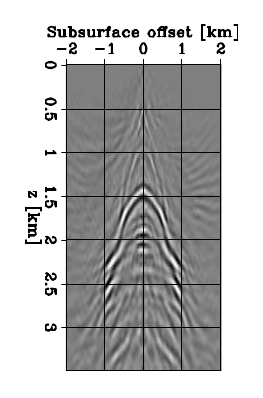}
    \caption{SOCIGs extracted at four horizontal positions from $\tilde{\mathbf{p}}_{\epsilon}^{opt}(\mathbf{m}_0)$. (a) $x=5.0$ km, (b) $x=8$ km, (c) $x=10$ km, and (d) $x=13$ km. All panels are displayed with the same grayscale.}
    \label{fig:offset_x_cig}
\end{figure}

\begin{figure}[t]
    \centering
    \label{fig:time_xcig5}\includegraphics[width=0.23\linewidth]{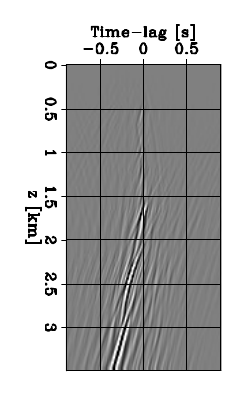}
    \label{fig:time_xcig8}\includegraphics[width=0.23\linewidth]{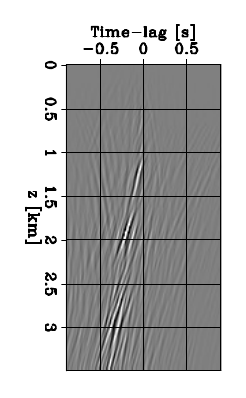}
    \label{fig:time_xcig10}\includegraphics[width=0.23\linewidth]{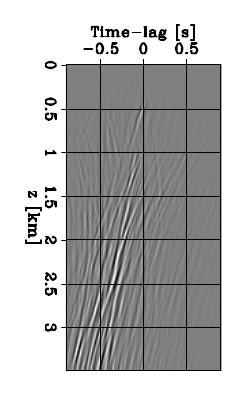}
    \label{fig:time_xcig13}\includegraphics[width=0.23\linewidth]{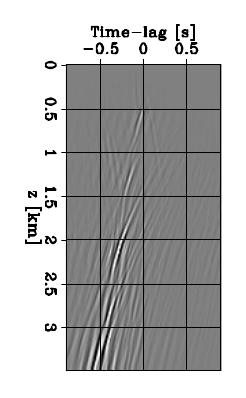}
    \caption{TLCIGs extracted at four horizontal positions from $\tilde{\mathbf{p}}_{\epsilon}^{opt}(\mathbf{m}_0)$. (a) $x=5.0$ km, (b) $x=8$ km, (c) $x=10$ km, and (d) $x=13$ km. All panels are displayed with the same grayscale.}
    \label{fig:time_x_cig}
\end{figure}

\begin{figure}[t]
    \centering
    \label{fig:offset_zcig1.5}\includegraphics[width=0.23\linewidth]{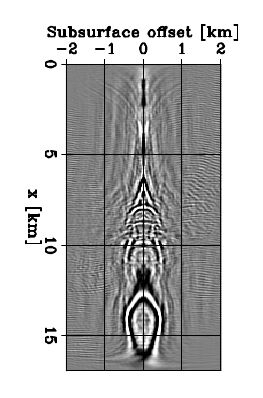}
    \label{fig:offset_zcig2}\includegraphics[width=0.23\linewidth]{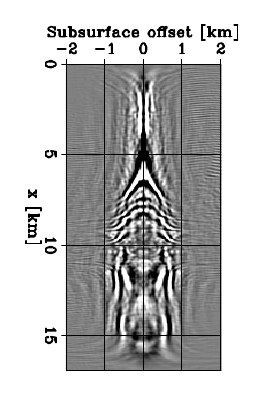}
    \label{fig:offset_zcig2.5}\includegraphics[width=0.23\linewidth]{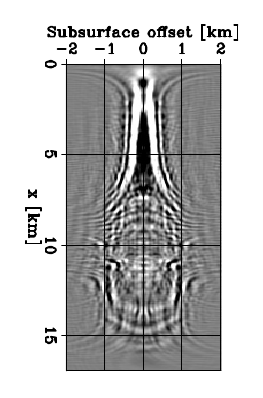}
    \label{fig:offset_zcig3}\includegraphics[width=0.23\linewidth]{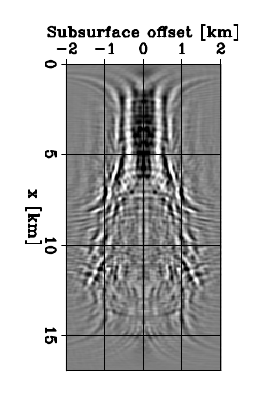}
    \caption{Constant-depth planes extracted from $\tilde{\mathbf{p}}_{\epsilon}^{opt}(\mathbf{m}_0)$ computed with a horizontal subsurface-offset extension $h_x$ at four depths. (a) $z = 1.5$ km, (b) $z=2$ km, (c) $z=2.5$ km, and (d) $z=3$ km. All panels are displayed with the same grayscale.}
    \label{fig:offset_z_cig}
\end{figure}

\begin{figure}[t]
    \centering
    \label{fig:time_zcig1.5}\includegraphics[width=0.23\linewidth]{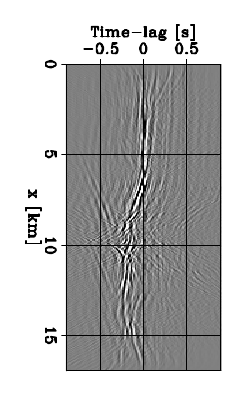}
    \label{fig:time_zcig2}\includegraphics[width=0.23\linewidth]{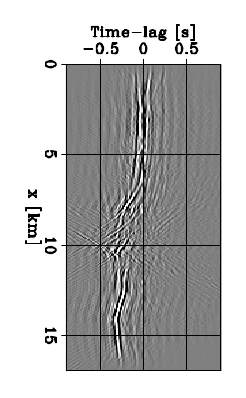}
    \label{fig:time_zcig2.5}\includegraphics[width=0.23\linewidth]{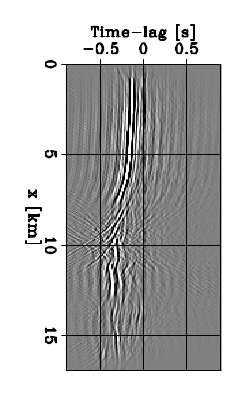}
    \label{fig:time_zcig3}\includegraphics[width=0.23\linewidth]{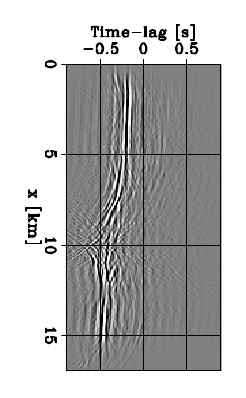}
    \caption{Constant-depth planes extracted from $\tilde{\mathbf{p}}_{\epsilon}^{opt}(\mathbf{m}_0)$ computed with a time-lag extension $\tau$ at four depths. (a) $z = 1.5$ km, (b) $z=2$ km, (c) $z=2.5$ km, and (d) $z=3$ km. All panels are displayed with the same grayscale.}
    \label{fig:time_z_cig}
\end{figure}

We re-visit the Marmousi2 example from the previous section. We examine the features of $\tilde{\mathbf{p}}_{\epsilon}^{opt}$ computed with $\epsilon = 0$ and with the initial velocity model $\mathbf{m}_0$ (Figure~\ref{fig:data_correcting_model_init}). Figures~\ref{fig:offset_x_cig} and \ref{fig:time_x_cig} show common image gathers (CIG) extracted from $\tilde{\mathbf{p}}_{\epsilon}^{opt}$ at four horizontal positions computed with a subsurface-offset extension (such CIG is then referred to as a subsurface-offset CIG or a SOCIG) and a time-lag extension (TLCIG), respectively. Clearly, $\tilde{\mathbf{p}}_{\epsilon}^{opt}(\mathbf{m}_0)$ possesses similar features as conventional extended images: for subsurface offsets and time lags, some coherent clusters of energy (corresponding to the mapping of events contained within $\mathbf{d}^{obs}-\mathbf{f}(\mathbf{m}_0)$) are located away from the physical plane, especially towards greater depths where the velocity error is the largest. Such events can easily be interpreted by examining SOCIGs (Figure~\ref{fig:offset_x_cig}), where the frowning moveout indicates that the velocity used for propagation is too low, a behavior commonly observed in conventional extended images \cite[]{biondi20063d}. A similar observation can be made by examining TLCIGs, where the energy clusters are positioned at negative time lags (Figure~\ref{fig:time_x_cig}). The compounding effect of the kinematic errors over depth can also be detected by analyzing constant-depth planes extracted from $\tilde{\mathbf{p}}_{\epsilon}^{opt}$ for both types of extensions, as shown in Figures~\ref{fig:offset_z_cig} and \ref{fig:time_z_cig}.

%CIGs offsets

% CIGs time

% z-CIGs offset 

% z-CIGs time 

%%%%%%%%%%%%%%%%%%%%%%%%%%%% The annihilating component %%%%%%%%%%%%%%%%%%%%%%%%%%%%%%%
\subsection{The annihilating component}
By adding a data-correcting term, we effectively created an enhanced non-physical modeling operator $\mathbf{f}(\mathbf{m}) + \tilde{\mathbf{B}} (\mathbf{m}) \mathbf{\tilde{p}}_{\epsilon}^{opt}(\mathbf{m})$
and we showed that when we set $\epsilon=0$, we could satisfy $\mathbf{f}(\mathbf{m}) + \tilde{\mathbf{B}} (\mathbf{m}) \mathbf{\tilde{p}}_{\epsilon}^{opt}(\mathbf{m}) \approx \mathbf{d}^{obs}$. With this new non-physical modeling operator, the predicted data are not cycle-skipped but there is no guarantee that the FWIME objective function possesses a wider basin of attraction about the global minimum. A necessary condition to guide the inversion to the optimal solution is that the contribution of the data-correcting term should be gradually reduced during the inversion process. This condition can be achieved by forcing the $L_2$-norm of $\mathbf{p}_{\epsilon}^{opt}$ to vanish, while still ensuring that equation~\ref{eqn:data.matching} remains satisfied.

\begin{figure}[t]
    \centering
    \label{fig:time_zcig1.5}\includegraphics[width=0.5\columnwidth]{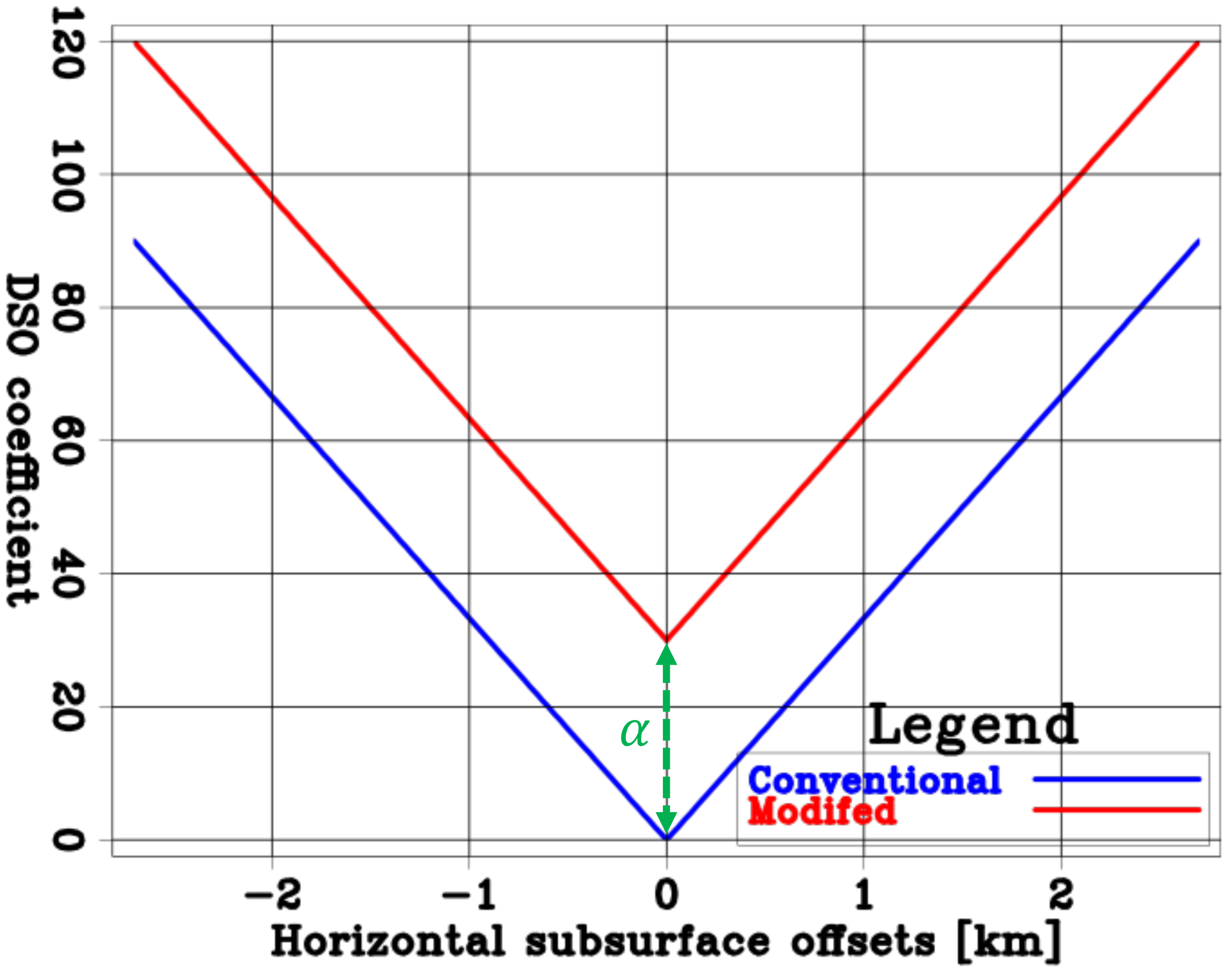}
    \caption{Absolute value of the penalty function for the conventional (blue curve) and modified (red curve) DSO operators applied to a fixed point $M(z_M, x_M, h_x)$, where $h_x$ is the horizontal subsurface-offset value. An analogous graph can be plotted for time-lag extensions.}
    \label{fig:dso_penalty}
\end{figure}

We add an annihilating component (second component on the right side of equation~\ref{eqn:fwime.obj}) which employs a modified form of the DSO operator $\mathbf{D}$, first proposed by \cite{symes1994inversion}. This operator has been extensively and successfully used for computing image residuals in MVA algorithms \cite[]{sava2004wave,symes2008migration}. It enhances features in the extended migrated images created by the presence of errors in the velocity model. For horizontal subsurface offset and time-lag extended images, it is a diagonal operator that multiplies each point of the extended image by a value proportional to its distance to the physical plane. Embedded into a MVA workflow, it rewards images having most of their energy focused in the vicinity of the zero-subsurface and zero time-lag planes. \cite{stolk2002smooth} demonstrated the DSO's ability to ``yield optima that are robust against large errors in the initial model estimates." Therefore, we take advantage of such beneficial properties to extract kinematic information from $\mathbf{p}_{\epsilon}^{opt}$ and guide the inversion towards the optimal solution. However, since the goal is not to obtain a well-focused image (but to make $\mathbf{p}_{\epsilon}^{opt}$ vanish), we modify the DSO operator by also penalizing energy located on the physical plane of $\mathbf{p}_{\epsilon}^{opt}$. Figure~\ref{fig:dso_penalty} shows the absolute value of the coefficients in the conventional (blue curve) and modified (red curve) DSO operators (for a subsurface-offset extension) at a given point $M$ within $\mathbf{p}_{\epsilon}^{opt}$ as function of its distance $|h_x|$ to the physical plane. An analogous penalty function is employed for the time-lag extension. The modification of operator $\mathbf{D}$ is important as it ensures that the FWIME and the conventional FWI objective functions share the same minimum (Appendix~\ref{sameMinimum}). 

% DSO penalty

%%%%%%%%%%%%%%%%%%%%%%%%%%%%%% The trade-off parameter %%%%%%%%%%%%%%%%%%%%%%%%%%%%%%%%
\subsection{The trade-off parameter $\epsilon$}
\label{tradeoff_parameter}
In FWIME, $\epsilon$ is an important hyper-parameter to select as its value affects the shape of the objective function. This hyper-parameter is set at the initial step and is fixed throughout the optimization process. It allows us to control the level of data fitting (with the data-correcting term) by adjusting the penalty applied to $\tilde{\mathbf{p}}_{\epsilon}^{opt}$. During the minimization of equations~\ref{eqn:fwime.obj} and \ref{eqn:vp.obj}, low $\epsilon$-values will impose less constraint on the annihilating component, thereby allowing more energy to be mapped into $\tilde{\mathbf{p}}_{\epsilon}^{opt}$, even far from the physical plane. This behavior allows the data-correcting term to match the data misfit $\mathbf{d}^{obs}-\mathbf{f}(\mathbf{m})$ (i.e., satisfy equation~\ref{eqn:data.matching}) with more accuracy. As we previously showed, by setting $\epsilon=0$, the FWIME objective function is constant and numerically close to zero for all numerically reasonable velocity models $\mathbf{m}$. Therefore, very low $\epsilon$-values are not optimal and may slow down the convergence. Conversely, high $\epsilon$-values will reward minimizing the annihilating component rather than the data-fitting component, thereby not mitigating the cycle-skipping effect, which may lead FWIME to converge to a local minimum. When $\epsilon$ tends to infinity, the $L_2$-norm of $\mathbf{p}_{\epsilon}^{opt}$ converges to zero and FWIME is mathematically equivalent to FWI (Appendix~\ref{fwimeToFwi}). In this paper, we do not develop a mathematical method to select an optimal $\epsilon$-value, but we use a trial and error approach based on examining a subset of the CIGs extracted from the optimal extended perturbation $\tilde{\mathbf{p}}_{\epsilon}^{opt}$ computed at the initial step. Fortunately, for 2D numerical tests and 3D field applications, we observe that our results are relatively insensitive to the choice of $\epsilon$ as long as the proper order of magnitude is determined. 

Additionally, \cite{fu2017discrepancy} show that adjusting the trade-off parameter throughout the inversion process can potentially increase its efficiency. In FWIME, we purposely choose to keep $\epsilon$ fixed as an effort to reduce the need for human input. As a result, FWIME is formulated in a compact and mathematically consistent manner that only requires a simple tuning of one hyper-parameter at the initial step. This feature makes our approach easier to apply, which can potentially impact a broader range of non-expert users.

%%%%%%%%%%%%%%%% Optimization %%%%%%%%%%%%%%%
\section{Optimization: a model-space multi-scale approach}
\label{optimization}
%%%%%%%%%%%%%%%%%%%%%%%%%%%%%%%%%%%%%%%%%%%%%%%%%%%%%%%%%%%%%%%%%%%%%
%%%%%%%%%%%%%%%%%%%%%%%%% Optimization: %%%%%%%%%%%%%%%%%%%%%%%%%%%%%
%%%%%%%%%%%%%%% A model-space multi-scale approach %%%%%%%%%%%%%%%%%%
%%%%%%%%%%%%%%%%%%%%%%%%%%%%%%%%%%%%%%%%%%%%%%%%%%%%%%%%%%%%%%%%%%%%%
We analyze the structure of the FWIME gradient. We show that for inaccurate initial models, its tomographic component is responsible for recovering the missing low wavenumbers during the initial stages of the inversion process, even when refracted and/or low-frequency energy is absent from the recorded data \cite[]{biondi2014simultaneous,barnier2018full}. Finally, we present a model-space multi-scale workflow, which is a key ingredient for the success of our method.

\subsection{FWIME gradient}
The gradient of the FWIME objective function (equation~\ref{eqn:fwime.obj}) with respect to the velocity model $\mathbf{m}$ is given by

\begin{eqnarray}
    \label{eqn:fwime.gradient}
    \nabla \Phi_{\epsilon}(\mathbf{m}) &=& \mathbf{M} \Big [ \mathbf{B}^*(\mathbf{m}) + \mathbf{T}^*(\mathbf{m}, \mathbf{\tilde{p}}_{\epsilon}^{opt}(\mathbf{m})) \Big ] \Big [ \mathbf{f}(\mathbf{m}) + \tilde{\mathbf{B}} (\mathbf{m}) \mathbf{\tilde{p}}_{\epsilon}^{opt}(\mathbf{m}) - \mathbf{d}^{obs} \Big ],
\end{eqnarray}

where $\mathbf{B}^*$ is the adjoint of the conventional (non-extended) Born modeling operator, $\mathbf{T}^*$ is the adjoint of the data-space tomographic operator \cite[]{biondi2014simultaneous}, and $\mathbf{M}$ is a masking operator that may be used to prevent the gradient from updating certain regions of the model (e.g., the water layer). Equation~\ref{eqn:fwime.gradient} (whose derivation is shown in Appendix~\ref{fwimeGradient}) can be expressed as the sum of two terms, 

\begin{eqnarray}
    \label{eqn:fwime.gradient.split}
    \nabla \Phi_{\epsilon}(\mathbf{m}) &=& \mathbf{M} \left ( \nabla \Phi_{\epsilon}^{B} + \nabla \Phi_{\epsilon}^{T} \right ),
\end{eqnarray}

with

\begin{eqnarray}
    \label{eqn:fwime.gradient.born}
    \nabla \Phi_{\epsilon}^{B}(\mathbf{m}) &=& \mathbf{B}^*(\mathbf{m}) \mathbf{r}^{\epsilon}_d (\mathbf{m}), \\
    \label{eqn:fwime.gradient.tomo}
    \nabla \Phi_{\epsilon}^{T}(\mathbf{m}) &=& \mathbf{T}^*(\mathbf{m},\mathbf{\tilde{p}}_{\epsilon}^{opt}(\mathbf{m})) \mathbf{r}^{\epsilon}_d (\mathbf{m}),
\end{eqnarray}

where the adjoint source $\mathbf{r}^{\epsilon}_d (\mathbf{m})$ is the argument of the FWIME data-fitting component. Its expression is given by

\begin{eqnarray}
    \label{eqn:fwime.adjoint.source}
    \mathbf{r}^{\epsilon}_d (\mathbf{m}) &=& \mathbf{f}(\mathbf{m}) + \tilde{\mathbf{B}} (\mathbf{m}) \mathbf{\tilde{p}}_{\epsilon}^{opt}(\mathbf{m}) - \mathbf{d}^{obs}.
\end{eqnarray}

$\nabla \Phi_{\epsilon}^{B}$ is referred to as the ``Born" gradient of FWIME and shares kinematic similarities with the conventional FWI gradient (they employ the same operator but use different adjoint sources). The second component $\nabla \Phi_{\epsilon}^{T}$ is referred to as the ``tomographic" (or ``WEMVA") gradient. This term arises from the differentiation of the data-correcting term (with respect to the velocity model $\mathbf{m}$) and is essential for the FWIME workflow to recover the missing low-wavenumber components of the velocity model at early stages of the optimization process \cite[]{barnier2018full}. The FWIME gradient can be summarized by

\begin{eqnarray}
    \boxed {
        \begin{array}{rcl}   
            \label{eqn:fwime.gradient.wemva.fwi}
            \nabla^{\rm FWIME} &=& \nabla^{\rm FWI} + \nabla^{\rm WEMVA}.
        \end{array}
    }       
\end{eqnarray}

After conducting many numerical tests, we observe that this structure produces three regimes throughout the inversion workflow. The first stage can been seen as a tomographic or WEMVA regime, where the low-wavenumber components of the velocity model are recovered. If the initial velocity model is very inaccurate, we observe that the tomographic gradient tends to dominate (in terms of amplitude), and FWIME mainly relies on the contribution of this term to avoid converging to local minima. As the inversion progresses, FWIME enters an intermediate regime where both gradient components share similar amplitudes and contribute equally to the total search direction. Finally, when the inverted model is close to the expected solution, FWIME enters what we refer to as the ``linear regime": the search direction is mainly guided by the Born component which primarily updates the high-wavenumber features of the velocity. At this point, conventional FWI is able to converge towards the global solution without any extension strategy. 

One of the main advantages of FWIME is its ability to automatically manage the transitions between the three different regimes without the need to apply any scale mixing or manual enhancement of the gradient components, in contrast with the method proposed by \cite{biondi2014simultaneous}. This ability is achieved by the use of two ingredients in the inversion workflow. First, the variable projection method allows $\mathbf{\tilde{p}}_{\epsilon}^{opt}$ to handle the coupling between both tomographic and Born gradients. Second, a model-space multi-scale strategy is applied to gradually increase the wavenumber content of the model updates.

\subsection{A closer look at the tomographic operator}
The tomographic component of the gradient is crucial for the success of FWIME. It is computed by applying the adjoint of the data-space tomographic operator $\mathbf{T}$ to the FWIME data-residuals $\mathbf{r}^{\epsilon}_d$ (equation~\ref{eqn:fwime.gradient.tomo}). Mathematically, $\mathbf{T}: \mathbb{R}^{N_m} \mapsto \mathbb{R}^{N_d}$ is the Jacobian operator of the data-correcting term with respect to the velocity model $\mathbf{m}$, given a fixed optimal extended perturbation $\mathbf{\tilde{p}}_{\epsilon}^{opt}$. That is, 

\begin{eqnarray}
    \label{eqn:tomo_operator_definition}
    \mathbf{T}(\mathbf{m}, \mathbf{\tilde{p}}_{\epsilon}^{opt}) &=& \frac{\partial \big ( \tilde{\mathbf{B}} (\mathbf{m}) \mathbf{\tilde{p}}_{\epsilon}^{opt} \big )}{\partial \mathbf{m}} \Big |_{\mathbf{\tilde{p}}_{\epsilon}^{opt}}.
\end{eqnarray}

For a small velocity perturbation $\Delta \mathbf{m}$, operator $\mathbf{T}$ is a linear mapping from $\Delta \mathbf{m}$ to changes in the Born-modeled data, $\Delta \mathbf{d}^{Born}$, assuming a fixed extended reflectivity $\mathbf{\tilde{p}}_{\epsilon}^{opt}$. 

\begin{figure}[t]
    \centering
    \subfigure[]{\label{fig:rabbit_true_vel}\includegraphics[width=0.45\linewidth]{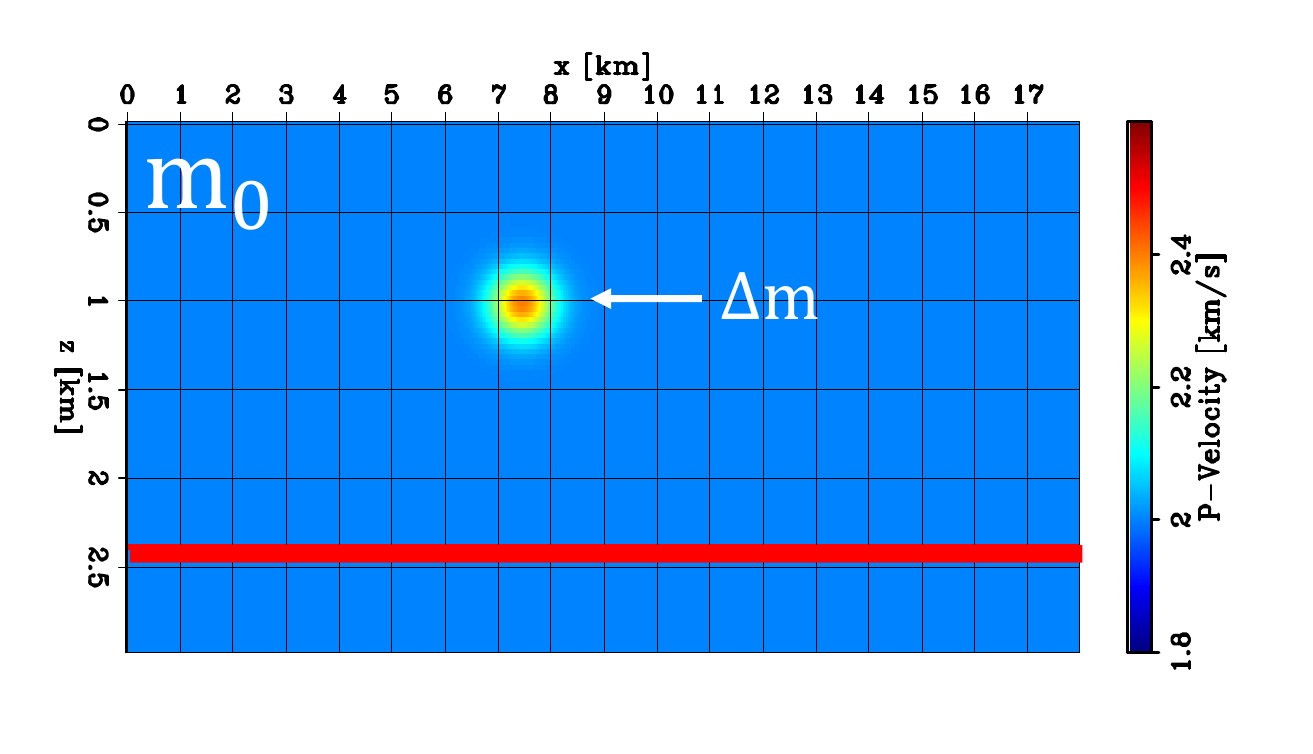}}
    \subfigure[]{\label{fig:rabbit_background_vel}\includegraphics[width=0.45\linewidth]{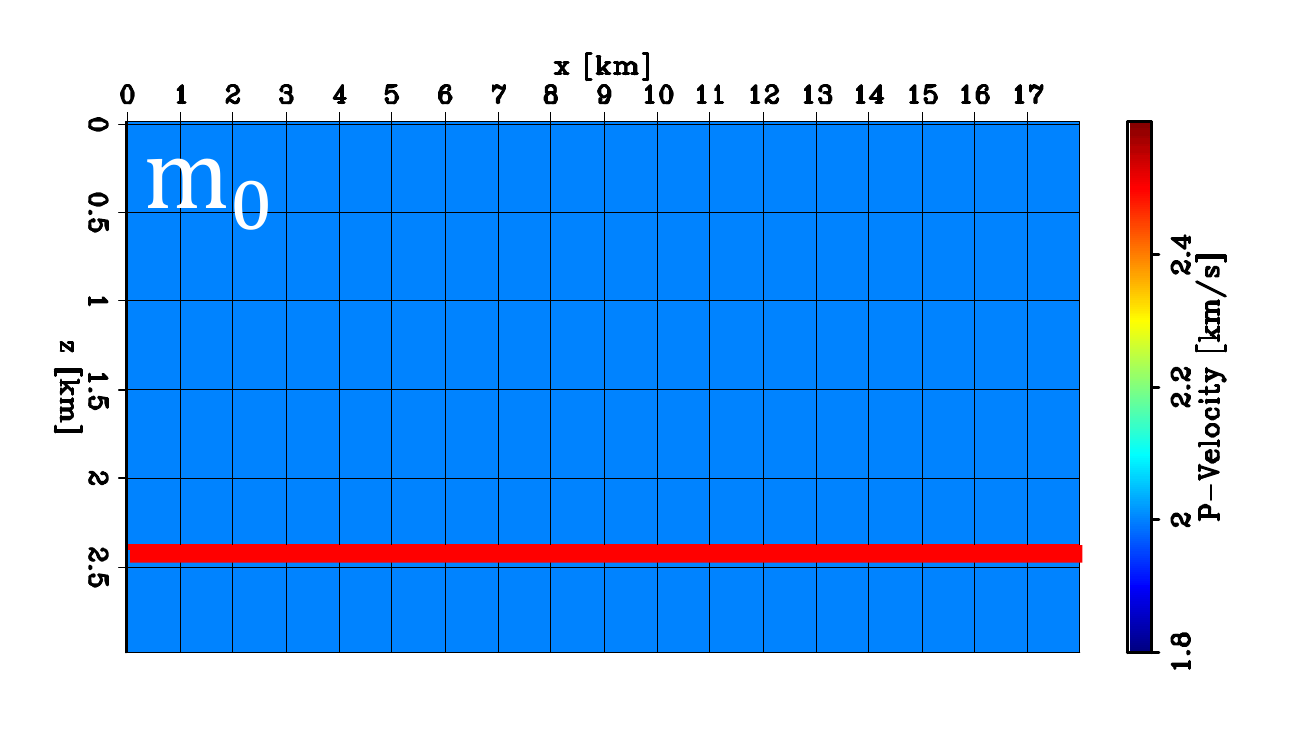}}\\
    \caption{2D panels of velocity models. The red horizontal line represents the reflector playing the role of $\mathbf{\tilde{p}}_{\epsilon}^{opt}$ (non-extended for this specific example). (a) True model $\mathbf{m}_0+\Delta \mathbf{m}$. (b) Background model $\mathbf{m}_0$ .}
    \label{fig:rabbit_vel}
\end{figure}

We show the characteristics of $\mathbf{T}$ on a simple numerical example. We compute the output of $\mathbf{T}$ by applying it to a small velocity perturbation $\Delta \mathbf{m}$ embedded in a homogeneous background $\mathbf{m}_0$ containing one horizontal reflector, which plays the role of $\mathbf{\tilde{p}}_{\epsilon}^{opt}$ (non-extended for the sake of this example) (Figure~\ref{fig:rabbit_true_vel}). The background model $\mathbf{m}_0$ is set to 2 km/s, the horizontal reflector is located at a depth of $z=2.4$ km (Figure~\ref{fig:rabbit_background_vel}). The Gaussian positive velocity perturbation $\Delta \mathbf{m}$ reaches a maximum value of 0.5 km/s. We place 150 sources and 600 receivers at the surface every 120 m and 30 m, and we first compute $\Delta \mathbf{d}^{tomo} = \mathbf{T}(\mathbf{m}_0, \mathbf{\tilde{p}}_{\epsilon}^{opt}) \Delta \mathbf{m}$ (not shown here). Examining the output of the forward mapping of $\mathbf{T}$ is challenging to interpret, but the adjoint mapping $\mathbf{T}^*: \mathbb{R}^{N_d} \mapsto \mathbb{R}^{N_m}$ provides better physical insight on the properties of the tomograpic gradient as it linearly relates data-space perturbations to velocity-model perturbations. We now apply $\mathbf{T}^*$ to $\Delta \mathbf{d}^{tomo}$, and we obtain the following velocity perturbation $\Delta \mathbf{m}_{tomo}$,

\begin{eqnarray}
    \label{eqn:tomo.fwd.adj}
    \Delta \mathbf{m}_{tomo} &=& \mathbf{T}^*(\mathbf{m}_0,\mathbf{\tilde{p}}_{\epsilon}^{opt}) \;  \mathbf{T}(\mathbf{m}_0,\mathbf{\tilde{p}}_{\epsilon}^{opt}) \; \Delta \mathbf{m}.
\end{eqnarray}

\begin{figure}[t]
    \centering
    \subfigure[]{\label{fig:rabbit_tomo_adj_shot2}\includegraphics[width=0.45\linewidth]{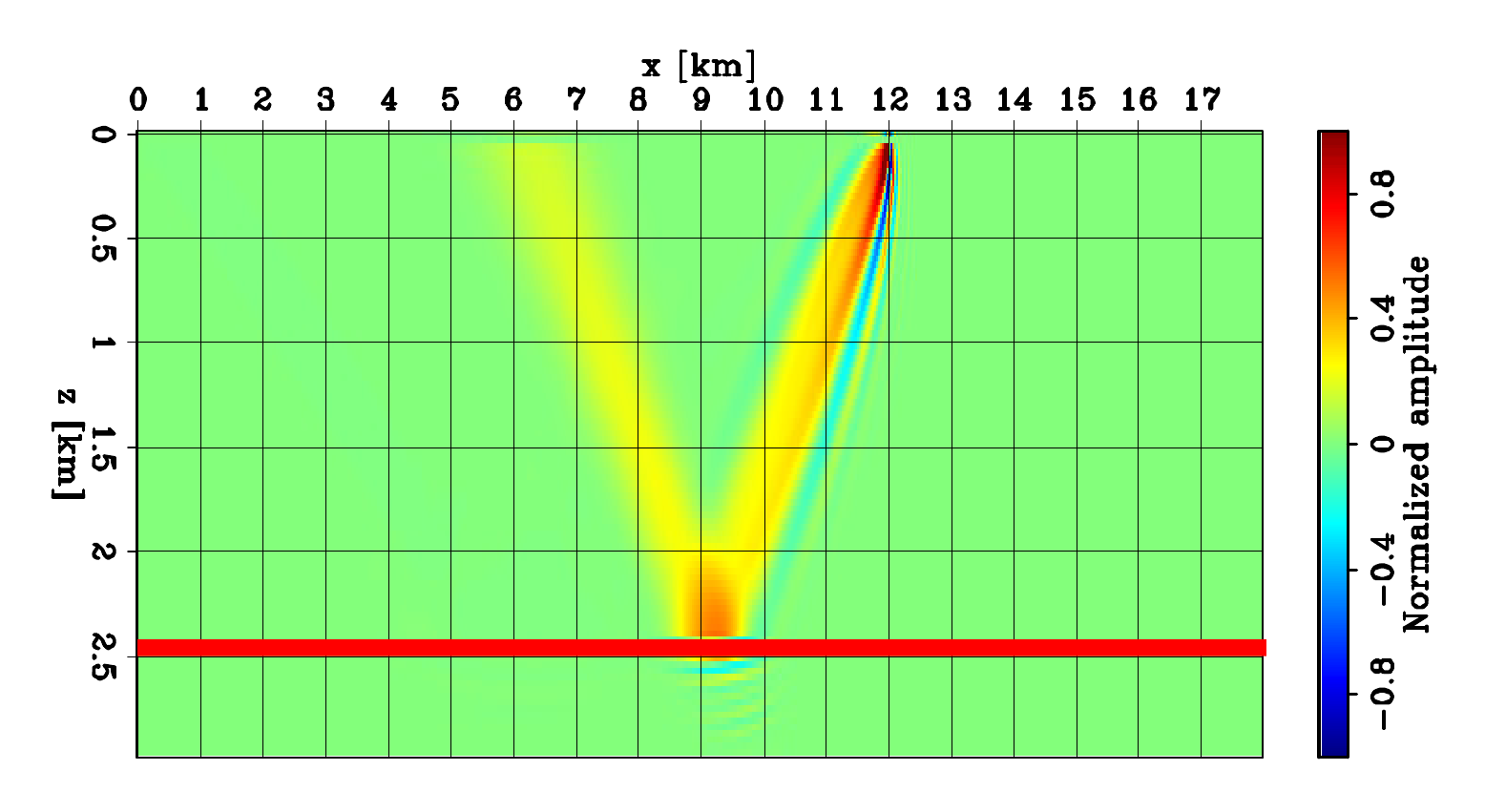}}
    \subfigure[]{\label{fig:rabbit_tomo_adj_shot1}\includegraphics[width=0.45\linewidth]{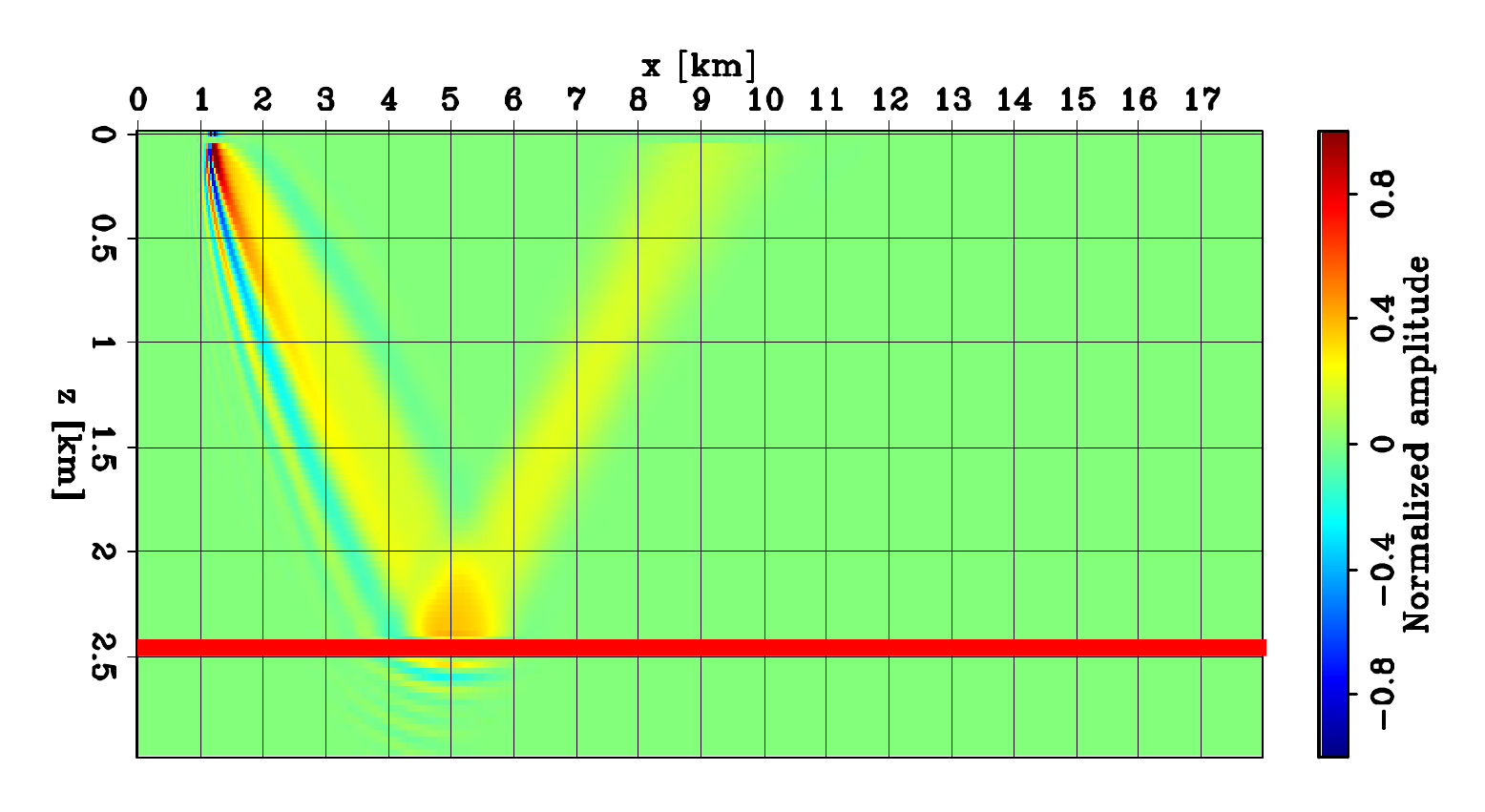}}\\    
    \subfigure[]{\label{fig:rabbit_tomo_adj_full}\includegraphics[width=0.45\linewidth]{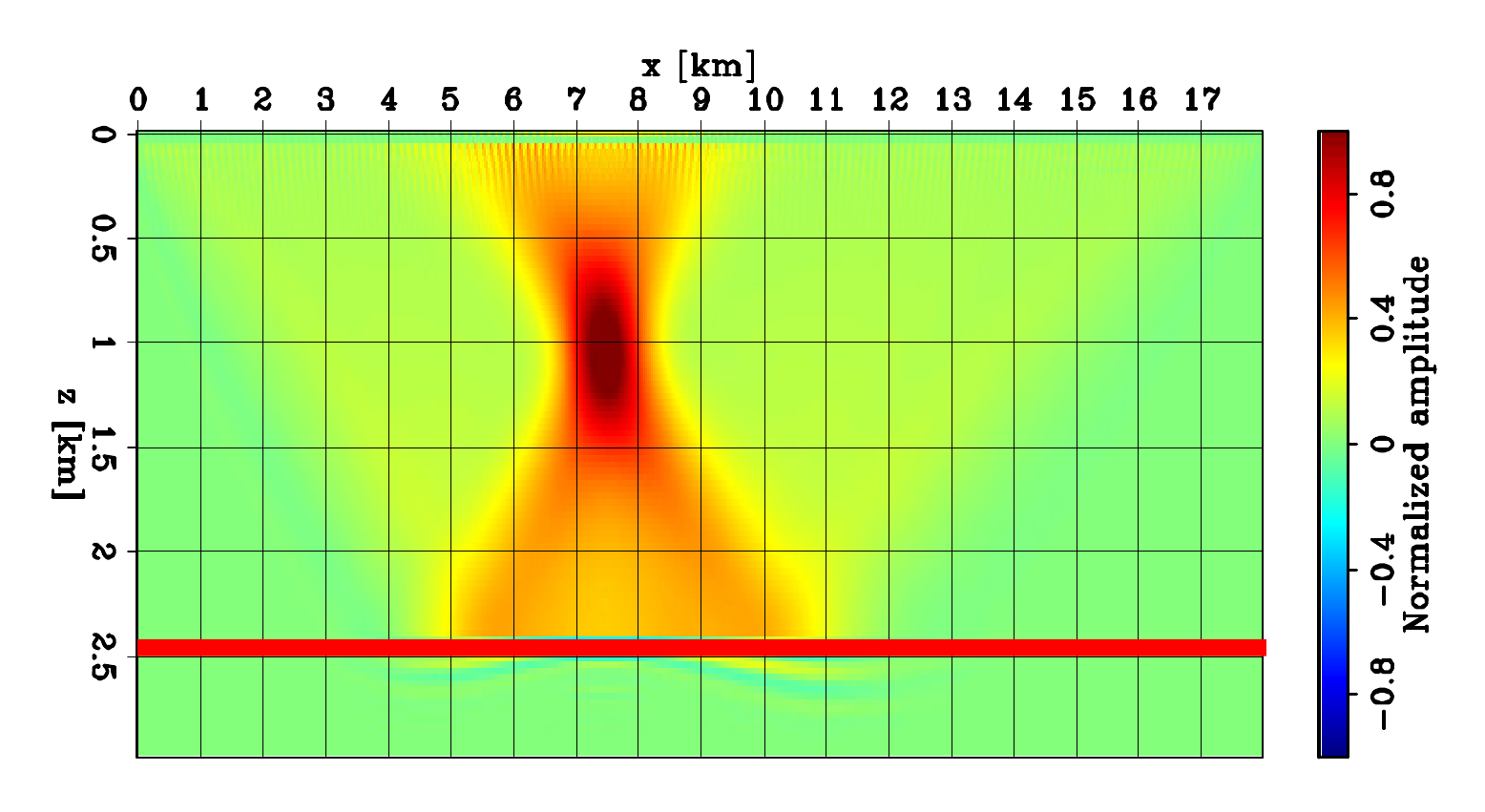}}
    \subfigure[]{\label{fig:rabbit_tomo_adj_true}\includegraphics[width=0.45\linewidth]{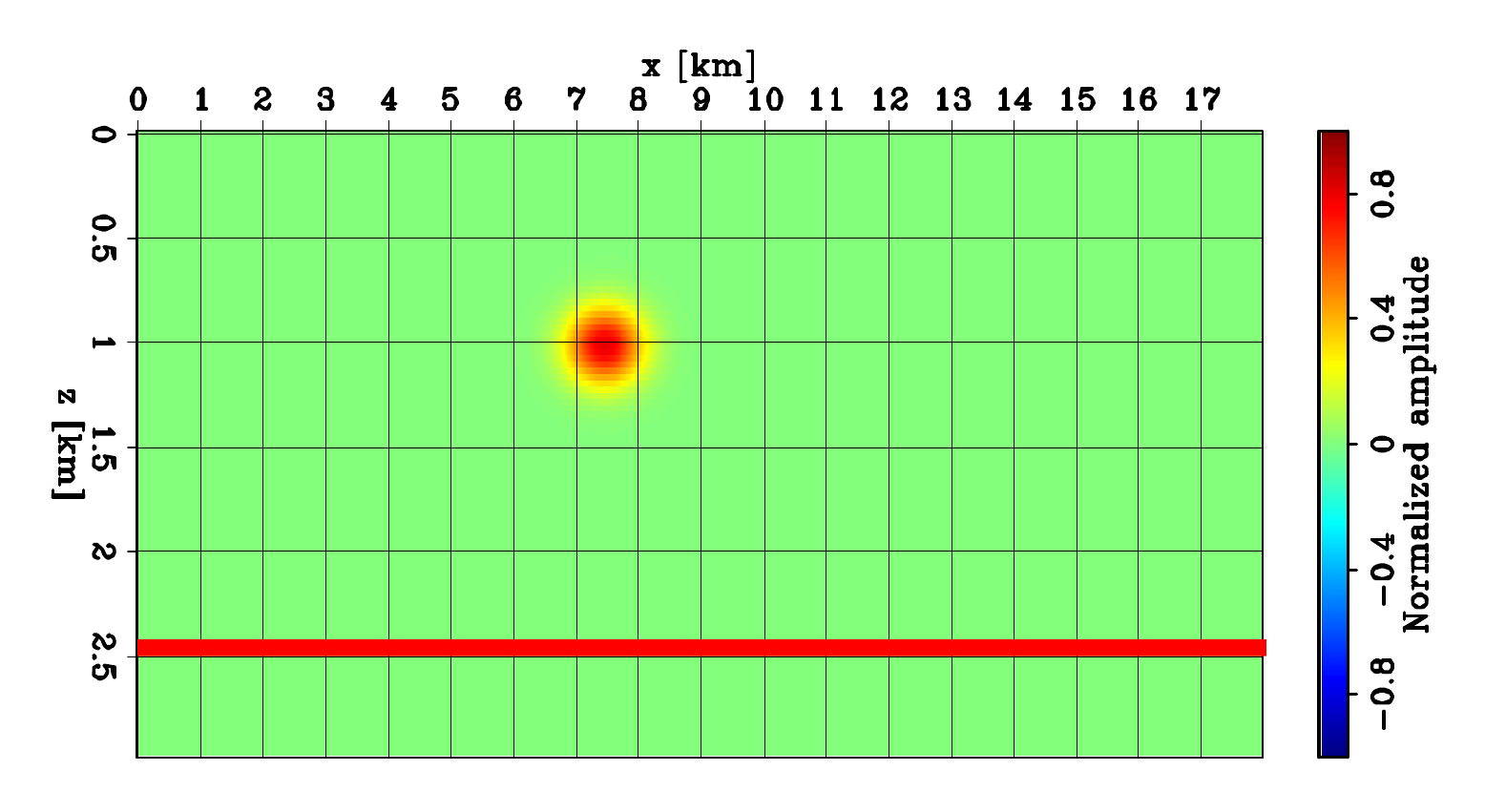}}\\    
    \caption{Panels showing velocity perturbations computed by the application of $\mathbf{T}^* (\mathbf{m}_0, \mathbf{\tilde{p}}_{\epsilon}^{opt}) \mathbf{T} (\mathbf{m}_0, \mathbf{\tilde{p}}_{\epsilon}^{opt})$ to panel (d). (a) Velocity perturbation computed with a single source placed at $x=1$ km. (b) Velocity perturbation computed with a single source placed at $x=12$ km. (c) Velocity perturbation computed with all available sources. (d) True velocity perturbation.}
    \label{fig:rabbit_tomo_adj}
\end{figure}

Figures~\ref{fig:rabbit_tomo_adj_shot2} and \ref{fig:rabbit_tomo_adj_shot1} show $\Delta \mathbf{m}_{tomo}$ computed according to equation~\ref{eqn:tomo.fwd.adj} for sources placed at $x=1$ km and $x=12$ km, respectively, which result in low-wavenumber (smooth) but accurate updates in the velocity model, even with the use of pure reflection data. 
$\Delta \mathbf{m}_{tomo}$ displays a pattern commonly known as the ``rabbit ears" updates in conventional FWI, and demonstrate the ability of the FWIME tomographic gradient to reconstruct the long-wavelength features of deeper targets that may not be illuminated by refracted energy, such as diving waves. Figure~\ref{fig:rabbit_tomo_adj_full} shows the analogous map computed for the entire collection of available source/receiver pairs (equation~\ref{eqn:tomo.fwd.adj}). By comparing it to the true perturbation  in Figure~\ref{fig:rabbit_tomo_adj_true}, the velocity update lacks vertical resolution but the recovered anomaly seems to be accurately positioned. 

As we illustrate with numerical examples in the last section of this paper, the FWIME tomographic gradient possesses similar features as the ones obtained with reflection full waveform inversion (RFWI) \cite[]{xu2012full,brossier2015velocity,zhou2015full}, but one of the main differences between the two approaches is the fact that FWIME employs an extended reflectivity. Even though this model extension increases the computational cost of our method, the additional degrees of freedom it provides allows the tomographic gradient to recover accurate model updates even if the velocity error falls outside of the linearization approximation, thereby making FWIME more robust.  
\subsection{FWIME inversion workflow}
We summarize the main steps of the FWIME workflow (minimization of equation~\ref{eqn:fwime.obj}) in algorithm~\ref{alg:fwime.workflow}. The FWIME scheme begins by the selection of an extension type. As first noticed by \cite{biondi2014simultaneous}, extending $\mathbf{\tilde{p}}_{\epsilon}^{opt}$ with time lags allows the data-correcting term to efficiently capture large time shifts for both reflected and refracted waves. Additionally, for 3D field applications, the time-lag extension requires a single additional axis (compared to two axes for space lags), thereby reducing the computational cost and memory footprint of the method. In algorithm~\ref{alg:fwime.workflow}, the full data bandwidth is simultaneously inverted from the start, and all events in the data are employed (including direct arrivals and reflected energy) to potentially produce long-wavelength updates (as shown in the previous section by the analysis of operator $\mathbf{T}^*$). Hence, data-space multi-scale approaches used in conventional FWI such as the one proposed by \cite{bunks1995multiscale} are not suited for FWIME. In the next section, we present an alternate multi-scale strategy that enables the simultaneous inversion of the full data bandwidth and gradually increases the resolution of the model updates while maintaining robust convergence properties. 

\begin{algorithm}
    \caption{FWIME}
    \label{alg:fwime.workflow}
    \begin{itemize}
        \item Select the initial model $\mathbf{m}_0$
        \item Select the extension type and the length of the extended axis
        \item Select the hyperparameter $\epsilon$ (fixed throughout the optimization process)
        \item For $i=0, \dots, n_{iter}-1$
        \begin{enumerate}
            \item Compute $\mathbf{d}^{obs}-\mathbf{f}(\mathbf{m}_i)$
            \item Compute $\tilde{\mathbf{p}}_{\epsilon}^{opt}(\mathbf{m}_i) = \underset{\tilde{\mathbf{p}}}{\mathrm{argmin}} \; \dfrac{1}{2} \left \| \tilde{\mathbf{B}}(\mathbf{m}_i)\mathbf{\tilde{p}} - \left ( \mathbf{d}^{obs} - \mathbf{f}(\mathbf{m}_i) \right ) \right \|^2_2 + \dfrac{\epsilon^2}{2} \left \| \mathbf{D} \tilde{\mathbf{p}} \right \|^2_2$
            \label{var.pro.algo}
            \item Compute objective function value $\Phi_{\epsilon}(\mathbf{m}_i)$
            \item Set $\mathbf{r}^{\epsilon}_d (\mathbf{m}_i) = \mathbf{f}(\mathbf{m}_i) + \tilde{\mathbf{B}} (\mathbf{m}_i) \mathbf{\tilde{p}}_{\epsilon}^{opt}(\mathbf{m}_i) - \mathbf{d}^{obs}$
            \item Compute FWIME gradient $\nabla_{\mathbf{m}} \Phi_{\epsilon}(\mathbf{m}_i) = \mathbf{M} \left [ \mathbf{B}^*(\mathbf{m}_i) + \mathbf{T}^*(\mathbf{m}_i) \right ] \mathbf{r}^{\epsilon}_d (\mathbf{m}_i)$
            \item Compute search direction $\mathbf{s}_i$ 
            \item Compute step length $\gamma_i$
            \item Update model $\mathbf{m}_{i+1}=\mathbf{m}_i + \gamma_i \mathbf{s}_i$
        \end{enumerate}
    \end{itemize}
\end{algorithm}

\subsection{The need for a multi-scale approach}
For any waveform inversion, it seems crucial to accurately recover the missing long spatial-wavelength components at early stages, and then gradually increase the resolution of the model updates \cite[]{alkhalifah2014scattering}. In fact, it has been observed that the extent of the basin of attraction of FWI about the optimal minimum increases when the low-frequency component of the data is inverted in a data-space multi-scale manner \cite[]{bunks1995multiscale,fichtner2010full}. Additionally, \cite{mora1989inversion} shows the connection between the propagation direction of the source and receiver wavefields and the wavenumber updates introduced by their cross-correlation (i.e., the model scale that is updated at each iteration). Finally, \cite{sirgue2004efficient} extend this discussion and describe the connection between the data frequency content and the model updates and propose a method to select the frequency band to be inverted. 

Unlike conventional methods, FWIME accepts the presence (and the simultaneous inversion) of the full data-bandwidth, which may include all wave types and all available frequencies from the start. We develop a new workflow where the data-space multi-scale strategy is substituted by a model-space multi-scale one. This process is achieved by considering a velocity re-parametrization on spatially adjustable non-uniform grids with the use of basic-splines (B-splines) basis functions \cite[]{de1986b,SplineNotes}, instead of a finite-difference grid. However, all wavefield modeling and propagation are still conducted on conventional finite-difference grids. The FWIME workflow should begin by recovering the low-wavenumber components by using a coarse-grid model representation. As the inversion progresses, the grid sampling is gradually refined and the inverted model on a given grid is then used as the initial guess for the following inversion performed on a finer/denser grid. This process is repeated until an accurate solution is successfully recovered. The benefit of this approach is that the spline parameterization and its refinement rate provide the ability to control and gradually increase the resolution of the model-updates with iterations.

\begin{figure}[t]
    \centering
    \subfigure[]{\label{fig:oneLayer3_true_mod}\includegraphics[width=0.45\linewidth]{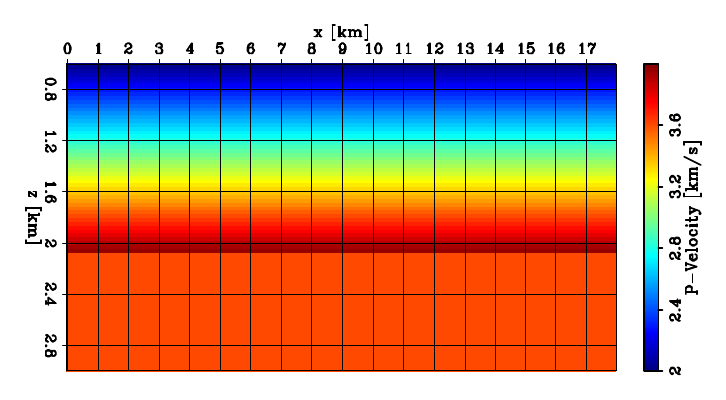}}
    \subfigure[]{\label{fig:oneLayer3_init_mod}\includegraphics[width=0.45\linewidth]{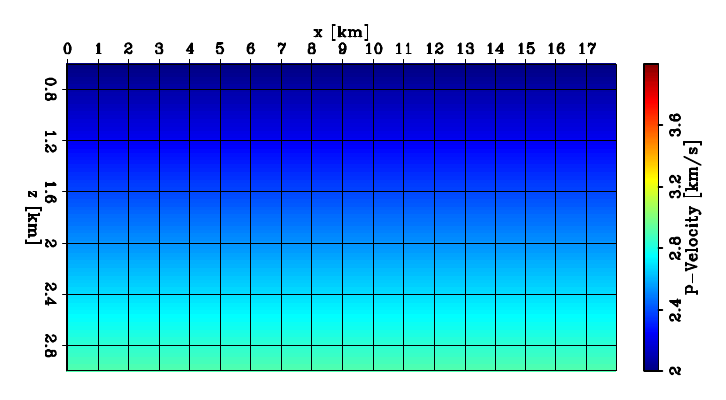}}
    \caption{2D panels of (a) the true and (b) initial velocity models.}
    \label{fig:oneLayer3_mod}
\end{figure}

\begin{figure}[t]
    \centering
    \subfigure[]{\label{fig:oneLayer3_data_true}\includegraphics[width=0.3\linewidth]{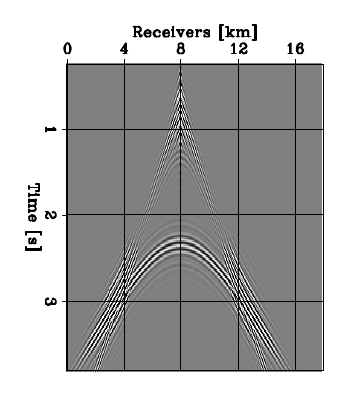}}
    \subfigure[]{\label{fig:oneLayer3_data_true_taper}\includegraphics[width=0.3\linewidth]{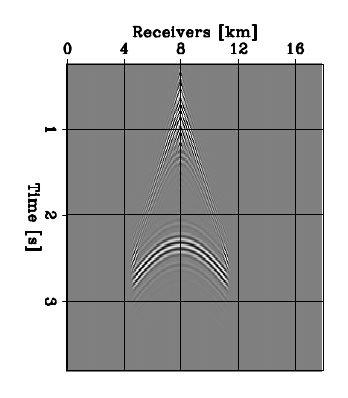}}
    \subfigure[]{\label{fig:oneLayer3_data_init_diff}\includegraphics[width=0.3\linewidth]{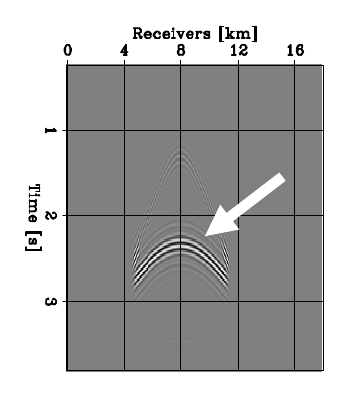}}
    \caption{Representative shot gathers generated by a source located at $x = 8$ km. (a) Observed data with no muting applied, $\mathbf{d}^{obs}$. (b) Observed data after muting, $\mathbf{M}_d \mathbf{d}^{obs}$. (c) Initial data difference after muting, $\mathbf{M}_d (\mathbf{d}^{obs}-\mathbf{f}(\mathbf{m}_0))$. All panels are displayed with the same grayscale.}
    \label{fig:oneLayer3_dat}
\end{figure}

We illustrate the importance of this new multi-scale approach on a numerical example where we compute the initial FWIME search direction. We design a 18 km-wide and 3 km-deep laterally invariant velocity model $\mathbf{m}_{true}$ composed of a shallow homogeneous layer, a linear $v(z)$ velocity gradient, and a sharp horizontal reflector at a depth of 2.1 km (Figure~\ref{fig:oneLayer3_true_mod}). The initial velocity model $\mathbf{m}_0$ (Figure~\ref{fig:oneLayer3_init_mod}) is also laterally invariant and composed of the same shallow homogeneous layer, but the velocity gradient in the deeper region is chosen to be inaccurate enough for conventional FWI to converge to a non-physical solution (not shown here). In addition, $\mathbf{m}_0$ does not contain any reflector. Figure~\ref{fig:oneLayer3_mod_1d} shows the velocity profiles of the true (blue curve) and initial (red curve) models. At the surface, we place 150 source every 120 m and 600 receivers every 30 m. We generate noise-free pressure data with a two-way acoustic modeling operator and a source containing energy restricted to the 9-18 Hz range. For this numerical example, we propose to solely use reflected energy and we apply a data-muting mask $\mathbf{M}_d$ on all shot records to mute events occurring at source-receiver offsets greater than $3.0$ km. Figure~\ref{fig:oneLayer3_dat} shows a representative shot gather of the raw observed data $\mathbf{d}^{obs}$ (left panel) and the muted observed data $\mathbf{M}_d \mathbf{d}^{obs}$ (middle panel) for a shot located at $x = 8$ km. Figure~\ref{fig:oneLayer3_data_init_diff} shows the muted initial data difference, $\mathbf{M}_d \left ( \mathbf{d}^{obs}-\mathbf{f}(\mathbf{m}_0)\right)$. 

\begin{figure}[t]
    \centering
    \label{fig:oneLayer3_mod_1d}\includegraphics[width=0.45\linewidth]{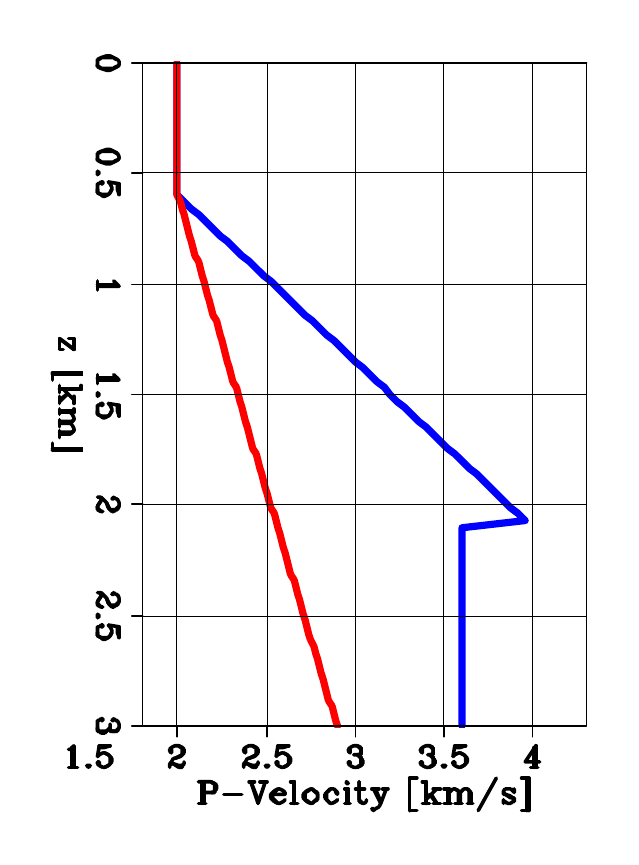}
    \caption{1D profiles of the true model (blue curve), and the initial model (red curve).}
    \label{fig:oneLayer3_mod_1d}
\end{figure}

\begin{figure}[t]
    \centering
    \label{fig:oneLayer3_cig_init.5}\includegraphics[width=0.35\linewidth]{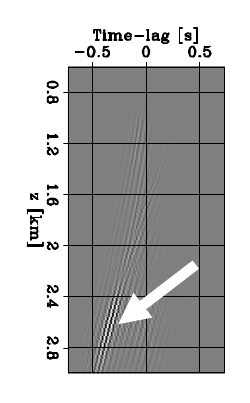}
    \caption{Time-lag common image gather (TLCIG) extracted at $x=9$ km from $\mathbf{\tilde{p}}_{\epsilon}^{opt} (\mathbf{m}_0)$ computed at the initial stage of the FWIME workflow.}
    \label{fig:oneLayer3_cig_init}
\end{figure}

We conduct the variable projection step of FWIME (step 2 of algorithm~\ref{alg:fwime.workflow}) by minimizing objective function~\ref{eqn:vp.obj} with 60 iterations of linear conjugate gradient and $\epsilon=2.5 \times 10^{-7}$. We use a time-lag extension $\tau$ for $\mathbf{\tilde{p}}_{\epsilon}^{opt}$ with a total of 91 points sampled at 16 ms, allowing $\tau$ to range from $-0.72$ s to $0.72$ s. Figure~\ref{fig:oneLayer3_cig_init} shows a TLCIG extracted at $x=9$ km from $\mathbf{\tilde{p}}_{\epsilon}^{opt}(\mathbf{m}_0)$. The event with strong energy located at negative time lags (white arrow) corresponds to the mapping of the reflection from the sharp horizontal interface (white arrow in Figure~\ref{fig:oneLayer3_data_init_diff}) into the extended space of $\mathbf{\tilde{p}}_{\epsilon}^{opt}(\mathbf{m}_0)$. As expected, the position of its maximum energy is shifted away from the physical axis (where $\tau=0$ s), which in this example indicates that the initial velocity is lower than the true velocity. Figure~\ref{fig:oneLayer3_cig_init} illustrates how $\mathbf{\tilde{p}}_{\epsilon}^{opt}$ contains crucial kinematic information and it shows the importance of using an extended perturbation with a large-enough extension (otherwise the information would be lost). 

\begin{figure}[t]
    \centering
    \subfigure[]{\label{fig:oneLayer3_fwime_BornGrad}\includegraphics[width=0.45\linewidth]{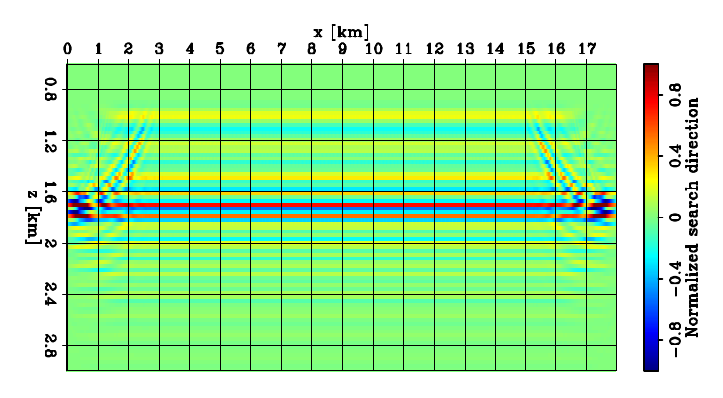}}
    \subfigure[]{\label{fig:oneLayer3_fwime_tomoGrad}\includegraphics[width=0.45\linewidth]{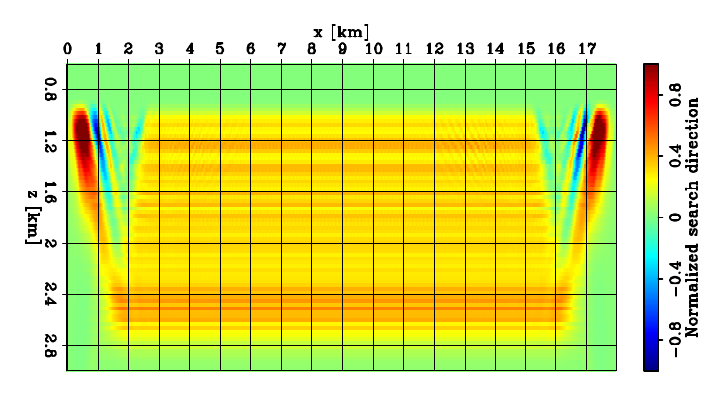}}\\    
    \subfigure[]{\label{fig:oneLayer3_fwime_grad}\includegraphics[width=0.45\linewidth]{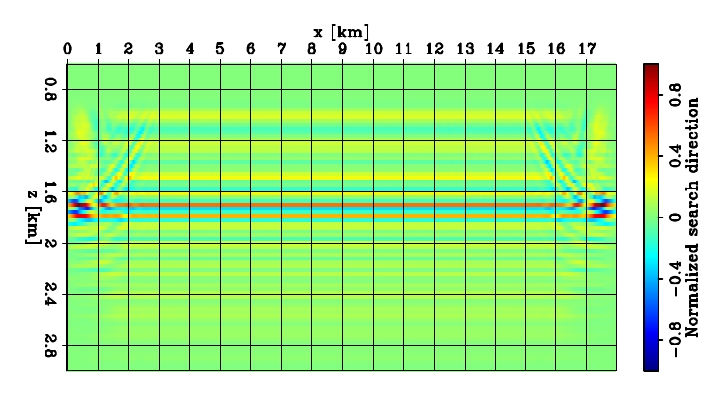}}
    \subfigure[]{\label{fig:oneLayer3_fwi_grad}\includegraphics[width=0.45\linewidth]{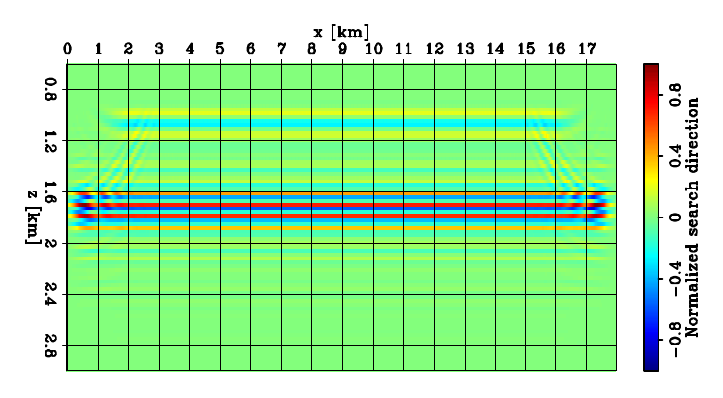}}\\
    \subfigure[]{\label{fig:oneLayer3_true_grad}\includegraphics[width=0.45\linewidth]{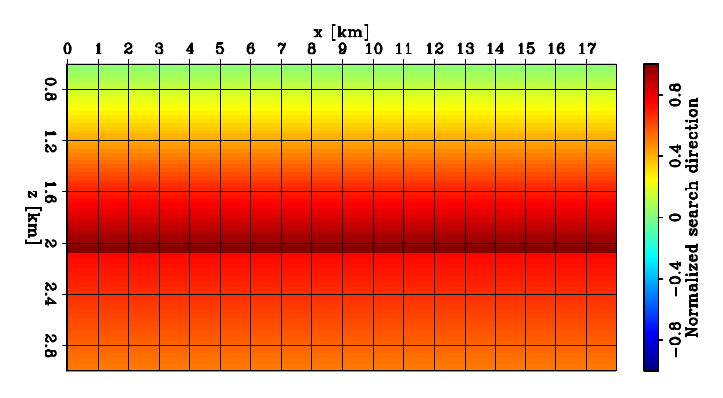}}
    \caption{2D panels of initial search directions. (a) FWIME Born component. (b) FWIME tomographic component. (c) FWIME total search direction, $\mathbf{s}_{total}$  (obtained by summing (a) and (b)). (d) Conventional FWI search direction. (e) True search direction, $\mathbf{s}_{true}$. For display purposes, panel (b) is normalized with a smaller value than in panel (a), and the amplitude of (b) is much smaller than the one of (a).}
    \label{fig:oneLayer3_grad}
\end{figure}

Figure~\ref{fig:oneLayer3_fwime_BornGrad} shows the Born component of the initial FWIME search direction. As expected, it is similar to the initial FWI search direction (Figure~\ref{fig:oneLayer3_fwi_grad}). In both cases, the position of the sharp interface is too shallow (due to the velocity error within the initial model). Figure~\ref{fig:oneLayer3_fwime_tomoGrad} displays the initial tomographic search direction, which seems promising by comparing it to the true search direction $\mathbf{s}_{true}$ shown in Figure~\ref{fig:oneLayer3_true_grad}. Its amplitude, however, is much smaller than the Born component (Figures~\ref{fig:oneLayer3_fwime_BornGrad} and \ref{fig:oneLayer3_fwime_tomoGrad} are normalized by different values for display purposes). Finally, Figure~\ref{fig:oneLayer3_fwime_grad} represents the total FWIME search direction $\mathbf{s}_{total}$, obtained by summing the two panels on the first row. Even though the tomographic component manages to accurately recover the missing low-resolution features of the velocity, its amplitude is overwhelmed by the Born update which will likely lead the optimization scheme to a non-physical solution. 

One potential solution would be to manually control the relative amplitudes of the two components by assigning more weight to the tomographic gradient at early stages, and gradually adjusting the relative weights throughout the inversion process. However, this approach would be very user intensive and challenging to automatize. Alternatively, one could apply a spatial smoothing filter to manually limit the spatial resolution of the velocity updates \cite[]{biondi2014simultaneous}. While this rather standard filtering approach may be valid in many practical situations, it is not consistent in the context of optimization. We prefer to consider a re-parametrization of the velocity model with B-spline basis functions which naturally filters high-wavenumber effects and allows a natural and more flexible refinement, thanks to the subdivision property of B-splines.

\subsection{A model-space multi-scale approach for FWIME}
\subsubsection{Velocity parametrization using B-spline basis functions}
B-spline basis functions are commonly used in computer-aided design and graphic to draw smooth curves and surfaces passing in the vicinity of a set of control points, also referred to as ``spline nodes" \cite[]{SplineNotes}. We employ these functions to represent seismic velocity models on a coarse and potentially spatially non-uniform grid (referred to as the spline grid) and we create a linear operator (referred to as the spline operator) that maps a spline grid onto a finer grid (the finite-difference propagation grid). This mapping does not require to fit the control points exactly, and is therefore technically not an interpolation method. However, releasing this constrain provides great flexibility for the spline grid positioning and provides the ability to take into account prior geological knowledge of the Earth's subsurface. Unlike radial basis functions (RBF), B-spline functions have very limited support (i.e., they are non-zero only locally), which makes them computationally efficient for 3D applications. 

We follow the theory described in \cite{SplineNotes} and we modify it for our application. The spline operator of order $p$ is the linear mapping $\mathbf{S}^{(p)}$ defined by

\begin{eqnarray}
    \label{eqn:mapping1}
    \mathbf{S}^{(p)}: \; \mathbb{R}^{N_{m_c}} &\mapsto& \mathbb{R}^{N_{m_f}} \\
    \mathbf{m}^c &\mapsto& \mathbf{m}^f = \mathbf{S}^{(p)} \; \mathbf{m}^c\nonumber.
\end{eqnarray}

$\mathbf{m}^f \in \mathbb{R}^{N_{m_f}}$ represent the seismic velocity model parametrized on a ``fine" uniform finite-difference grid, where $N_{m_f} = N_z \times N_x \times N_y$ represents the total number of finite-difference grid points. $\mathbf{m}^c \in \mathbb{R}^{N_{m_c}}$ is a representation of the velocity model on a predefined coarse and (potentially) non-regularly spaced grid, and $N_{m_c}$ is the number of points on the coarse grid (i.e., the number of spline nodes or control points). The adjoint of the spline operator is therefore a mapping from the finite-difference grid into the coarse grid, 

\begin{eqnarray}
    \label{eqn:mapping1_adj}
    \left (\mathbf{S}^{(p)} \right )^*: \; \mathbb{R}^{N_{m_f}} &\mapsto& \mathbb{R}^{N_{m_c}} \\
    \mathbf{m}^f &\mapsto& \mathbf{m}^c = \left ( \mathbf{S}^{(p)} \right )^* \; \mathbf{m}^f \nonumber.
\end{eqnarray}

We denote by $\mathcal{G}_c \subset \mathbb{R}^{N_{m_c}}$ and $\mathcal{G}_f \subset \mathbb{R}^{N_{m_f}}$ the spline and finite-difference grids, respectively. The entries of the spline operator $\mathbf{S}^{(p)} \in \mathbb{R}^{N_{m_f} \times N_{m_c}}$ are computed using B-spline basis functions of order $p$ whose expressions are given by the Cox-de Boor recursion formula \cite[]{de1986b}. They depend both on the interpolation technique (i.e., the type of basis functions employed) and on the way the coarse grid is arranged. In the following, we set $p=3$ to ensure that the reconstructed functions in the output space are $\mathcal{C}^2$ within the area of interest (where wavefields are modeled). To simplify notations, and we do not explicitly write the dependency of the operators on $p$. 

In the inversion scheme, $\mathbf{m}^c$ is now the set of unknown parameters we wish to recover, and its entries should be seen as weights rather than seismic velocity values. Hence, their actual magnitudes/units are not directly physically interpretable. However, once $\mathbf{m}^c$ is known, the corresponding velocity field can be inferred by simply applying the forward spline operator $\mathbf{S}$ to map the inverted model onto the finite-difference grid. To gain better insight on this mapping, we express the velocity value at the $i^{th}$ point on the finite-difference grid as a function of the model values at the spline nodes. For that, we examine the $i^{th}$ row of equation~\ref{eqn:mapping1}, which is given by

\begin{eqnarray}
    \label{eqn:iRow}
    \mathbf{m}^f_i &=& \sum_{k=1}^{N_{m_c}} S_{ik} \, \mathbf{m}^c_k.
\end{eqnarray}

Equation~\ref{eqn:iRow} simply indicates that the velocity value at the $i^{th}$ point on the finite-difference grid can be expressed as a linear combination of the weights at each spline node. For example, $S_{ik}$ is the contribution (or weight) of spline node $k$ to the velocity value computed at the $i^{th}$ point on the finite-difference grid. Since B-spline functions possess a compact support,  $\mathbf{S}$ is very sparse. For a given point on the finite-difference grid, only a maximum of six (2D-case) and nine (3D-case) coefficients are non-zero while still ensuring that the reconstructed velocity function (on the finite-difference grid) is $\mathcal{C}^2$ (when $p=3$). This implies that at most nine terms from the sum in the right-side of in equation~\ref{eqn:iRow} will contribute to the computation of $\mathbf{m}^f_i$. More stringent conditions on the level of smoothness and continuity (i.e., for $p>3$) will increase the number of non-zero coefficients. 

\begin{figure}[t]
    \centering
    \subfigure[]{\label{fig:mesh_regular}\includegraphics[width=0.45\linewidth]{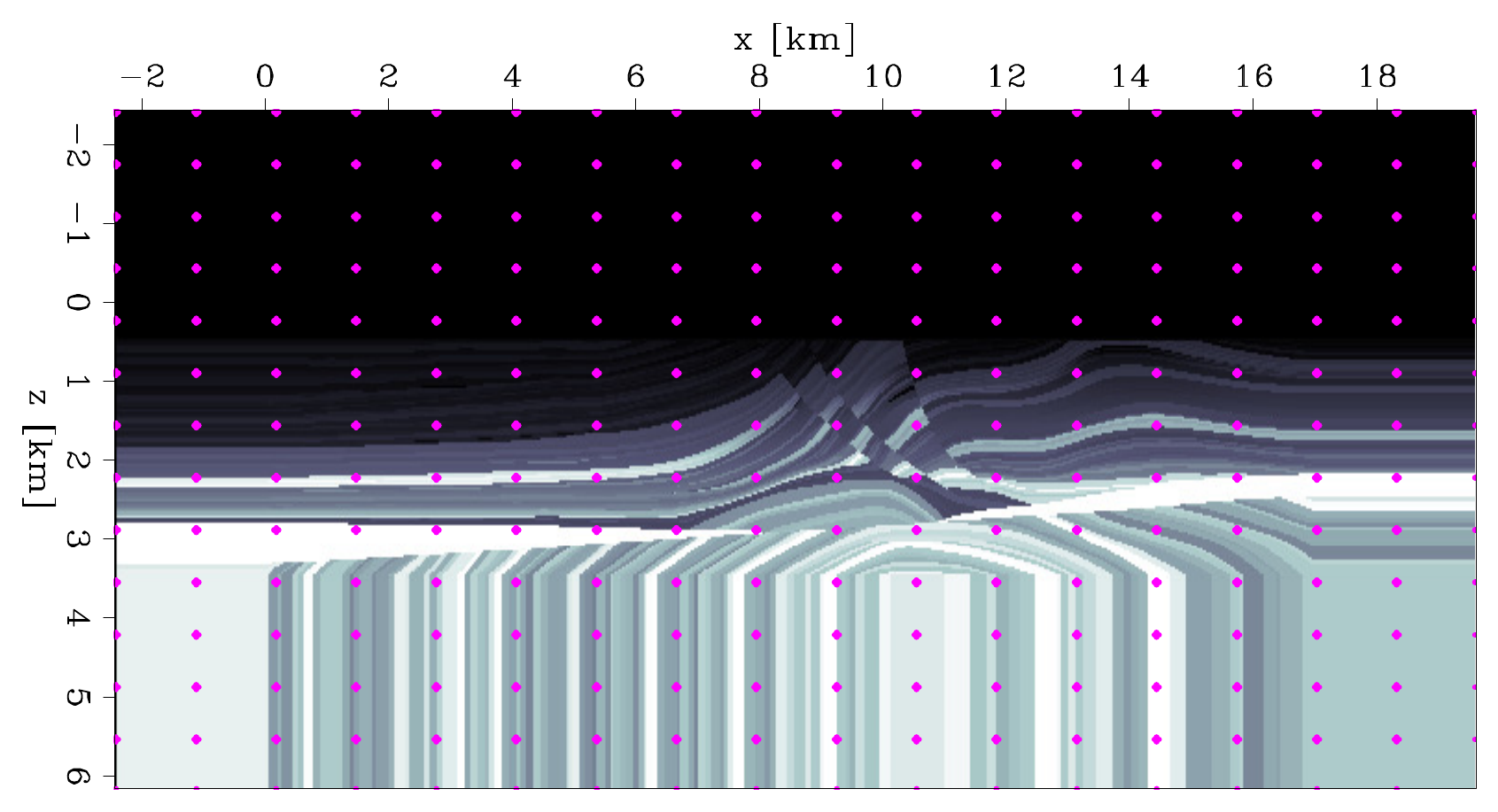}} \hspace{5mm}
    \subfigure[]{\label{fig:mesh_irregular}\includegraphics[width=0.45\linewidth]{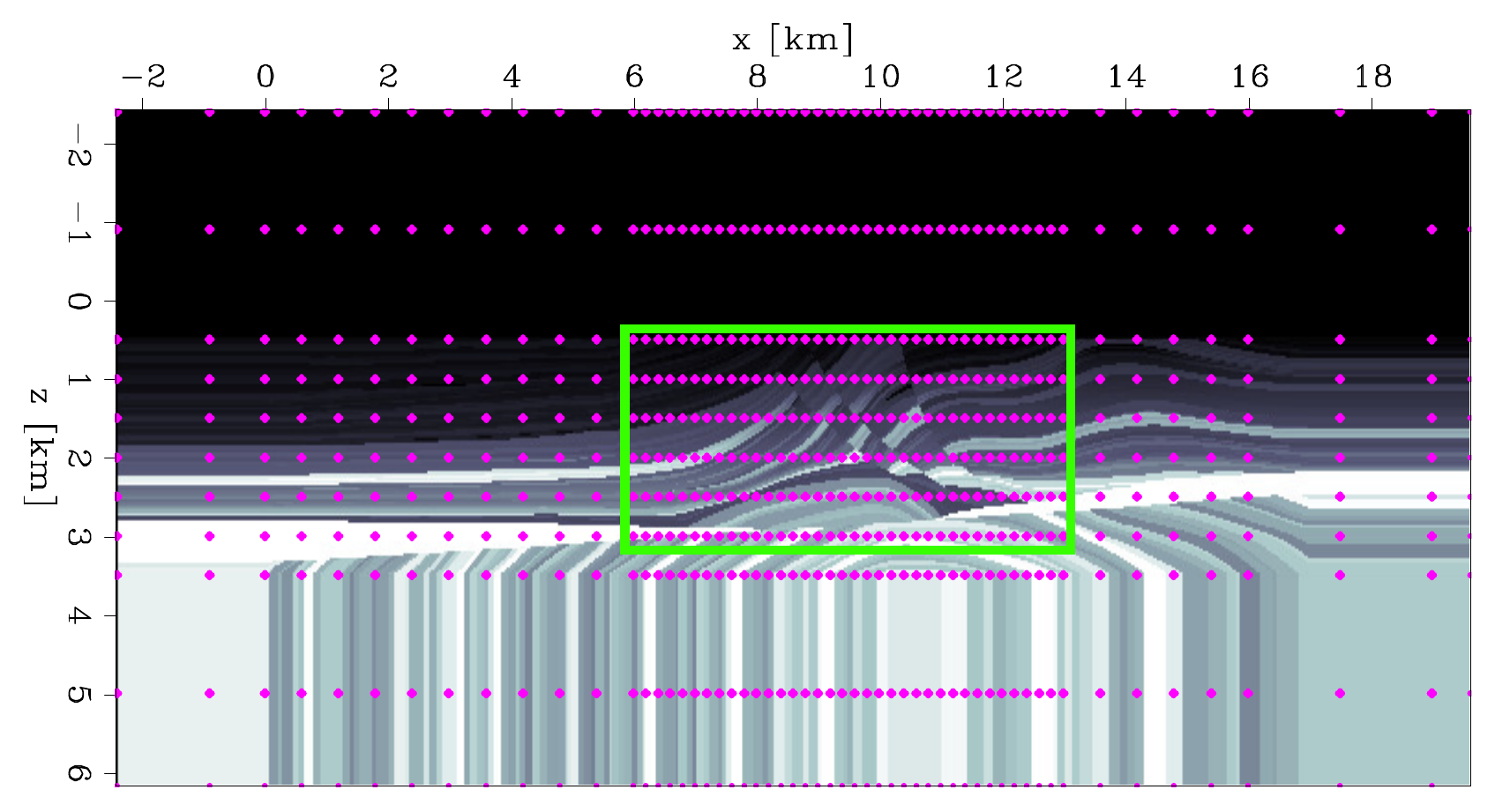}}\\    
    \caption{2D panels of spline meshes overlaid on the Marmousi2 velocity model. (a) Regular spline mesh. (b) Irregular spline mesh. The pink dots indicate the position of spline nodes.}
    \label{fig:mesh}
\end{figure}

We illustrate the application and the properties of the spline operator $\mathbf{S}$ by parametrizing a 2D velocity field based on the Marmousi2 model with two different spline grids. The underlying assumption for such model representation is that denser spline grids will be able to better represent higher-resolution features from the velocity field, while coarser spline grids will tend to spatially smooth the velocity model. Figure~\ref{fig:mesh} shows two spline grid dispositions (the pink dots correspond to the spline nodes) overlaid on the Marmousi2 velocity model displayed with the absorbing boundaries used for the finite-difference propagation. If no prior geological information is known, we can simply choose to represent the velocity model on a regularly sampled spline grid, as shown in Figure~\ref{fig:mesh_regular}. For this regular mesh, the distance between two consecutive nodes is set to 0.7 km and 1.2 km in the vertical and horizontal directions, respectively. We can also use the fact that there will likely be very little model updates in the absorbing boundaries and therefore reduce the spline grid density in this region of the model. Furthermore, we can leverage prior geological information if the velocity model is likely to contain high-wavenumber features within a specific region, and adapt the spline mesh by increasing the node density within the zone of interest, as shown by the green box in Figure~\ref{fig:mesh_irregular}. The spline nodes cannot be arbitrarily placed and must be arranged in a net disposition. However, each direction can have its own irregular sampling, which gives plenty of flexibility. 

\begin{figure}[t]
    \centering
    \subfigure[]{\label{fig:mesh_velocity_raw}\includegraphics[width=0.45\linewidth]{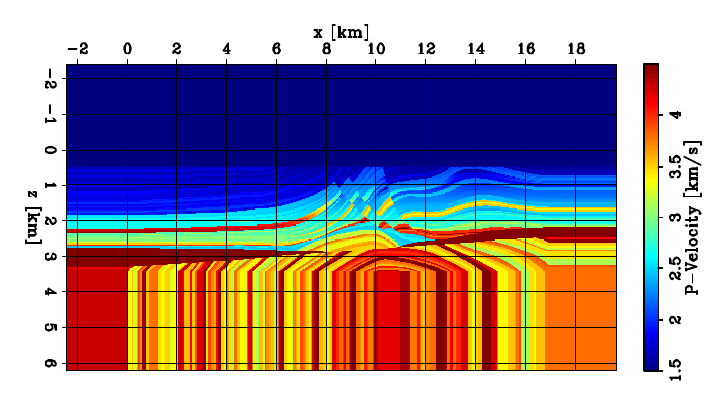}} \\
    \subfigure[]{\label{fig:mesh_velocity_reg_coarse}\includegraphics[width=0.45\linewidth]{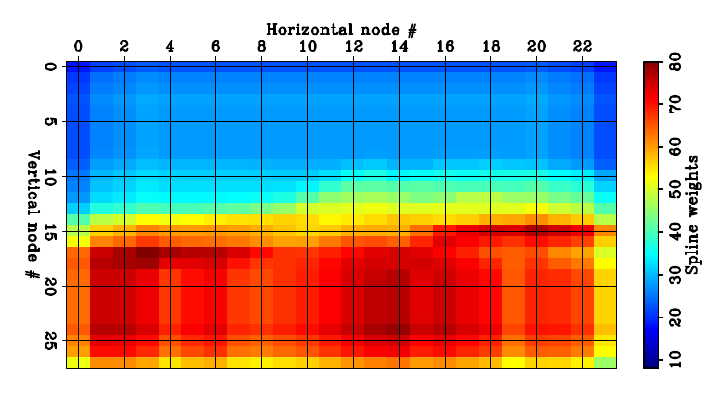}}
    \subfigure[]{\label{fig:mesh_velocity_irreg_coarse}\includegraphics[width=0.45\linewidth]{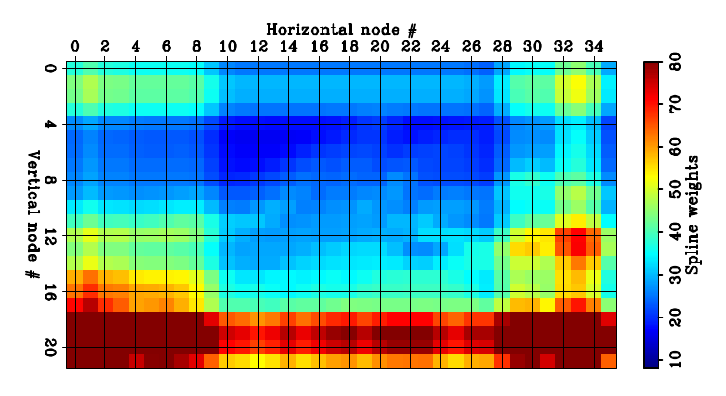}}\\
    \subfigure[]{\label{fig:mesh_velocity_reg_fine}\includegraphics[width=0.45\linewidth]{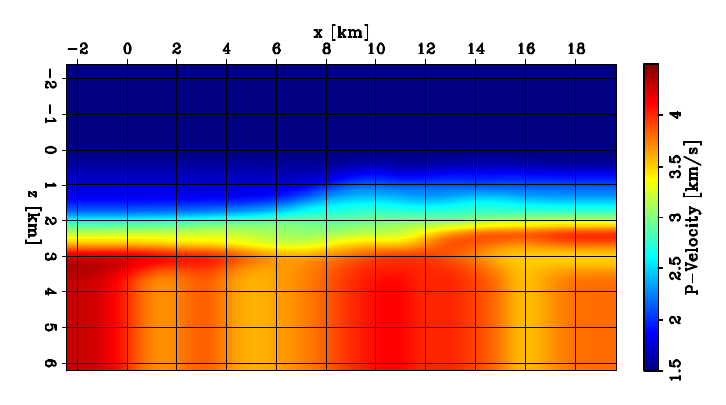}}
    \subfigure[]{\label{fig:mesh_velocity_irreg_fine}\includegraphics[width=0.45\linewidth]{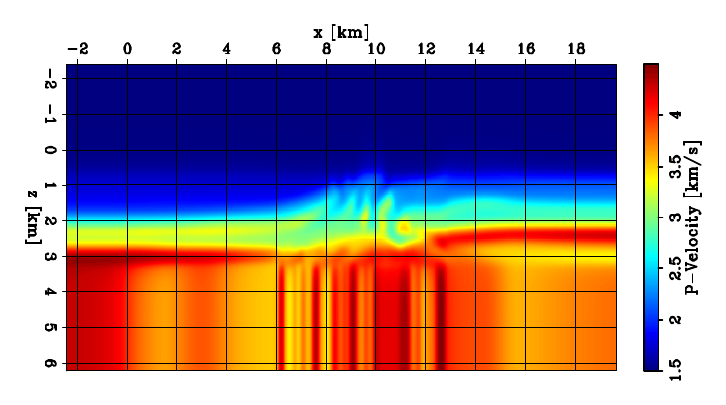}}     
    \caption{2D panels of the Marmousi2 velocity model after sequential applications of the adjoint spline operator $\mathbf{S}^*$ and then $\mathbf{S}$. (a) True velocity model $\mathbf{m}_{true}$. (b) Application of $\mathbf{S}^*$ on $\mathbf{m}_{true}$ for the spline mesh shown in Figure~\ref{fig:mesh_regular}. (c) Application of $\mathbf{S}^*$ on $\mathbf{m}_{true}$ for the spline mesh shown in Figure~\ref{fig:mesh_irregular}. (d) Application of $\mathbf{S}$ on panel (b). (e) Application of $\mathbf{S}$ on panel (c). }
    \label{fig:spline_mapping_velocity}
\end{figure}

Figure~\ref{fig:spline_mapping_velocity} shows the spatial smoothing effect resulting from sequentially applying operator $\mathbf{S}^*$ and then $\mathbf{S}$ to the Marmousi2 velocity model $\mathbf{m}_{true}$ shown in Figure~\ref{fig:mesh_velocity_raw}. Figures~\ref{fig:mesh_velocity_reg_coarse} and \ref{fig:mesh_velocity_irreg_coarse} show the application of $\mathbf{S}^*$ on $\mathbf{m}_{true}$ for the regular mesh (Figure~\ref{fig:mesh_regular}) and irregular mesh (Figure~\ref{fig:mesh_irregular}), respectively. The vertical and horizontal labels correspond to the spline node indices, and the actual value at each grid point can not be interpreted as a velocity field. Figures~\ref{fig:mesh_velocity_reg_fine} and \ref{fig:mesh_velocity_irreg_fine} show the application of the forward mapping $\mathbf{S}$ on the panels shown in Figures~\ref{fig:mesh_velocity_reg_coarse} and \ref{fig:mesh_velocity_irreg_coarse}, respectively. As expected, the effect of mapping the velocity model onto the ``regular" spline grid and then back to the finite-difference grid introduces a strong spatially-homogeneous smoothing effect (Figure~\ref{fig:mesh_velocity_reg_fine}). Moreover, increasing the grid density in certain regions of the model allows the mapping to preserve high-resolution features, as shown in Figure~\ref{fig:mesh_velocity_irreg_fine} (central region of the model). Therefore, one of the main advantages of the B-spline parametrization (compared to more conventional smoothing methods based on wavenumber-domain filtering) is its ability to easily and efficiently apply non-uniform  spatial smoothing for different regions of the velocity model.

\subsubsection{Embedding a spline parametrization within FWIME}
Another advantage of this general framework is that it may also be easily and elegantly implemented in many types of waveform inversion techniques \cite[]{barnier2019waveform}, as well as for different parametrization methods (e.g., RBFs). Here, we apply it to FWIME. In the following, we assume we have already constructed a coarse spline grid $\mathcal{G}_c$, a finite-difference grid $\mathcal{G}_f$, and the corresponding spline operator $\mathbf{S}: \; \mathcal{G}_c \mapsto \mathcal{G}_f$. Recall that the FWIME objective function defined on $\mathcal{G}_f$ is given by equation~\ref{eqn:fwime.obj}

\begin{eqnarray}
    \label{eqn:fwime.obj.fine.grid}
    \Phi_{\epsilon}(\mathbf{m}^f) &=& \dfrac{1}{2} \left\| \mathbf{f}(\mathbf{m}^f) + \tilde{\mathbf{B}} (\mathbf{m}^f) \mathbf{\tilde{p}}_{\epsilon}^{opt}(\mathbf{m}^f) - \mathbf{d}^{obs} \right\|^2_2 + \dfrac{\epsilon^2}{2} \left\| \mathbf{D}{\mathbf{\tilde p}}_{\epsilon}^{opt}(\mathbf{m}^f) \right\|^2_2,
\end{eqnarray}

where $\mathbf{m}^f$ is the velocity model represented on the finite-difference grid $\mathcal{G}_f$. We now modify equation~\ref{eqn:fwime.obj.fine.grid} by re-parametrizing $\Phi_{\epsilon}$ on the spline grid $\mathcal{G}_c$. We introduce the spline operator $\mathbf{S}$ and we substitute $\mathbf{m}^f$ by $\mathbf{S}\mathbf{m}^c$. The new FWIME objective function is now given by

\begin{eqnarray}
    \label{eqn:fwime.obj.coarse.grid}
    \tilde{\Phi}_{\epsilon}(\mathbf{m}^c) &=& \dfrac{1}{2} \left\| \mathbf{f}(\mathbf{S}\mathbf{m}^c) + \tilde{\mathbf{B}} (\mathbf{S}\mathbf{m}^c) \mathbf{\tilde{p}}_{\epsilon}^{opt}(\mathbf{S}\mathbf{m}^c) - \mathbf{d}^{obs} \right\|^2_2 + \dfrac{\epsilon^2}{2} \left\| \mathbf{D}{\mathbf{\tilde p}}_{\epsilon}^{opt}(\mathbf{S}\mathbf{m}^c) \right\|^2_2,
\end{eqnarray}

where $\tilde{\Phi}_{\epsilon}: \, \mathcal{G}_c \mapsto \mathbb{R}^+$, and 

\begin{eqnarray}
    \label{eqn:vp.obj.coarse.grid}
    \mathbf{\tilde{p}}_{\epsilon}^{opt}(\mathbf{S}\mathbf{m}^c) = \underset{\tilde{\mathbf{p}}}{\mathrm{argmin}} \; \dfrac{1}{2} \left \| \tilde{\mathbf{B}}(\mathbf{S}\mathbf{m}^{c})\mathbf{\tilde{p}} - \left ( \mathbf{d}^{obs} - \mathbf{f}(\mathbf{S}\mathbf{m}^{c}) \right ) \right \|^2_2 + \dfrac{\epsilon^2}{2} \left \| \mathbf{D} \tilde{\mathbf{p}} \right \|^2_2.
\end{eqnarray}

In equation~\ref{eqn:fwime.obj.coarse.grid}, the dimension of the search space (i.e., model space) has been reduced from $\mathbb{R}^{N_{m_f}}$ to $\mathbb{R}^{N_{m_c}}$ ($N_{m_c}$ is usually much smaller than $N_{m_f}$). It is important to note that in equations~\ref{eqn:fwime.obj.coarse.grid} and \ref{eqn:vp.obj.coarse.grid}, $\mathbf{\tilde{p}}_{\epsilon}^{opt}$ \textbf{is never parametrized on the spline grid}, but rather on the (finer) finite-difference grid because it may contain all wavenumber components at any stage of the inversion process. Recall that $\mathbf{\tilde{p}}_{\epsilon}^{opt}$ is the mapping into the extended space of all the events present in the observed data $\mathbf{d}^{obs}$ that our modeling $\mathbf{f}(\mathbf{S} \mathbf{m}^c)$ failed to predict. These events can (and will likely) include all types of waves, such as refractions and reflections. 

The gradient of the modified FWIME objective function is obtained by applying the chain rule, which results in mapping the conventional FWIME gradient from the finite-difference grid onto the spline grid,

\begin{eqnarray}
    \label{eqn:fwime.gradient.grid}
    \nabla_{\mathbf{m^{c}}} \tilde{\Phi}_{\epsilon}(\mathbf{\mathbf{m^{c}}}) &=& \mathbf{S}^* \; \nabla_{\mathbf{m}^f} \Phi(\mathbf{m}^{f}).
\end{eqnarray}

\begin{figure}[t]
    \centering
    \subfigure[]{\label{fig:oneLayer3_fwime_BornGrad_spline}\includegraphics[width=0.45\linewidth]{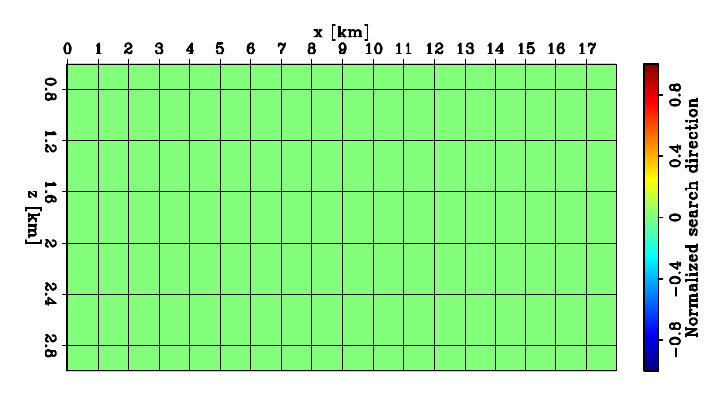}}
    \subfigure[]{\label{fig:oneLayer3_true_tomoGrad_spline}\includegraphics[width=0.45\linewidth]{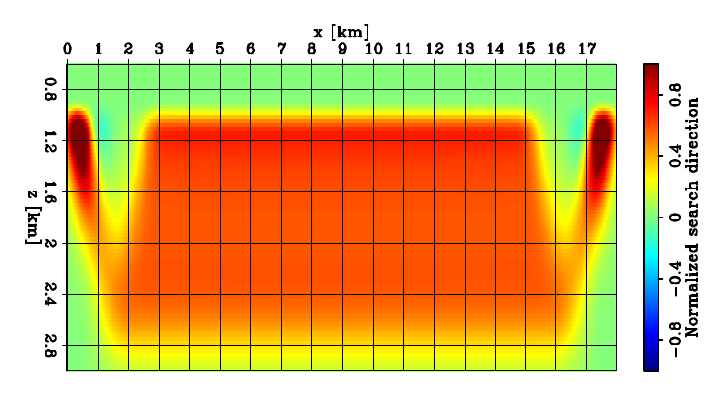}} \\
    \subfigure[]{\label{fig:oneLayer3_fwime_grad_spline}\includegraphics[width=0.45\linewidth]{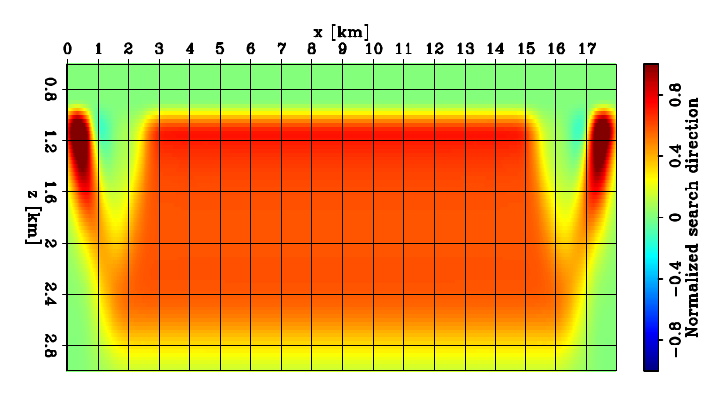}}
    \subfigure[]{\label{fig:oneLayer3_true_grad_spline}\includegraphics[width=0.45\linewidth]{Fig/oneLayer3-grad-true.pdf}}
    \caption{2D panels of initial search directions. (a) FWIME Born search direction obtained by applying $\mathbf{S} \mathbf{S}^*$ to the panel shown in Figure~\ref{fig:oneLayer3_fwime_BornGrad}. (b) FWIME tomographic search direction obtained by applying $\mathbf{S} \mathbf{S}^*$ to the panel shown in Figure~\ref{fig:oneLayer3_fwime_tomoGrad}. (c) FWIME search direction obtained by summing panels (a) and (b). (d) True search direction. Panels (a), (b) and (c) are normalized by the same value and displayed on the same color scale.}
    \label{fig:oneLayer3_grad_spline}
\end{figure}

To illustrate the usefulness of such strategy on the FWIME gradient/search direction, we re-visit the numerical example proposed in the previous section (Figure~\ref{fig:oneLayer3_true_mod}). We generate a spline mesh regularly sampled at $0.3$ km in both directions (whereas the finite-difference grid is sampled at $0.03 $ km in both directions), and we compute the new FWIME search direction according to equation~\ref{eqn:fwime.gradient.grid}. Figures~\ref{fig:oneLayer3_grad_spline}a-c show the new Born, tomographic, and total FWIME search directions displayed on the finite-difference grid, which are obtained by applying operator $\mathbf{S}\mathbf{S}^*$ to the panels shown in Figures~\ref{fig:oneLayer3_grad}a-c, respectively. The FWIME search direction is now solely guided by the tomographic component and accurately captures the missing low wavenumbers. In this numerical example, the spline parametrization behaves in a similar fashion as a high-cut filter (in the spatial frequency domain) and removes the undesired high-wavenumber features introduced by the Born update (Figure~\ref{fig:oneLayer3_fwime_BornGrad}). 

With this new formulation, we now have to construct an initial velocity model on the coarse grid, $\mathbf{m}_0^c \in \mathcal{G}_c$. Naturally, $\mathbf{m}_0^f$ must first be designed on $\mathcal{G}_f$, and then converted to $\mathcal{G}_c$. This is achieved by finding the unique minimizer $\mathbf{m}^c_0 \in \mathcal{G}_c$ that satisfies the following equation:

\begin{eqnarray}
    \label{eqn:fine.to.coarse}
    \mathbf{m}^c_{0} &=& \underset{\mathbf{m}^c \in \mathcal{G}_c}{\mathrm{argmin}} \; \dfrac{1}{2} \left \| \mathbf{S} \mathbf{m}^c - \mathbf{m}^f_{0} \right \|_2^2.
\end{eqnarray}

For a well-chosen grid pair $\left ( \mathcal{G}_c, \mathcal{G}_f \right)$, it can be shown that operator $\mathbf{S}^*\mathbf{S}$ is invertible, which is usually the case when the finite-difference grid is much more densely sampled than the spline grid. 

For a given/fixed spline grid, the workflow we employ to minimize equation~\ref{eqn:fwime.obj.coarse.grid} is summarized in algorithm~\ref{alg:fwime.workflow.spline}. Note that the operations in steps (f) through (i) are conducted on the spline grid. At the end our FWIME workflow, our final inverted model $\mathbf{m}^{c}_{opt}$ must be mapped onto the finite-difference grid for better interpretation/visualization,

\begin{eqnarray}
    \mathbf{m}^f_{opt} = \mathbf{S} \mathbf{m}^c_{opt}.
\end{eqnarray}   

\begin{algorithm}
    \caption{FWIME with B-spline parametrization} 
    \label{alg:fwime.workflow.spline}
    \begin{enumerate}
        \item Select a finite-difference grid $\mathcal{G}_f$
        \item Construct a coarse grid $\mathcal{G}_c$ and its mapping operator $\mathbf{S}$
        \item Select the extension type and the length of the extended axis
        \item Select the hyperparameter $\epsilon$ (fixed throughout the optimization process)        
        \item Design an initial model on the finite-difference grid, $\mathbf{m}^f_{0}$
        \item Compute the initial model on the coarse grid, $\mathbf{m}^c_{0} = \underset{\mathbf{m}^c \in \mathcal{G}_c}{\mathrm{argmin}} \; \dfrac{1}{2} \left \| \mathbf{S} \mathbf{m}^c - \mathbf{m}^f_{0} \right \|_2^2$
        \item For $i = 0,...,n_{iter}-1$
        \label{inversion.one.grid}
            \begin{enumerate}
                \item Map current model estimate onto the finite-difference grid, $\mathbf{m}^{f}_i = \mathbf{S} \, \mathbf{m}^{c}_i$
                \item Compute $\mathbf{d}^{obs} - \mathbf{f}(\mathbf{m}^{f}_i)$
                \item Compute $\tilde{\mathbf{p}}_{\epsilon}^{opt}(\mathbf{m}^{f}_i) = \underset{\tilde{\mathbf{p}}}{\mathrm{argmin}} \; \dfrac{1}{2} \left \| \tilde{\mathbf{B}}(\mathbf{m}^{f}_i)\mathbf{\tilde{p}} - \left ( \mathbf{d}^{obs} - \mathbf{f}(\mathbf{m}^{f}_i) \right ) \right \|^2_2 + \dfrac{\epsilon^2}{2} \left \| \mathbf{D} \tilde{\mathbf{p}} \right \|^2_2$
                \item Set $\mathbf{r}^{\epsilon}_d (\mathbf{m}^f_i) = \mathbf{f}(\mathbf{m}^f_i) + \tilde{\mathbf{B}} (\mathbf{m}^f_i) \mathbf{\tilde{p}}_{\epsilon}^{opt}(\mathbf{m}^f_i) - \mathbf{d}^{obs}$
                \item Compute objective function value $\Phi_{\epsilon}(\mathbf{m}^f_i)$
                \item Compute conventional FWIME gradient with respect to the model parametrized on the spline grid 
                \begin{eqnarray}
                    \nabla_{\mathbf{m}^c} \Phi_{\epsilon}(\mathbf{m}^c_i) = \mathbf{S}^* \mathbf{M} \left [ \mathbf{B}^*(\mathbf{m}^f_i) + \mathbf{T}^*(\mathbf{m}^f_i) \right ] \mathbf{r}^{\epsilon}_d (\mathbf{m}^f_i) \nonumber
                \end{eqnarray}                
                \item Compute search direction on the spline grid $\mathbf{s}^c_i \in \mathcal{G}_c$
                \item Compute step length $\gamma_i$
                \item Update model on the spline grid $\mathcal{G}_c$, $\mathbf{m}^c_{i+1} = \mathbf{m}^c_{i}+\gamma_i \mathbf{s}^c_i$
            \end{enumerate}
    \end{enumerate}
\end{algorithm}

The inversion scheme shown in algorithm~\ref{alg:fwime.workflow.spline} is incorporated into a model-space multi-scale approach where the spline grid is gradually refined throughout the optimization process (the finite-difference grid remains fixed). We start the FWIME workflow with a coarse spline grid $\mathcal{G}_{c_0}$ (along with its corresponding spline operator $\mathbf{S}_0$). We minimize the FWIME objective function for that particular spline grid (algorithm~\ref{alg:fwime.workflow.spline}), and the inverted model $\mathbf{m}^{c_0}_{opt}$ is then used as initial guess for the following inversion performed on the next denser grid $\mathcal{G}_{c_1}$. This multi-scale process is repeated $n_g$ times (where $n_g$ is the number of coarse grids) until the inverted model is satisfactory, or when the coarse grid $\mathcal{G}_{c_{n_g}}$ coincides with the finite-difference grid. We summarize this multi-scale process in algorithm~\ref{alg:fwime.workflow.spline.multi.scale}. 

\begin{algorithm}
    \caption{FWIME with a model-space multi-scale approach} 
    \label{alg:fwime.workflow.spline.multi.scale}
    \begin{enumerate}
        \item Select a finite-difference grid $\mathcal{G}_f$
        \item Construct a collection of $n_g$ spline grids $\{ \mathcal{G}_{c_i}  \}_{0 \leq i < n_g}$ and their respective spline operators $\{ \mathbf{S}_i \}_{0 \leq i < n_g}$
        \item Design an initial model on the finite-difference grid, $\mathbf{m}^f_{0}$
        \item Compute the initial model on the initial coarse grid $\mathcal{G}_{c_0}$, 
        \begin{eqnarray}
            \mathbf{m}^{c_0}_{0} = \underset{\mathbf{m}^{c_0} \in \mathcal{G}_{c_0}}{\mathrm{argmin}} \; \dfrac{1}{2} \left \| \mathbf{S}_0 \mathbf{m}^{c_0} - \mathbf{m}^f_{0} \right \|_2^2\nonumber
        \end{eqnarray}   
        \item Select the extension type and the length of the extended axis
        \item Select the hyperparameter $\epsilon$ (fixed throughout the optimization process)        
        \item For $i = 0,...,n_g-1$
            \begin{enumerate}
                \item Minimize the FWIME objective function on $\mathcal{G}_i$ using $\mathbf{m}^{c_i}_0$ as initial guess and by applying step \ref{inversion.one.grid} of algorithm~\ref{alg:fwime.workflow.spline}, and obtain $\mathbf{m}^{c_i}_{opt}$
                \item Map $\mathbf{m}^{c_i}_{opt}$ onto the finite-difference grid (for wave-propagation), $\mathbf{m}^{f}_{opt, i} = \mathbf{S}_i \, \mathbf{m}^{c_i}_{opt}$
                \item Convert the FWIME inverted model on spline grid $\mathcal{G}_i$ into a model parametrized on the new spline grid, $\mathcal{G}_{i+1}$:
                \begin{eqnarray}
                    \mathbf{m}^{c_{i+1}}_{0} = \underset{\mathbf{m}^{c_{i+1}}\in \mathcal{G}_{c_{i+1}}}{\mathrm{argmin}} \; \dfrac{1}{2} \left \| \mathbf{S}_{i+1} \mathbf{m}^{c_{i+1}} - \mathbf{m}^{f}_{opt,i} \right \|_2^2 \nonumber
                \end{eqnarray}                      
                \item Use $\mathbf{m}^{c_{i+1}}_{0}$ as initial guess for the inversion on $\mathcal{G}_{i+1}$
            \end{enumerate}
    \end{enumerate}
\end{algorithm}

\section{FWIME theory: summary}
The success of FWIME requires the presence of two ingredients: (1) our new loss function formulation, and (2) our model-space multi-scale strategy. The multi-scale strategy by itself is not sufficient to mitigate the presence of local minima because the gradient relies on the useful tomographic component, as illustrated by the numerical examples in this section and thoroughly studied in \cite{barnier2019waveform}. We also show that minimizing our new loss formulation without being able to control the resolution of the model updates can initially introduce spurious high-wavenumber features, which are detrimental. Therefore, in FWIME, there are two fundamental hyper-parameters to adjust: the trade-off parameter $\epsilon$ and the spline grid refinement rate. 

%%%%%%%%%%%% Numerical examples %%%%%%%%%%%%%
\section{Numerical examples}
%%%%%%%%%%%%%%%%%%%%%%%%%%%%%%%%%%%%%%%%%%%%%%%%%%%%%%%%%%%%%%%%%%%%%%%%%%%%%%%%%%%%%%
%%%%%%%%%%%%%%%%%%%%%%%%%%%% Numerical examples %%%%%%%%%%%%%%%%%%%%%%%%%%%%%%%%%%%%%%
%%%%%%%%%%%%%%%%%%%%%%%%%%%%%%%%%%%%%%%%%%%%%%%%%%%%%%%%%%%%%%%%%%%%%%%%%%%%%%%%%%%%%%
We design three 2D synthetic examples (modified from the ones proposed by \cite{barnier2018modified,barnier2018full}) where we illustrate FWIME's ability to accurately and automatically invert simple cycle-skipped datasets composed of one specific type of wave. Our goal is to carefully analyze and show the reader how each wave mode is inverted with the exact same algorithm, without the need to filter/select any specific event from the dataset. In each case, conventional data-space multi-scale FWI converges to unrealistic solutions. In the first example, we simulate a borehole experiment and the dataset solely contains transmitted waves. Then, we re-visit a similar experiment as the one proposed by \cite{mora1989inversion} where reflection data containing wavefront triplications are generated. Finally, we invert a dataset only composed of diving waves. For all three experiments, we generate and invert noise-free pressure data with the same two-way acoustic isotropic constant-density finite-difference propagator. 

%%%%%%%%%%%%%%%%%%%%%%%%%%%%%%%%%%% Borehole %%%%%%%%%%%%%%%%%%%%%%%%%%%%%%%%%%%%%%%%%
\subsection{Inversion of transmitted data}
\begin{figure}[t]
    \centering
    \subfigure[]{\label{fig:oned_data_true}\includegraphics[width=0.3\linewidth]{Fig/sep20-oned-fwime-2-35-slow2-trueData.pdf}}
    \subfigure[]{\label{fig:oned_data_pred_fwd}\includegraphics[width=0.3\linewidth]{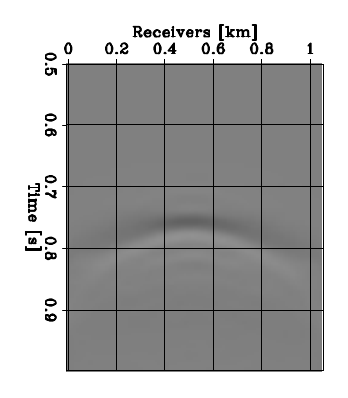}} 
    \subfigure[]{\label{fig:oned_data_diff_fwd}\includegraphics[width=0.3\linewidth]{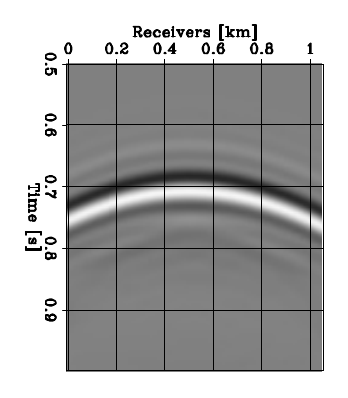}}
    \caption{Representative shot gathers generated by a source placed at $z=0.5$ km in the left borehole. (a) Observed data, $\mathbf{d}^{obs}$. (b) Predicted data computed with the final FWI inverted model, $\mathbf{f}(\mathbf{m}_{FWI})$. (c) Final FWI data residual, $\mathbf{d}^{obs}-\mathbf{f}(\mathbf{m}_{FWI})$. All panels are displayed with the same grayscale.}
    \label{fig:oned_data_fwi}
\end{figure}

We conduct a transmission experiment where we place 20 sources every 50 m inside a 1 km-deep vertical borehole, and 100 receivers every 10 m in a second identical borehole. The distance between the two boreholes is 1 km, and the true velocity model is uniform and set to 2.5 km/s. We generate the dataset with a finite-difference grid spacing of 10 m in both directions. The frequency spectrum of the source is strictly limited to the 9-35 Hz range. The initial velocity model $\mathbf{m}_0$ is uniform and set to 2.0 km/s. Figure~\ref{fig:oned_data_init} shows the observed data $\mathbf{d}^{obs}$, the initial data prediction $\mathbf{f}(\mathbf{m}_0)$, and the initial data difference $\Delta \mathbf{d}(\mathbf{m}_0)$ for a shot gather generated by a source placed at $z = 0.5$ km in the left borehole. As expected, conventional data-space multi-scale FWI converges to a local minimum, and the final FWI data-residuals, $\mathbf{d}^{obs}-\mathbf{f}(\mathbf{m}_{FWI})$ are cycle-skipped (Figures~\ref{fig:oned_data_fwi}b and c). 

\begin{figure}[t]
    \centering
    \subfigure[]{\label{fig:oned_data_true}\includegraphics[width=0.3\linewidth]{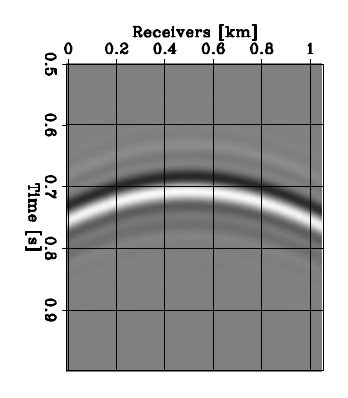}} 
    \subfigure[]{\label{fig:oned_data_init}\includegraphics[width=0.3\linewidth]{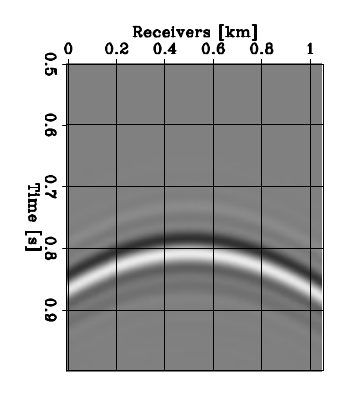}}
    \subfigure[]{\label{fig:oned_data_initDiff}\includegraphics[width=0.3\linewidth]{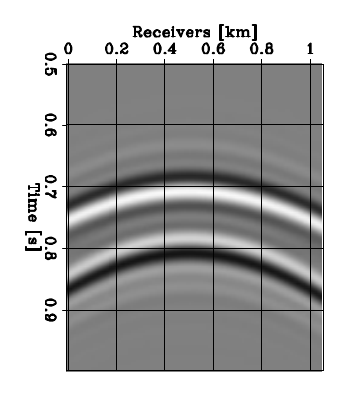}}
    \caption{Representative shot gathers generated by a source placed at $z=0.5$ km in the left borehole. (a) Observed data, $\mathbf{d}^{obs}$. (b) Initial prediction, $\mathbf{f}(\mathbf{m}_0)$. (c) Initial data difference, $\mathbf{d}^{obs}-\mathbf{f}(\mathbf{m}_0)$. All panels are displayed with the same grayscale.}
    \label{fig:oned_data_init}
\end{figure}

\begin{figure}[t]
    \centering
    \subfigure[]{\label{fig:oned_objective_functions_all}\includegraphics[width=0.45\linewidth]{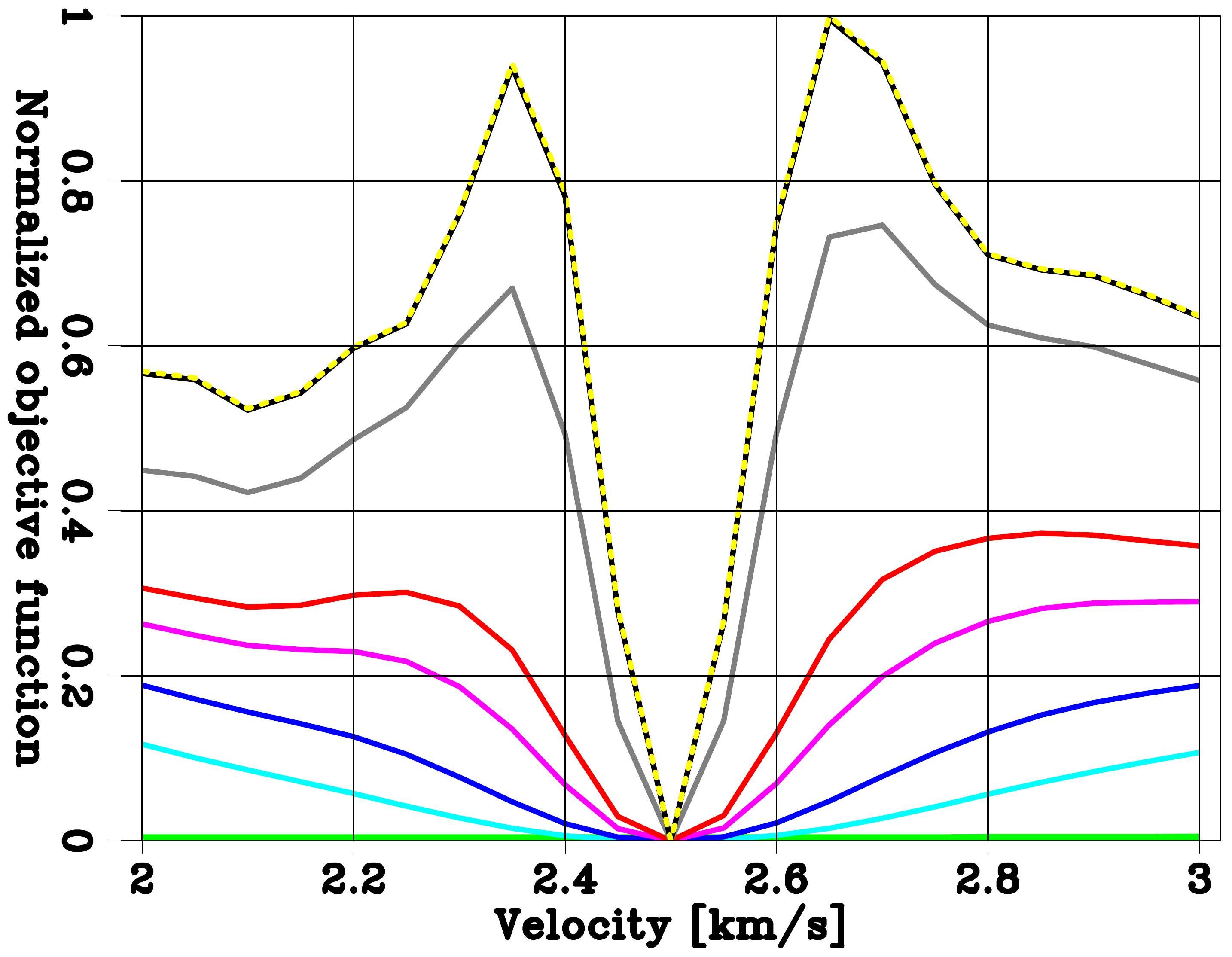}} \hspace{5mm}
    \subfigure[]{\label{fig:oned_objective_functions_eps3}\includegraphics[width=0.45\linewidth]{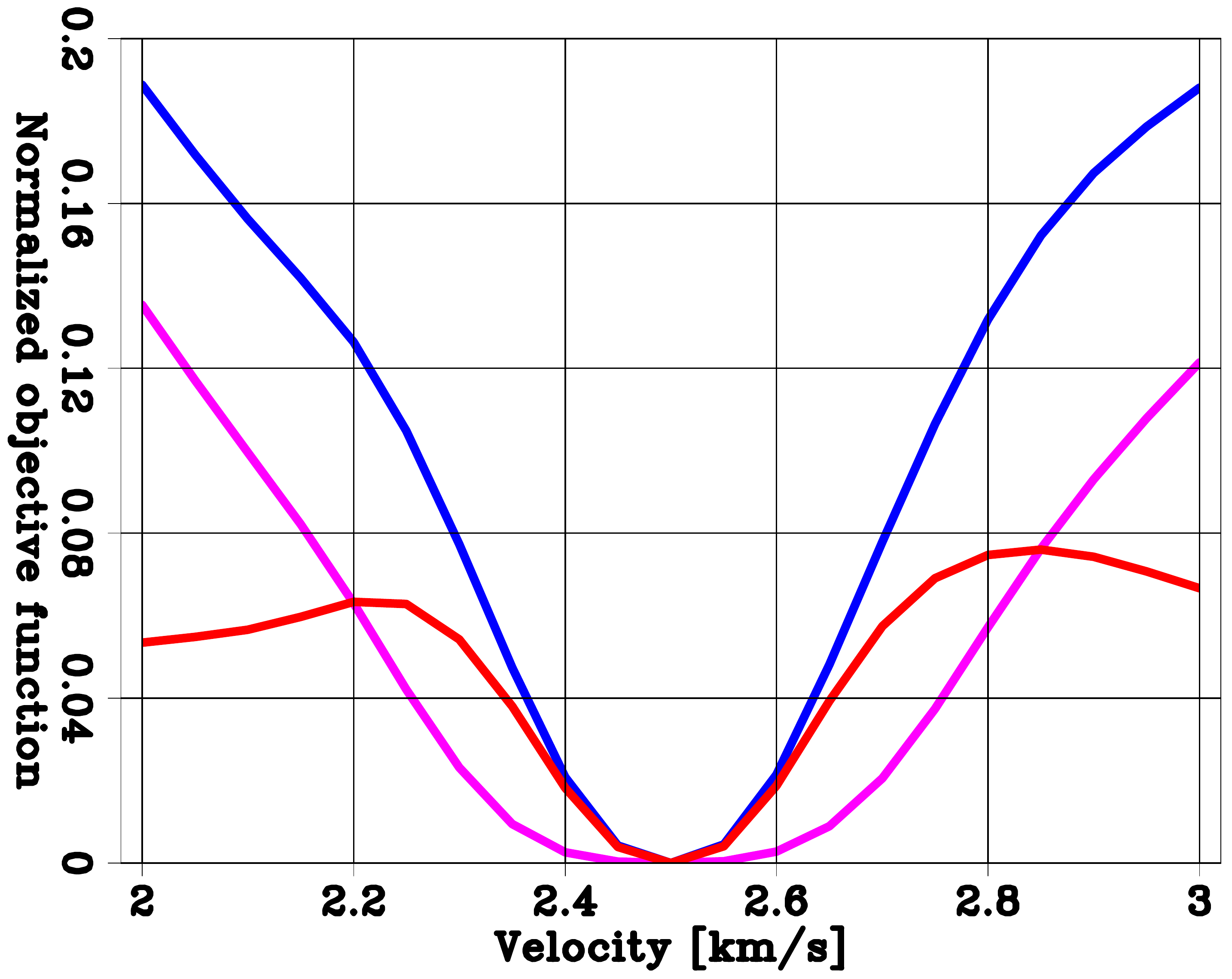}} 
    \caption{Normalized objective functions computed for homogeneous models ranging from 2.0 km/s to 3.0 km/s, using the full data bandwidth. (a) FWIME objective functions computed with increasing $\epsilon$-values: $\epsilon_1=0$ (green curve), $\epsilon_2 = 2.0 \times 10^{-7}$ (light-blue curve), $\epsilon_3=1.5 \times 10^{-6}$ (dark-blue curve), $\epsilon_4 = 1.0 \times 10^{-6}$ (pink curve), $\epsilon_5 = 2.0 \times 10^{-6}$ (red curve), $\epsilon_6 = 1.0 \times 10^{-5}$ (grey curve), and $\epsilon_7 = 1.0 \times 10^{-4}$ (solid black curve). The yellow dashed curve shows the conventional FWI objective function. (b) Components of the FWIME objective function (equation 4) computed with $\epsilon = \epsilon_3$ (dark blue curve in panel (a)). The blue curve shows the total FWIME objective function, the pink curve is the data-fitting component, and the red curve is the annihilating component scaled by $\displaystyle \frac{\epsilon^2}{2}$.}
    \label{fig:oned_objective_functions}
\end{figure}

We sample the FWI and FWIME objective functions (using the full data bandwidth for both cases) for uniform velocity models ranging from 2.0 km/s to 3.0 km/s by increments of 0.05 km/s, and for seven $\epsilon$-values ranging from $\epsilon_1 = 0$ to $\epsilon_7 = 1.0 \times 10^{-4}$ (Figure~\ref{fig:oned_objective_functions_all}). As expected, the FWI objective function presents local minima (yellow dashed curve in Figure~\ref{fig:oned_objective_functions_all}), but for certain $\epsilon$-values, the FWIME objective function is monotonically decreasing toward the global solution (dark- and light-blue curves in Figure~\ref{fig:oned_objective_functions_all}). For these $\epsilon$-values, the FWIME formulation managed to remove all local minima (for this range of models) and guarantees global convergence for gradient-based methods when inverting a scalar parameter. Figure~\ref{fig:oned_objective_functions_eps3} displays the three components of the FWIME objective function computed with $\epsilon = 1.5 \times 10^{-6}$, which corresponds to the dark-blue curve in Figure~\ref{fig:oned_objective_functions_all}. The data-fitting component has been convexified (pink curve), and the local minima are now carried by the annihilating component (red curve). However, the total objective function is now free of local minima. As expected, when $\epsilon = 0$, the FWIME objective function is approximately constant and equal to zero (green curve in Figure~\ref{fig:oned_objective_functions_all}): the data-correcting term satisfies equation~\ref{eqn:data.matching}, and the FWIME data-fitting term vanishes for all velocity models $\mathbf{m}$. Conversely, as the $\epsilon$-value increases, the FWIME objective function converges pointwise to the FWI objective function, which illustrates the property shown in Appendix~\ref{fwimeToFwi}. In fact, for $\epsilon_7 = 1.0 \times 10^{-4}$, the FWI and FWIME objective functions are already nearly identical (solid black curve and yellow dashed curve in Figure~\ref{fig:oned_objective_functions_all}). 

\begin{figure}[t]
    \centering
    \includegraphics[width=0.6\linewidth]{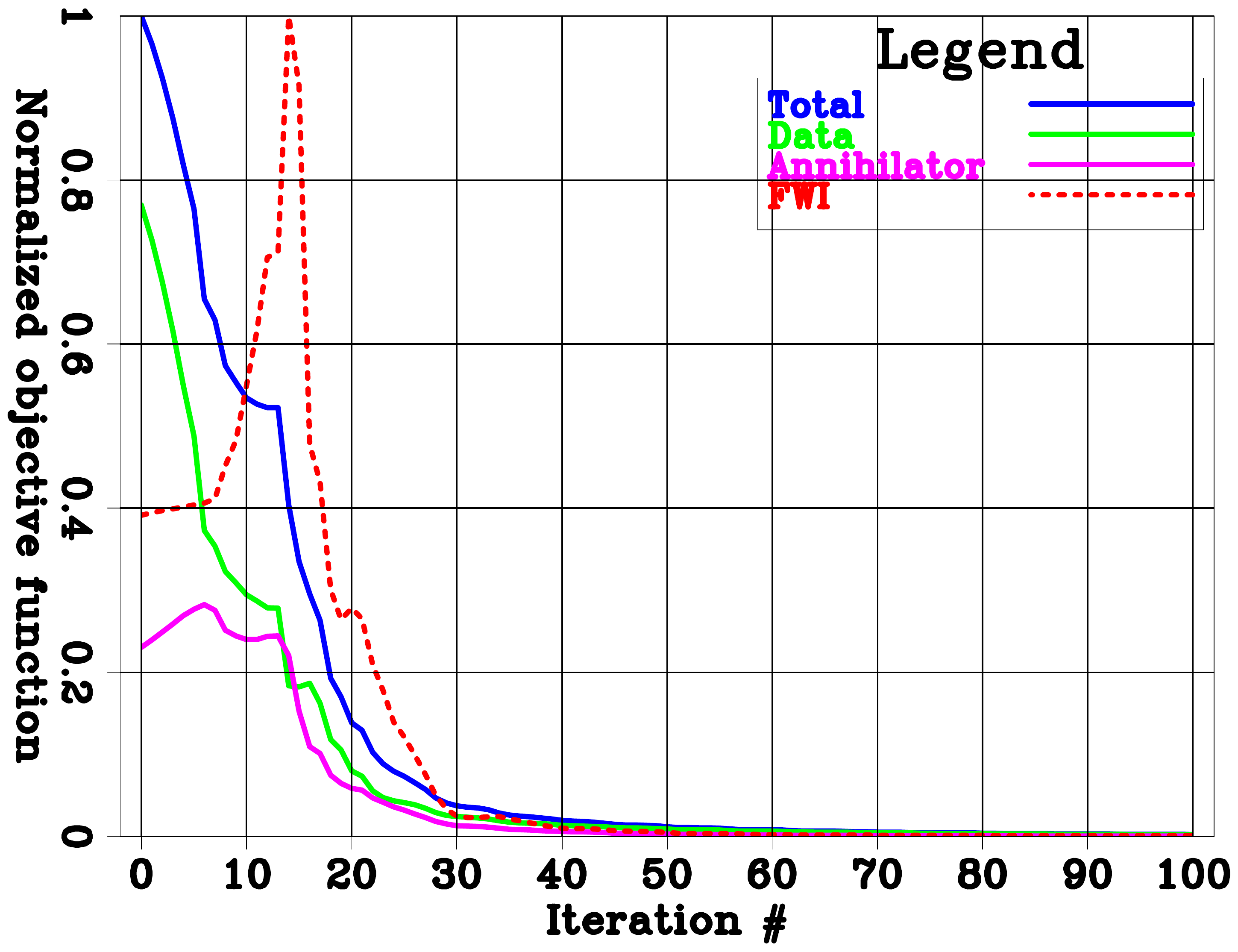}
    \caption{Convergence curves as a function of iterations. The solid curves correspond to the three components of the FWIME objective function obtained by simultaneously inverting the full bandwidth of the transmission dataset with $\epsilon = 1.5 \times 10^{-6}$. The blue curve corresponds to the total FWIME objective function, the red curve is the data-fitting component, and the pink curve is the annihilating component. The red dashed curve shows the value of the conventional FWI objective function evaluated at each inverted FWIME model (it is not the output of an optimization process).}
    \label{fig:oned_fwime_obj_total}
\end{figure}

\begin{figure}[t]
    \centering
    \label{fig:oned_data_true}\includegraphics[width=0.6\linewidth]{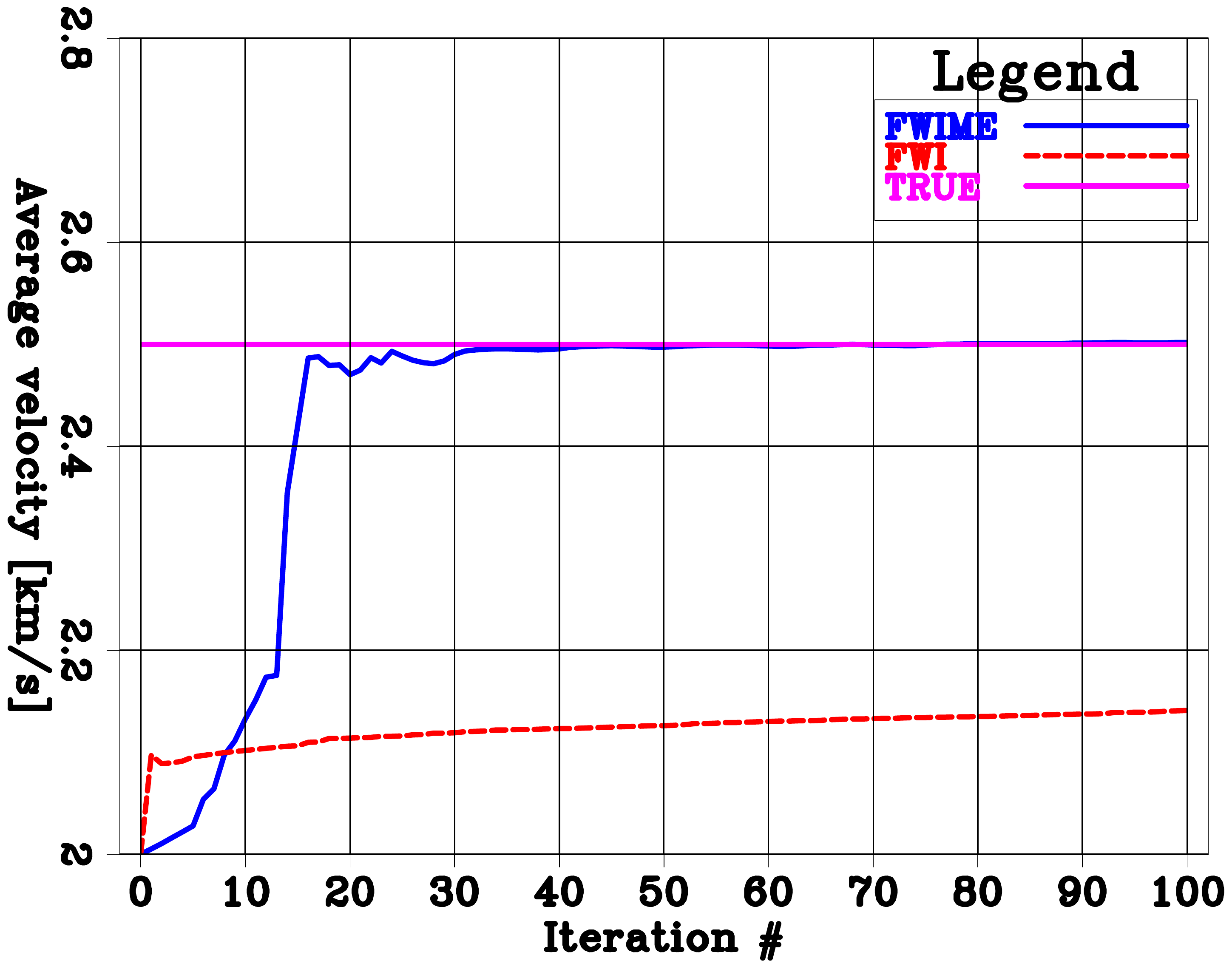}
    \caption{Inverted average velocity as a function of iterations for FWIME (solid blue curve) and conventional FWI (red dashed curve). The pink curve shows the true velocity value, 2.5 km/s.}
    \label{fig:oned_fwime_avg}
\end{figure}

We conduct 100 iterations of the FWIME workflow by simultaneously inverting the full data bandwidth, starting with the same uniform velocity model $\mathbf{m}_0$ set to 2.0 km/s. For this specific numerical example, FWIME successfully retrieves an accurate solution without the need to employ the model-space multi-scale strategy (the unknown velocity model is thus parametrized directly on the finite-difference grid from the start). We do not assume spatial uniformity of the velocity model and we invert for all $N_m = 10^{4}$ unknown velocity model parameters. We use a time-lag extended axis with 81 points sampled at $\Delta \tau = 4$ ms. The variable projection step is performed with 30 iterations of linear conjugate gradient. Figure~\ref{fig:oned_fwime_obj_total} shows the resulting FWIME convergence curves (solid lines). All three components of the objective function converge to zero, which indicates that the scheme has successfully converged to the global solution. On the same plot, we superimpose the FWI objective function evaluated at each iteration of FWIME (red dashed line). This curve is not monotonically decreasing and is not the result of an inversion process. It shows the values that the FWI objective function would have taken for this sequence of inverted models. These observations indicate that in this case, the FWIME optimization path is insensitive to the local minima present in the conventional FWI objective function. This analysis is also confirmed by the average velocity of the inverted models from the two optimization schemes (Figure~\ref{fig:oned_fwime_avg}). We choose this model metric because of the inherent uncertainty in the conventional traveltime tomography problem \cite[]{squires1994borehole,barnier2018modified}. 

\begin{figure}[t]
    \centering
    \subfigure[]{\label{fig:oned_data_res_fwime0}\includegraphics[width=0.18\linewidth]{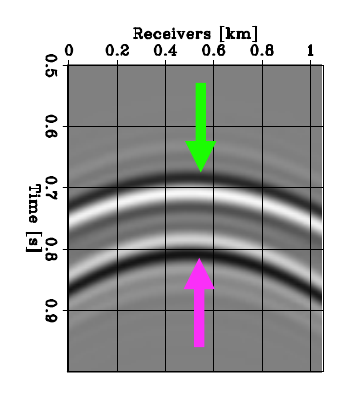}}
    \subfigure[]{\label{fig:oned_data_res_fwime10}\includegraphics[width=0.18\linewidth]{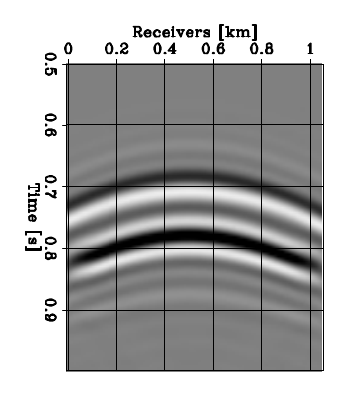}}
    \subfigure[]{\label{fig:oned_data_res_fwime15}\includegraphics[width=0.18\linewidth]{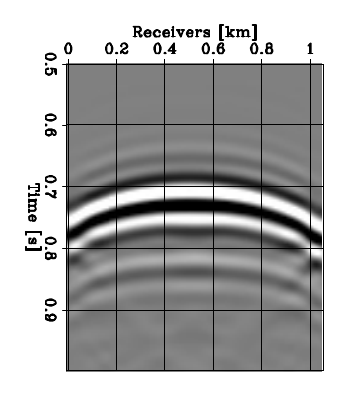}}  
    \subfigure[]{\label{fig:oned_data_res_fwime20}\includegraphics[width=0.18\linewidth]{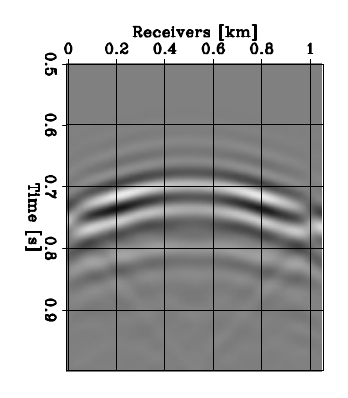}}
    \subfigure[]{\label{fig:oned_data_res_fwime100}\includegraphics[width=0.18\linewidth]{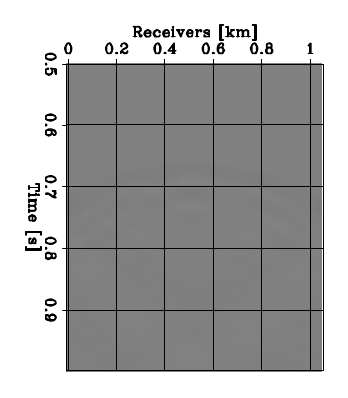}}         
    \caption{Shot gathers generated by a source placed at $z= 0.5$ km in the left borehole. Each panel corresponds to the difference between observed data and predicted data computed with the inverted FWIME model $\mathbf{m}_i$ at five stages of the optimization workflow, $\Delta \mathbf{d} (\mathbf{m}_i) = \mathbf{d}^{obs} - \mathbf{f}(\mathbf{m}_i)$. (a) Initial step, (b) iteration 10, (c) iteration 15, (d) iteration 20, and (e) iteration 100 (last iteration). All panels are displayed with the same grayscale.}
    \label{fig:one_data_fwime}
\end{figure}

Figure~\ref{fig:one_data_fwime} shows the difference between observed and predicted data (i.e., $\Delta \mathbf{d} (\mathbf{m}_i) = \mathbf{d}^{obs} - \mathbf{f}(\mathbf{m}_i)$) computed with the FWIME inverted models at iterations $i=$ 0, 10, 15, 20, and 100. In Figure~\ref{fig:oned_data_res_fwime0}, we can see that initial model largely underestimates the true velocity value (i.e., $\mathbf{m}_0 < \mathbf{m}_{true}$) and the data are cycle-skipped. The green and pink arrows correspond to the observed data $\mathbf{d}^{obs}$ and predicted data $\mathbf{f}(\mathbf{m}_0)$, respectively. At iteration 10, the velocity model has been updated in the correct direction and the time-shift between predicted and observed data has shrunk (Figure~\ref{fig:oned_data_res_fwime10}). At iteration 15, the two events begin to overlap with a misaligned phase, which corresponds to the increase in the FWI objective function in Figure~\ref{fig:oned_objective_functions_all} (yellow dashed curve) and Figure~\ref{fig:oned_fwime_obj_total} (red dashed curve). This effect begins to disappear at iteration 20 as the phase of the predicted and observed data start to align (Figure~\ref{fig:oned_data_res_fwime20}). At the last iteration, the predicted and observed data are almost identical (Figure~\ref{fig:oned_data_res_fwime100}), which indicates that the FWIME workflow has converged to the global solution.

\begin{figure}[t]
    \centering
    \subfigure[]{\label{fig:oned_fwime_cig_init}\includegraphics[width=0.18\linewidth]{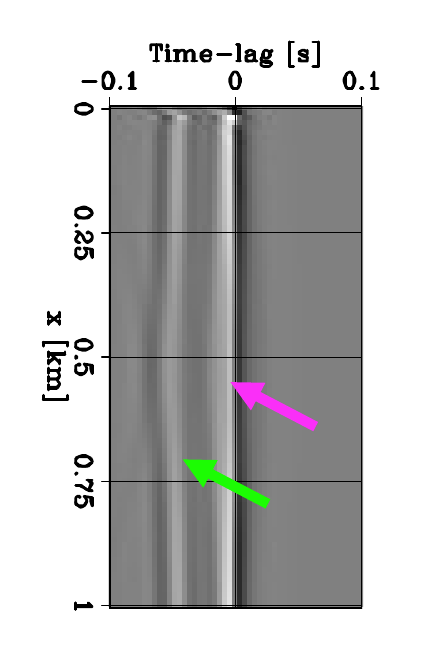}}
    \subfigure[]{\label{fig:oned_fwime_cig_it10}\includegraphics[width=0.18\linewidth]{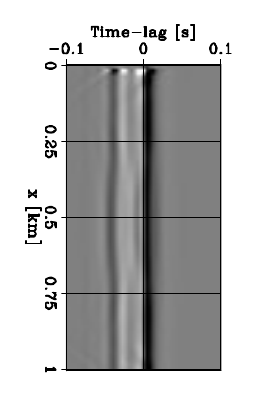}}
    \subfigure[]{\label{fig:oned_fwime_cig_it15}\includegraphics[width=0.18\linewidth]{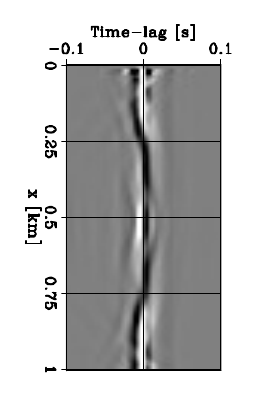}}    
    \subfigure[]{\label{fig:oned_fwime_cig_it20}\includegraphics[width=0.18\linewidth]{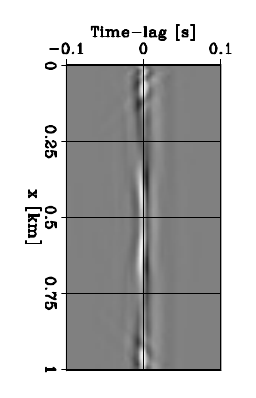}}
    \subfigure[]{\label{fig:oned_fwime_cig_it100}\includegraphics[width=0.18\linewidth]{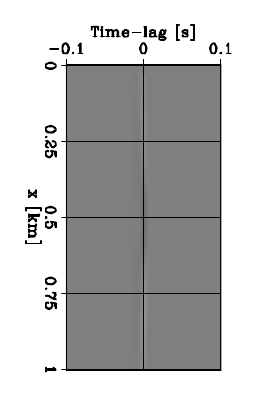}}    
    \caption{TLCIGs extracted at $z = 0.5$ km from $\mathbf{\tilde{p}}_{\epsilon}^{opt}(\mathbf{m}_i)$ at five stages of the FWIME process. (a) Initial step, (b) iteration 10, (c) iteration 15, and (d) iteration 20, and (e) iteration 100 (last iteration). All panels are displayed with the same grayscale.}
    \label{fig:oned_fwime_cig}
\end{figure}

The simplicity of this dataset allows us to easily identify each event and provides us with better insight on the connection between data and extended space. We conduct an analogous step-by-step analysis of $\mathbf{\tilde{p}}_{\epsilon}^{opt}(\mathbf{m}_i)$. Figure~\ref{fig:oned_fwime_cig} shows the evolution of a TLCIG extracted at $z = 0.5$ km from $\mathbf{\tilde{p}}_{\epsilon}^{opt}(\mathbf{m}_i)$ computed at iterations $i=$ 0, 10, 15, 20, and 100 of the FWIME inversion process. At the initial step (Figure~\ref{fig:oned_fwime_cig_init}), we observe the presence of two separate vertical clusters of energy, which correspond to the mapping (by minimizing equation~\ref{eqn:vp.obj}) of the two events from the data space (green and pink arrows in Figure~\ref{fig:oned_data_res_fwime0}) into $\mathbf{\tilde{p}}_{\epsilon}^{opt}(\mathbf{m}_{0})$. First, the event corresponding to $\mathbf{f}(\mathbf{m}_0)$ (pink arrow in Figure~\ref{fig:oned_data_res_fwime0}) is mapped into $\mathbf{\tilde{p}}_{\epsilon}^{opt}(\mathbf{m}_{0})$ at $\tau=0$ s: no extension is needed to generate such an event because all modeled wavefields propagate with $\mathbf{m}_0$. The second vertical cluster of energy is located away from the physical axis (green arrow in Figure~\ref{fig:oned_fwime_cig_init}) and corresponds the mapping of the observed data $\mathbf{d}^{obs}$ (green arrow in Figure~\ref{fig:oned_data_res_fwime0}) into $\mathbf{\tilde{p}}_{\epsilon}^{opt}(\mathbf{m}_{0})$. In this case, an extension is required to generate an event with an apparent propagation velocity $\mathbf{m}_{true}$. Moreover, the fact that the energy focuses at negative values of $\tau$ confirms that our velocity model $\mathbf{m}_0$ is too slow. Note that all 100 shot gathers such as the one displayed in Figure~\ref{fig:oned_data_res_fwime0} are employed to compute $\mathbf{\tilde{p}}_{\epsilon}^{opt}(\mathbf{m}_{0})$ when minimizing equation~\ref{eqn:vp.obj}. As the optimization progresses and the velocity model becomes more accurate, the energy within $\mathbf{\tilde{p}}_{\epsilon}^{opt}(\mathbf{m}_i)$ begins to diminish (starting from time lags with larger magnitude) and gradually focuses toward the physical axis, as shown in Figures~\ref{fig:oned_fwime_cig}b-d. At the end of the FWIME workflow (iteration 100), the energy within $\mathbf{\tilde{p}}_{\epsilon}^{opt}(\mathbf{m}_{100})$ completely vanishes (Figure~\ref{fig:oned_fwime_cig_it100}). 

%%%%%%%%%%%%%%%%%%%%%%%%%%%%%%%%%%%%% Mora %%%%%%%%%%%%%%%%%%%%%%%%%%%%%%%%%%%%%%%%%%%
\subsection{Inversion of reflection data}
\begin{figure}[t]
    \centering
    \subfigure[]{\label{fig:Mora_init}\includegraphics[width=0.45\linewidth]{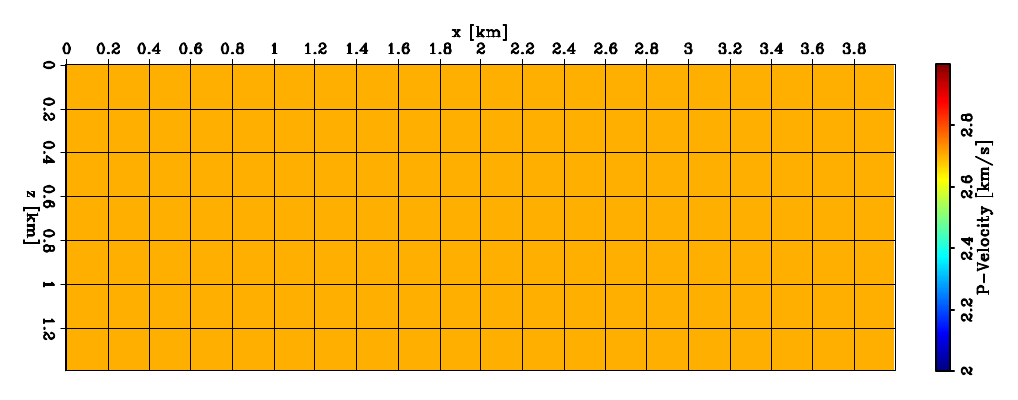}} 
    \subfigure[]{\label{fig:Mora_fwi}\includegraphics[width=0.45\linewidth]{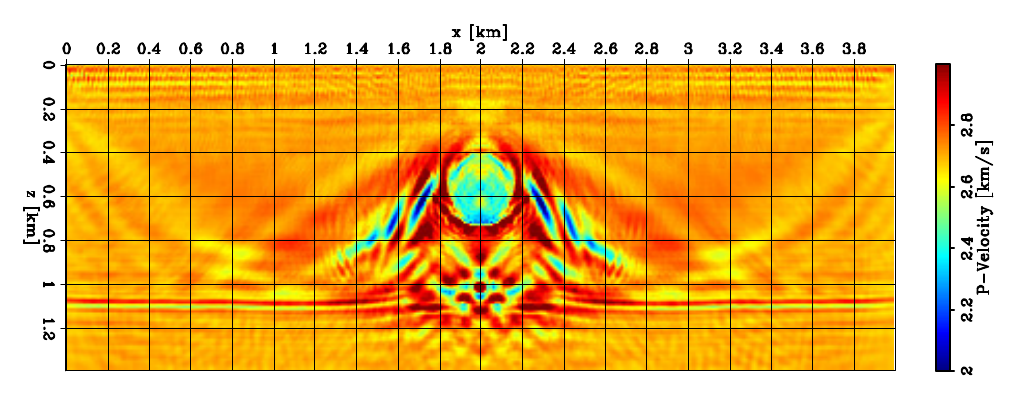}}\\
    \subfigure[]{\label{fig:Mora_fwime_no_spline}\includegraphics[width=0.45\linewidth]{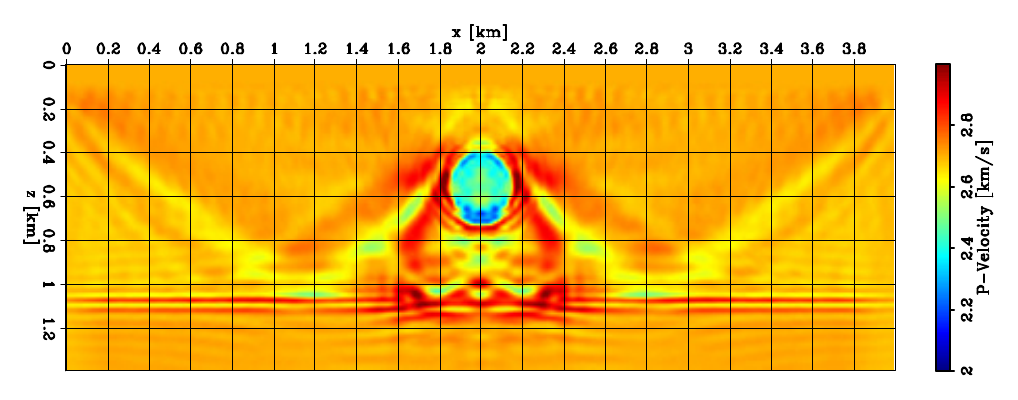}}        
    \subfigure[]{\label{fig:Mora_true}\includegraphics[width=0.45\linewidth]{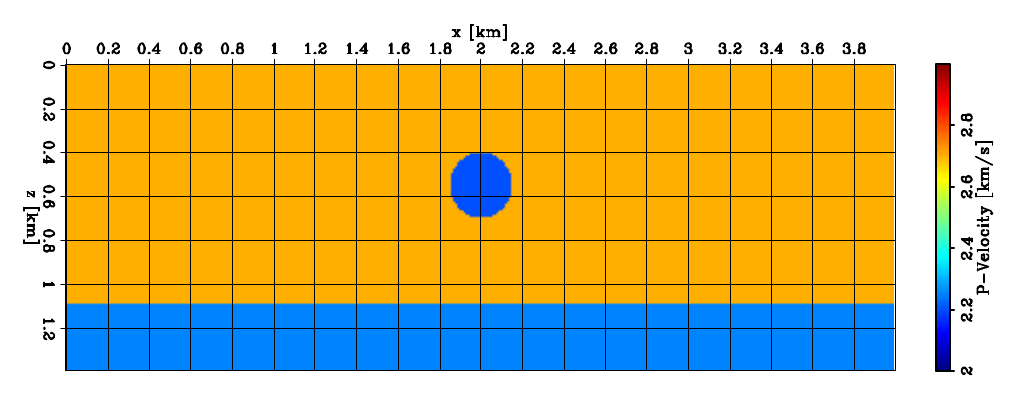}}    
    \caption{2D panels of velocity models (based on \cite{mora1989inversion}). (a) Initial model. (b) Inverted model obtained by conducting conventional data-space multi-scale FWI (four frequency bands). (c) FWIME inverted model without any multi-scale strategy. (d) True model.}
    \label{fig:Mora_mod}
\end{figure}

We test FWIME on a reflection-dominated dataset generated by a model similar to the one shown in \cite{mora1989inversion}. We use this numerical example to show the benefits of combining FWIME with our model-space multi-scale approach. The true model is 4 km wide and is composed of two homogeneous horizontal layers with velocity values of 2.7 km/s and 2.25 km/s, respectively. The interface between the two horizontal layers is located at a depth of 1.1 km. In the top layer, we embed a circular-shaped low-velocity anomaly with sharp contours and a velocity value of 2.2 km/s, which is $19\%$ lower than the top-layer velocity value (Figure~\ref{fig:Mora_true}). The initial model $\mathbf{m}_0$ is homogeneous and set to 2.7 km/s (Figure~\ref{fig:Mora_init}). 

\begin{figure}[t]
    \centering
    \subfigure[]{\label{fig:Mora_datDiffInit}\includegraphics[width=0.3\linewidth]{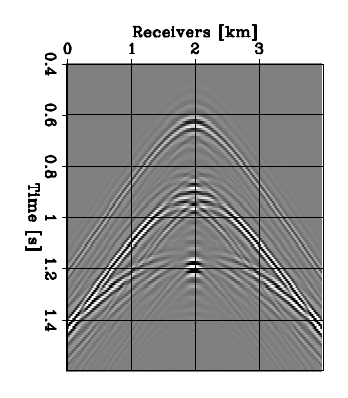}}
    \subfigure[]{\label{fig:Mora_datDiffFwi}\includegraphics[width=0.3\linewidth]{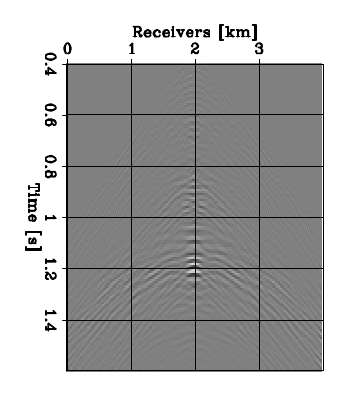}}
    \subfigure[]{\label{fig:Mora_datDiffFwime}\includegraphics[width=0.3\linewidth]{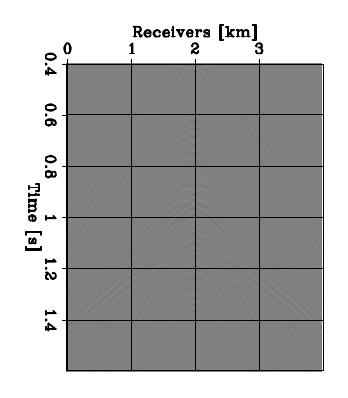}}    
    \caption{Shot gathers generated with a source located at $x = 2$ km showing the difference between observed and predicted data, $\Delta \mathbf{d}(\mathbf{m})=\mathbf{d}^{obs} - \mathbf{f}(\mathbf{m})$, computed with different velocity models. (a) Initial model (Figure~\ref{fig:Mora_init}). (b) FWI inverted model (Figure~\ref{fig:Mora_fwi}). (c) FWIME inverted model with a model-space multi-scale approach (Figure~\ref{fig:Mora_fwime_spline3}). All panels are displayed with the same grayscale.}
    \label{fig:Mora_data}
\end{figure}

The noise-free pressure data are generated using a finite-difference grid spacing of $\Delta z = \Delta x = 10$ m, and with a source containing energy strictly restricted to the 20-50 Hz frequency range. We choose this unrealistic frequency range to ensure that conventional multi-scale FWI fails to retrieve a physical solution. We set 40 sources and 400 receivers at the surface with a spacing of 100 m and 10 m, respectively.  Figure~\ref{fig:Mora_datDiffInit} shows the initial data difference, $\Delta \mathbf{d}(\mathbf{m}_0) = \mathbf{d}^{obs}-\mathbf{f}(\mathbf{m}_0)$ for a source placed at $x=2$ km. Besides reflected energy, triplications stemming from the presence of the low-velocity anomaly can also be observed in the recorded events. 

We first conduct a conventional data-space multi-scale FWI workflow using four frequency bands spanning the 20-50 Hz frequency range. The inverted model, shown in Figure~\ref{fig:Mora_fwi}, indicates that FWI has converged to a local minimum. Figure~\ref{fig:Mora_datDiffFwi} displays the data-residual computed at the last iteration of FWI, which confirms that the inverted model is unable to accurately predict the complex waveform (i.e., the triplications in the wavefield) generated by the presence of the low-velocity anomaly.

\begin{figure}[t]
    \centering
    \subfigure[]{\label{fig:Mora_grad_born}\includegraphics[width=0.45\linewidth]{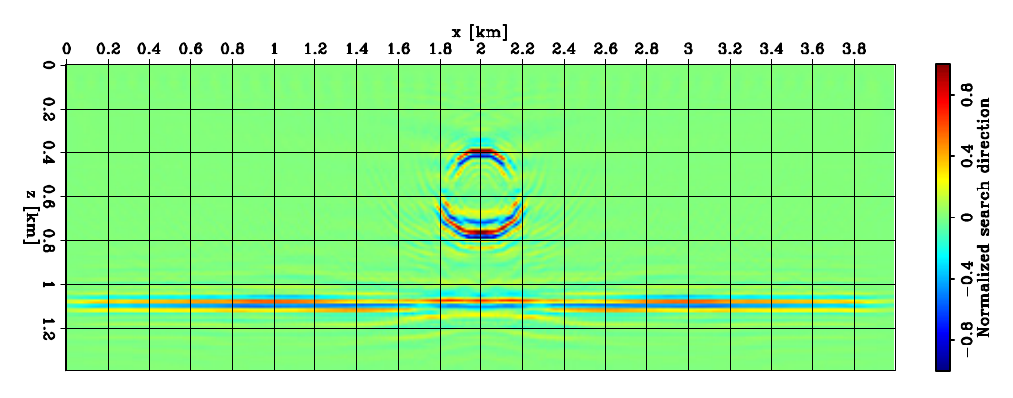}}
    \subfigure[]{\label{fig:Mora_grad_tomo}\includegraphics[width=0.45\linewidth]{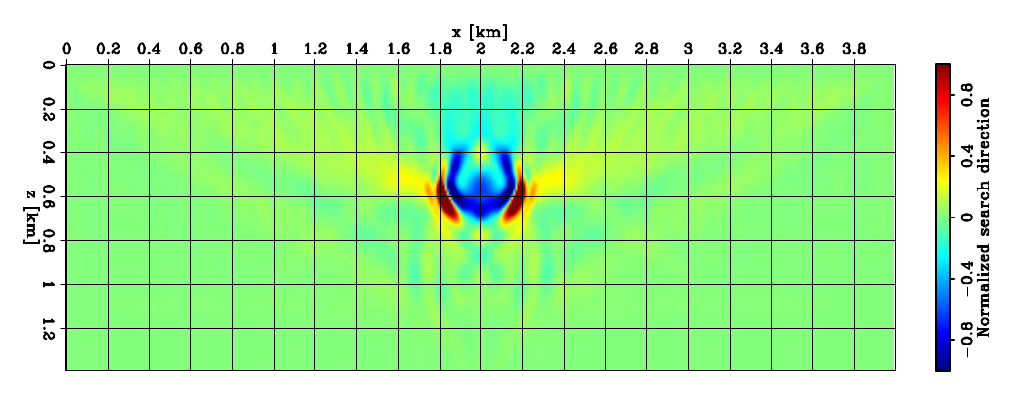}} \\
    \subfigure[]{\label{fig:Mora_grad_total}\includegraphics[width=0.45\linewidth]{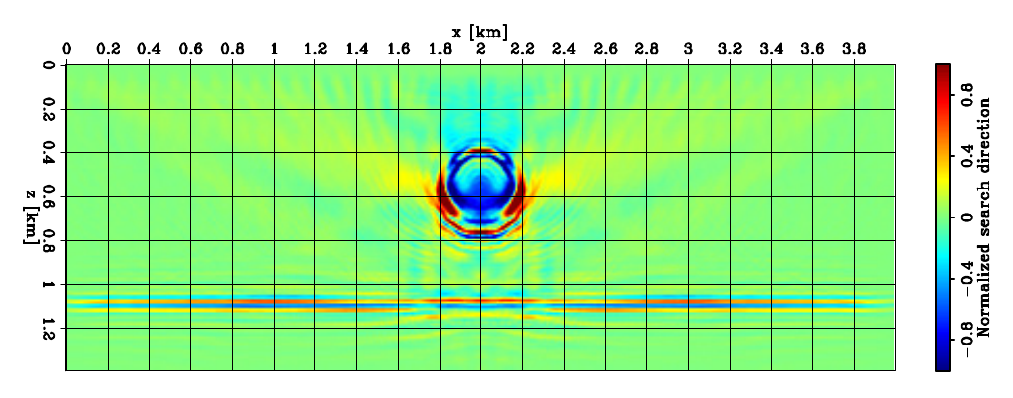}}
    \subfigure[]{\label{fig:Mora_grad_true}\includegraphics[width=0.45\linewidth]{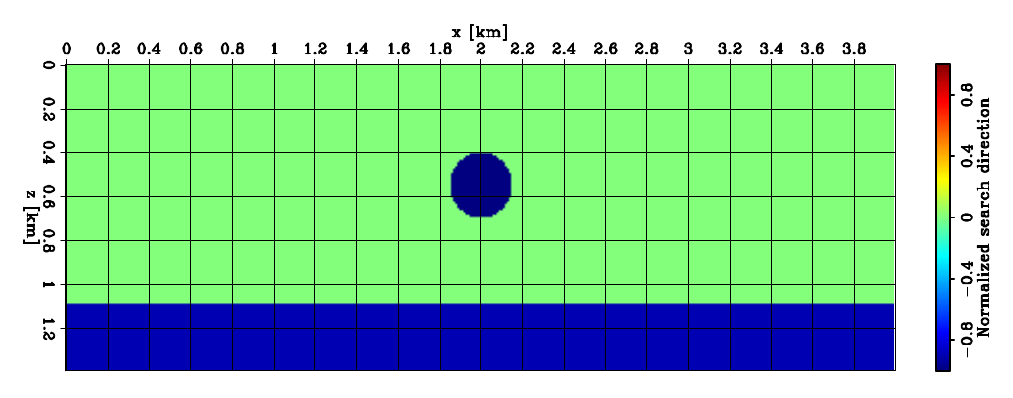}}
    \caption{2D panels of normalized initial search directions. (a) FWIME Born search direction. (b) FWIME tomographic search direction. (c) FWIME total search direction. (d) Ideal search direction. Panels (a), (b), and (c) are normalized by the same value and displayed on the same color scale.}
    \label{fig:Mora_grad}
\end{figure}

\begin{figure}[t]
    \centering
    \subfigure[]{\label{fig:Mora_grad_fwi_spectrum}\includegraphics[width=0.30\linewidth]{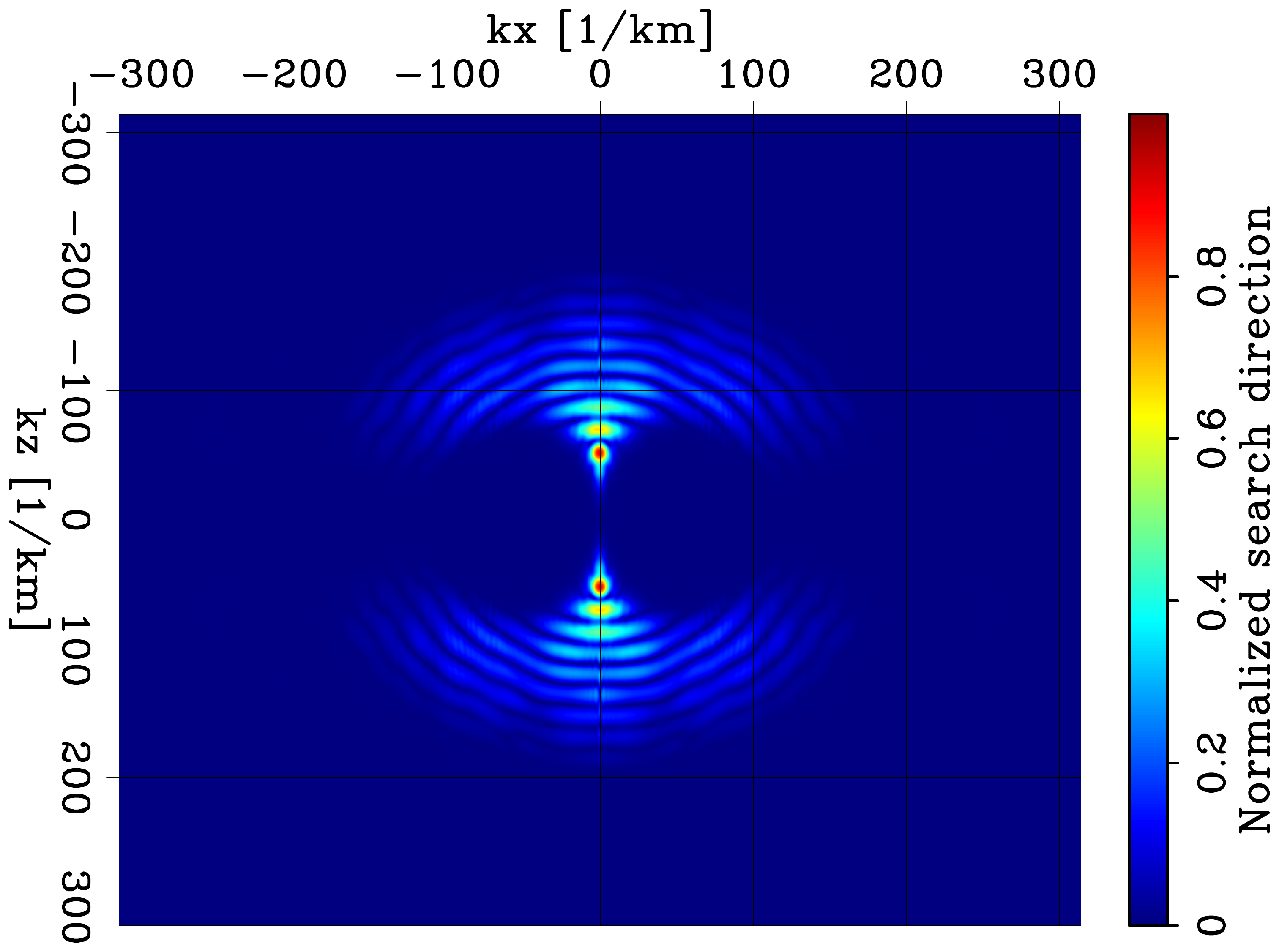}} \hspace{5mm}     
    \subfigure[]{\label{fig:Mora_grad_born_spectrum}\includegraphics[width=0.30\linewidth]{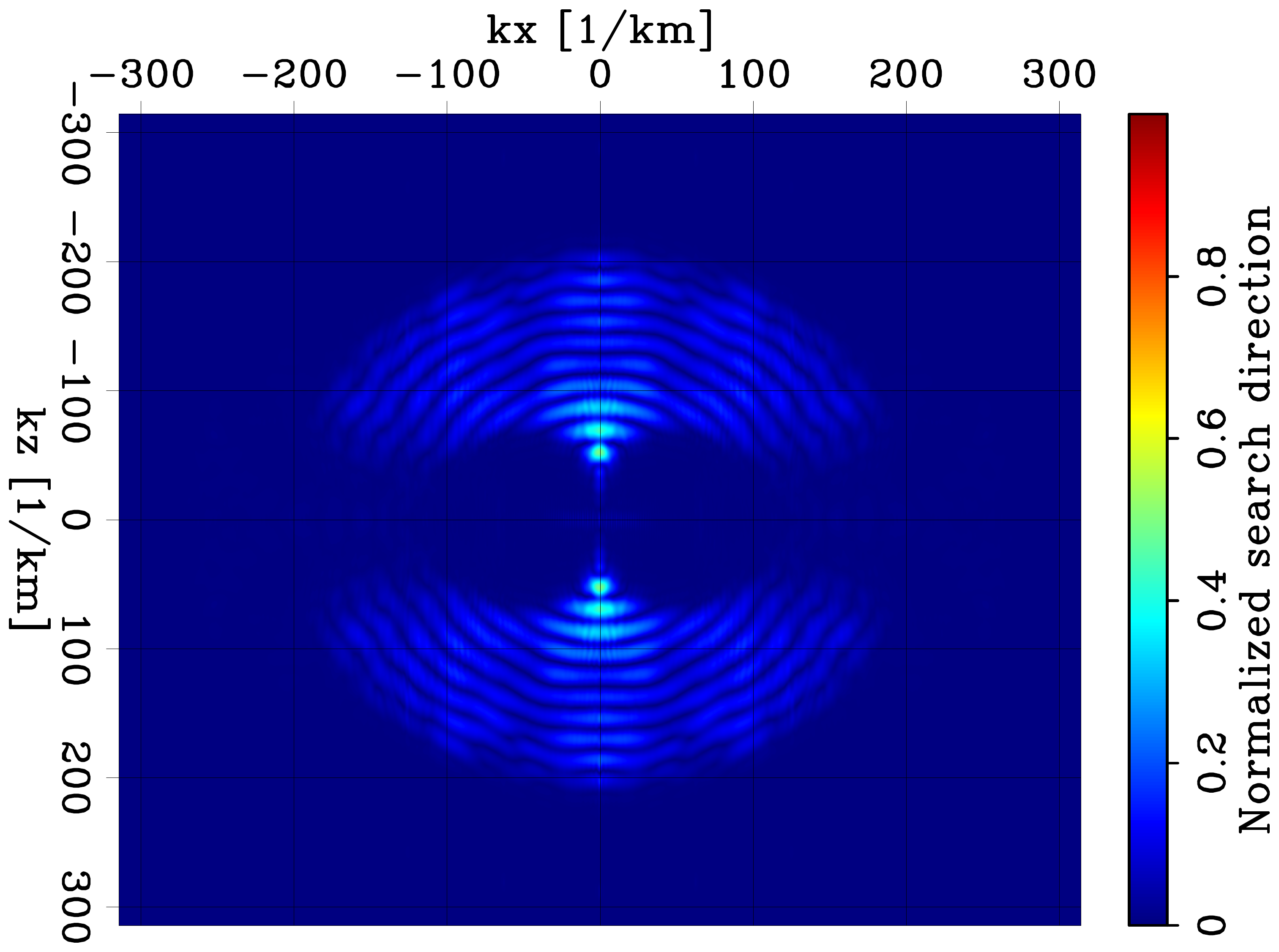}}
    \subfigure[]{\label{fig:Mora_grad_tomo_spectrum}\includegraphics[width=0.30\linewidth]{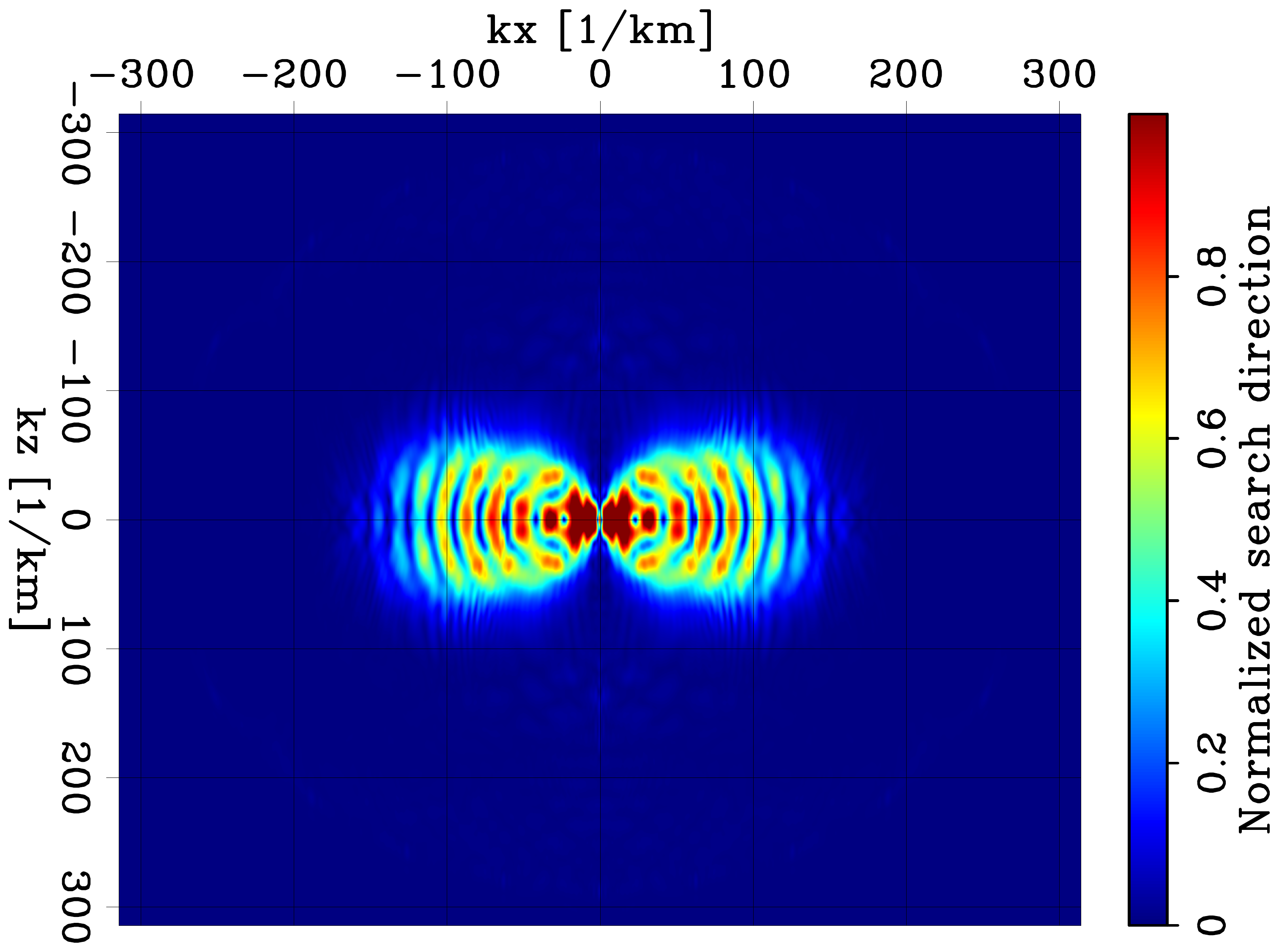}}\\
    \subfigure[]{\label{fig:Mora_grad_total_spectrum}\includegraphics[width=0.30\linewidth]{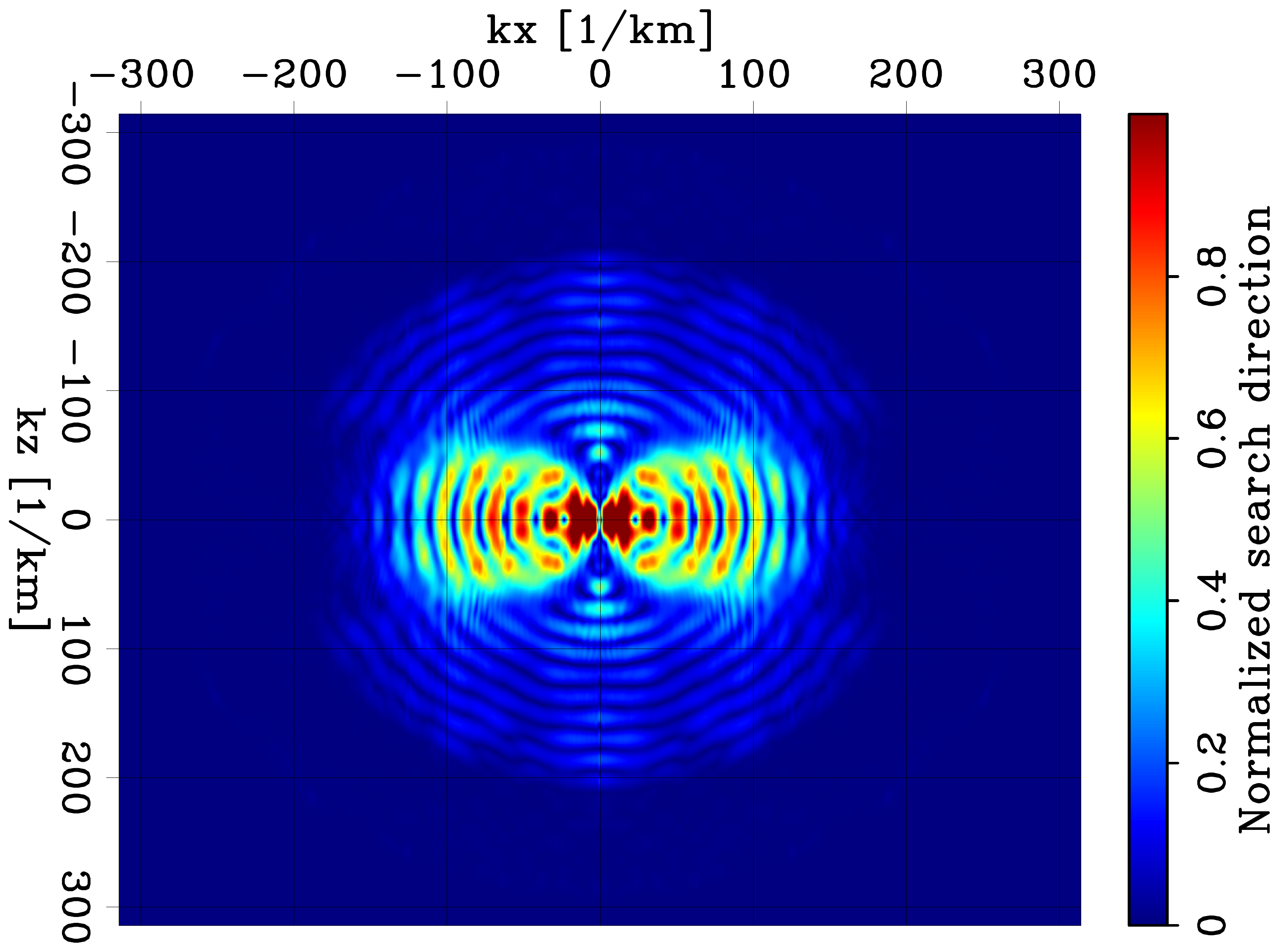}} 
    \subfigure[]{\label{fig:Mora_grad_true_spectrum}\includegraphics[width=0.30\linewidth]{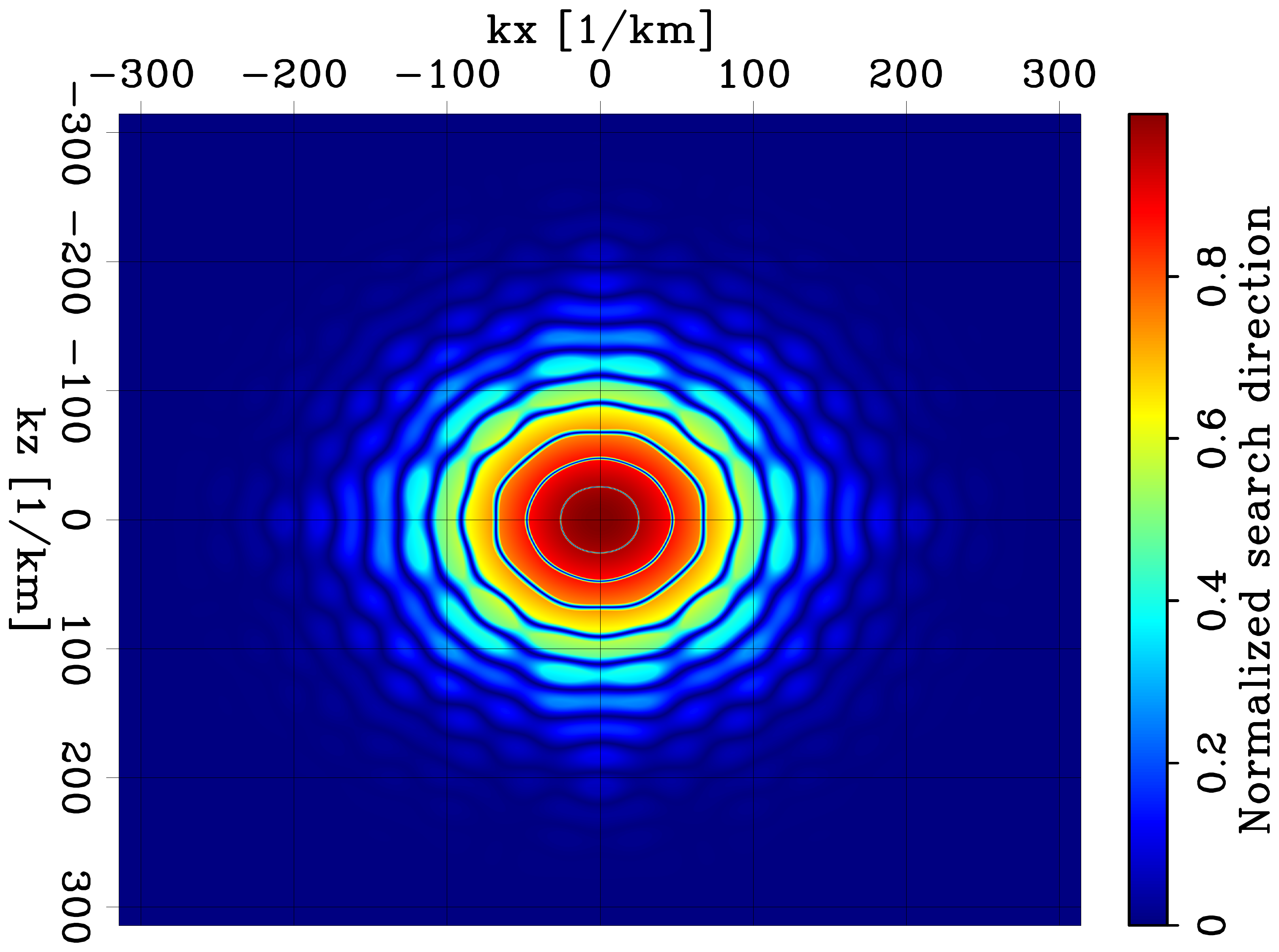}}    
    \caption{Amplitude spectra of the spatial Fourier transforms of initial search directions. (a) Conventional FWI. (b) FWIME Born component. (c) FWIME tomographic component. (d) FWIME total initial search direction (sum of panels (b) and (c)). (e) Ideal search direction. Panels (b), (c) and (d) are normalized by the same value and displayed on the same color scale.}
    \label{fig:Mora_grad_spectrum}
\end{figure}

For the FWIME workflow, the full 20-50 Hz data bandwidth is inverted at once. $\mathbf{\tilde{p}}_{\epsilon}^{opt}$ is extended in time-lags with 101 points sampled at $\Delta \tau = 8$ ms, and the variable projection step is performed with 50 iterations of linear conjugate gradient. Figure~\ref{fig:Mora_grad} shows the different components of the initial FWIME search direction computed on the finite-difference grid (without any spline re-parametrization). As expected, the Born component (Figure~\ref{fig:Mora_grad_born}) is similar to the conventional initial FWI search direction (not shown here): the reflections from the dataset are mapped as high-wavenumber migration isochromes into the model space \cite[]{zhou2015full}. This is confirmed by examining the amplitude spectra of the spatial Fourier transforms of the initial FWI search direction (Figure~\ref{fig:Mora_grad_fwi_spectrum}), and the FWIME Born component (Figure~\ref{fig:Mora_grad_born_spectrum}). Both update directions are missing the low-wavenumber information present in the ideal search direction (Figure~\ref{fig:Mora_grad_true_spectrum}). Moreover, since the initial background velocity model is inaccurate (absence of the low-velocity circular anomaly), these migration isochromes are initially misplaced and will likely guide the inversion to a nonphysical solution, especially in the zone between the bottom of the anomaly and the horizontal interface. The tomographic update (Figures~\ref{fig:Mora_grad_tomo} and \ref{fig:Mora_grad_tomo_spectrum}) is more promising and recovers regions of the spectrum that were not captured by the FWI nor the Born component. Nevertheless, the total search directions (Figures~\ref{fig:Mora_grad_total} and \ref{fig:Mora_grad_total_spectrum}) are contaminated by the migration isochromes from the Born component. Therefore, if no multi-scale strategy is employed, FWIME also converges to a local minimum, as shown in Figure~\ref{fig:Mora_fwime_no_spline}. 

\begin{figure}[t]
    \centering
    \subfigure[]{\label{fig:Mora_grad_born_spline}\includegraphics[width=0.45\linewidth]{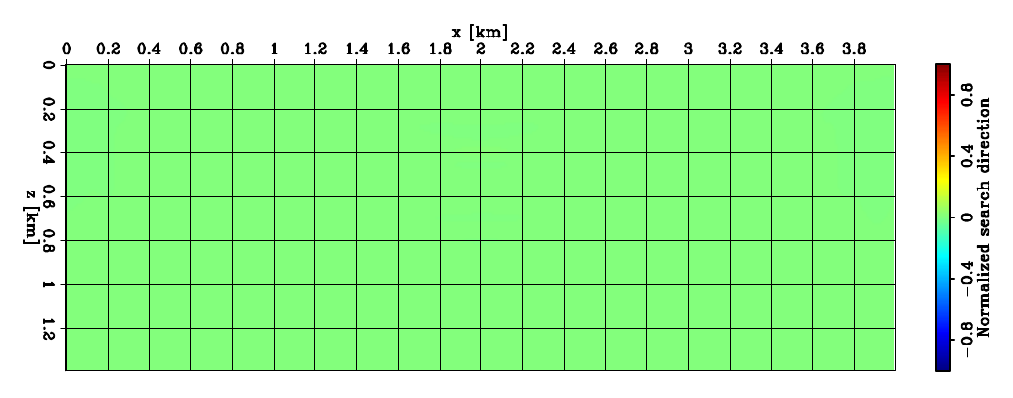}}
    \subfigure[]{\label{fig:Mora_grad_tomo_spline}\includegraphics[width=0.45\linewidth]{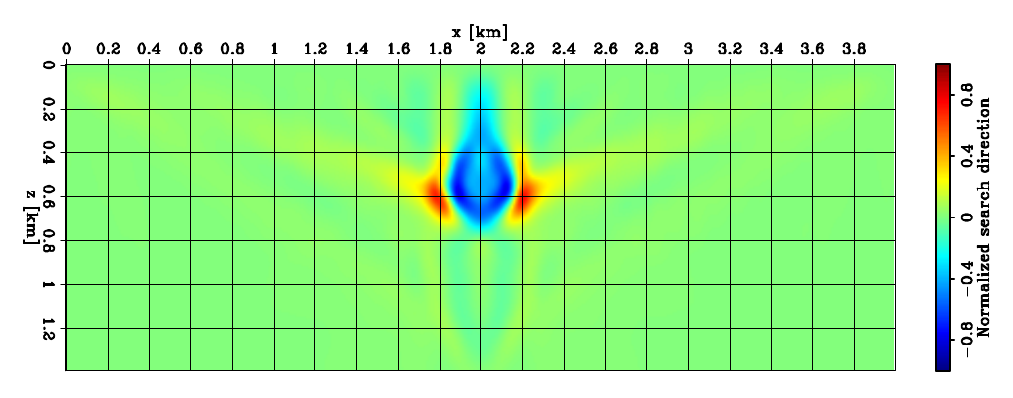}} \\
    \subfigure[]{\label{fig:Mora_grad_total_spline}\includegraphics[width=0.45\linewidth]{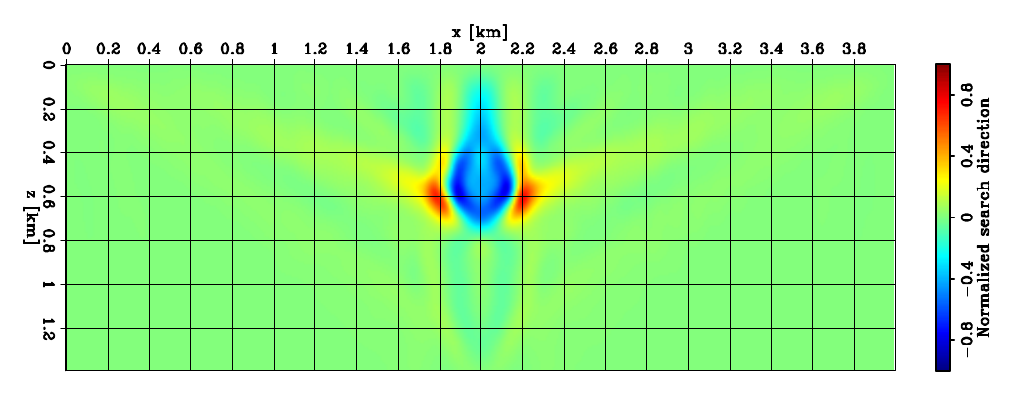}}
    \subfigure[]{\label{fig:Mora_grad_true}\includegraphics[width=0.45\linewidth]{Fig/Mora-true-gradient-2d.pdf}}
    \caption{2D panels of normalized initial search directions after their mapping onto the first spline grid (after applying operator $\mathbf{S}\mathbf{S}^*$ to the analogous panels in Figure~\ref{fig:Mora_grad}). (a) FWIME Born search direction. (b) FWIME tomographic search direction. (c) FWIME total search direction. (d) Ideal search direction (no spline mapping was applied to this panel). Panels (a), (b), and (c) are normalized by the same value and displayed on the same color scale.}
    \label{fig:Mora_grad_spline}
\end{figure}

To overcome this issue, we use a sequence of three spatially-uniform spline grids sampled at 50 m, 20 m, and 10 m, respectively (the third grid coincides with the finite-difference propagation grid). Figures~\ref{fig:Mora_grad_spline}a-c show the initial FWIME search directions after applying operator $\mathbf{S}\mathbf{S}^*$ to Figures~\ref{fig:Mora_grad}a-c, where $\mathbf{S}$ is the spline operator for the initial grid. By examining Figure~\ref{fig:Mora_grad_spline}, we can see that the amplitude of the Born component is now much smaller than the amplitude of the tomographic update, the spurious high-wavenumber features have been removed, and the total search direction is improved. 

\begin{figure}[t]
    \centering
    \subfigure[]{\label{fig:Mora_fwime_spline1}\includegraphics[width=0.45\linewidth]{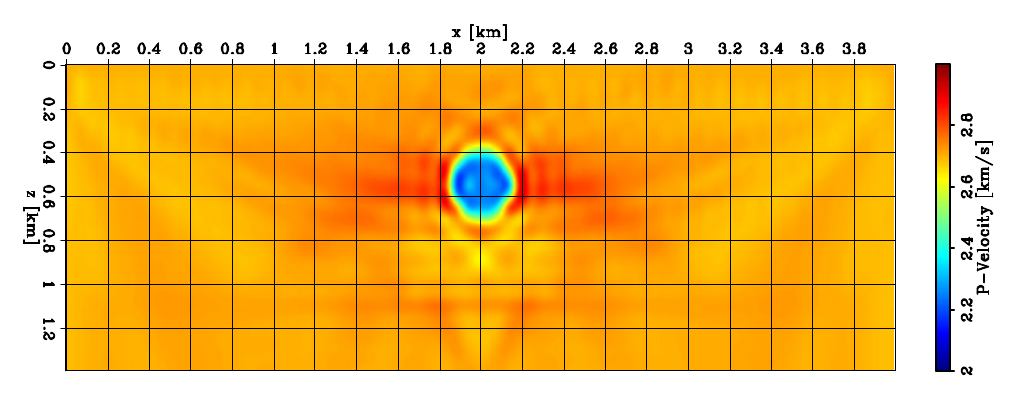}}
    \subfigure[]{\label{fig:Mora_fwime_spline2}\includegraphics[width=0.45\linewidth]{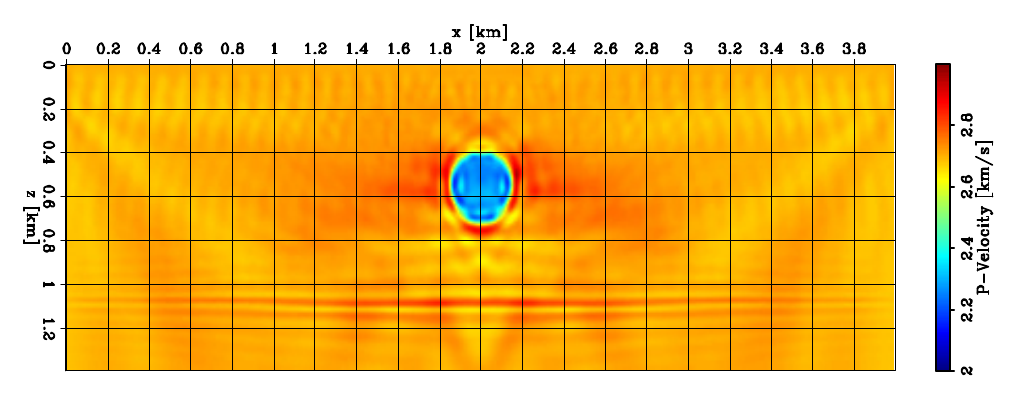}} \\
    \subfigure[]{\label{fig:Mora_fwime_spline3}\includegraphics[width=0.45\linewidth]{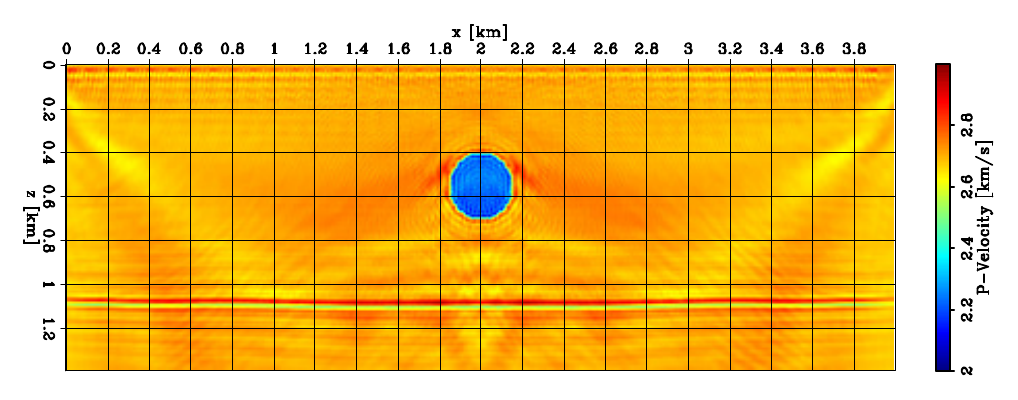}} 
    \subfigure[]{\label{fig:Mora_true_spline}\includegraphics[width=0.45\linewidth]{Fig/Mora-true-mod-2d.pdf}} 
    \caption{2D panels of inverted velocity models obtained at different stages of the FWIME model-space multi-scale workflow. (a) Inverted model after 32 iterations on first spline grid. (b) Inverted model after 20 iterations on second spline grid. (c) Final inverted model after 30 iterations on the finite-difference grid. (d) True model. A total of 82 L-BFGS iterations of FWIME were used to obtain the result in panel (c).}
    \label{fig:Mora_mod_spline}
\end{figure}

\begin{figure}[t]
    \centering
    \subfigure[]{\label{fig:Mora_fwime_z1d}\includegraphics[width=0.45\linewidth]{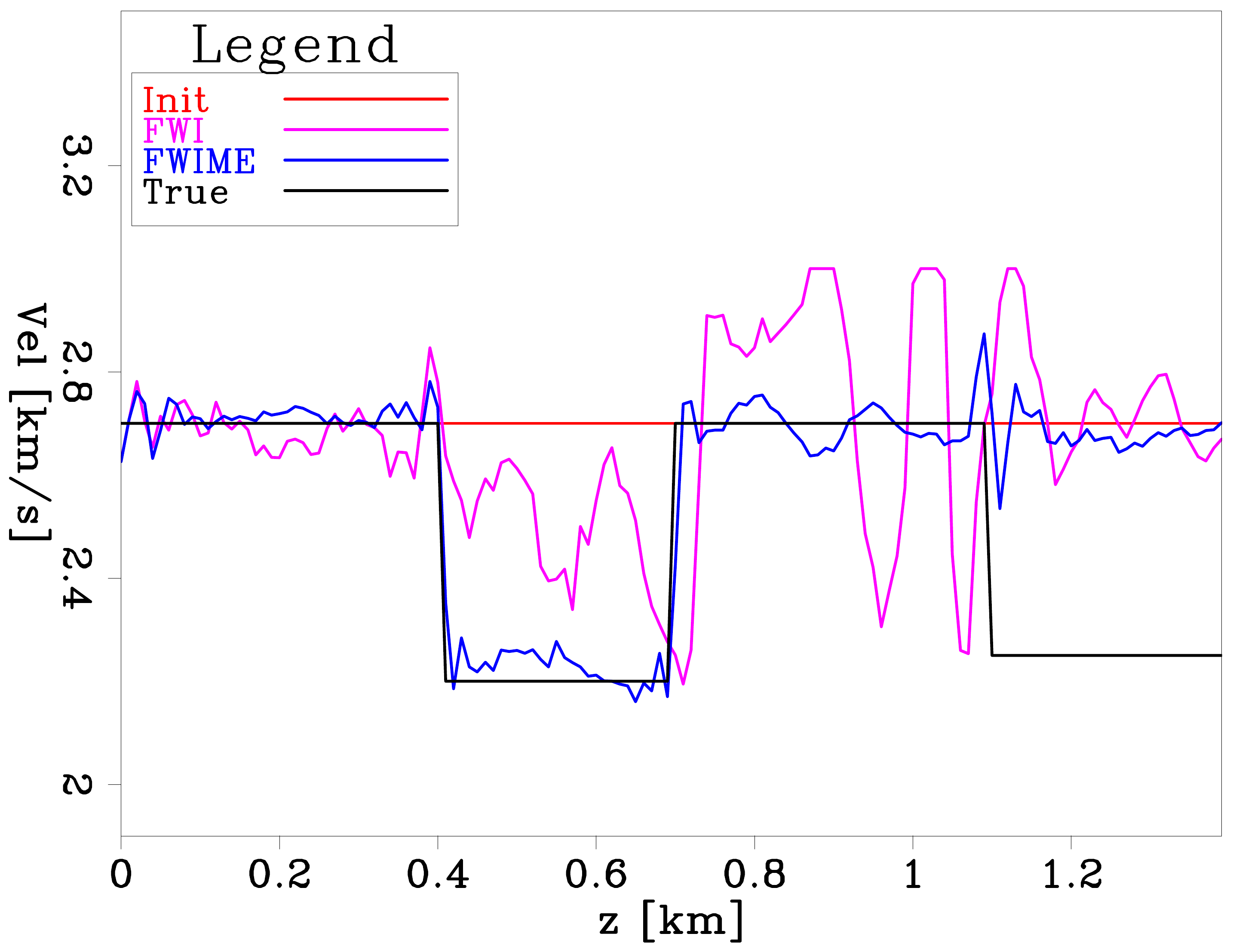}} \hspace{5mm}
    \subfigure[]{\label{fig:Mora_fwime_x1d}\includegraphics[width=0.45\linewidth]{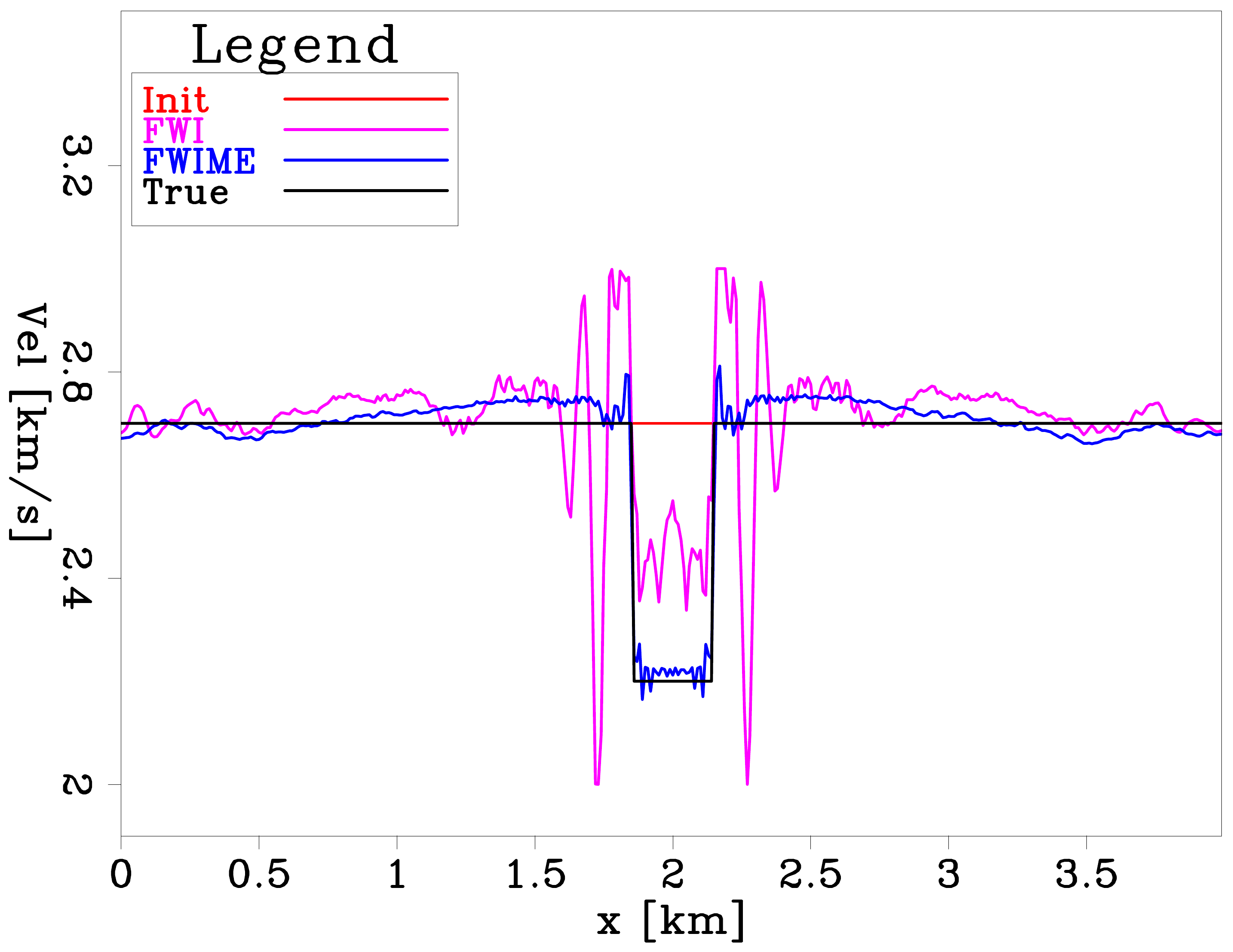}}
    \caption{Velocity profiles of the true (black curve), initial (red curve), conventional FWI (pink curve), and FWIME (blue curve) inverted models. (a) Depth velocity profiles extracted at $x = 2$ km. (b) Horizontal velocity profiles extracted at $z = 0.6$ km.}
    \label{fig:Mora_mod_1d}
\end{figure}

We can now successfully apply the multi-scale FWIME workflow. We use a fixed $\epsilon$-value of $1.7 \times 10^{-7}$ throughout the entire process. Each spline grid refinement is automatically triggered when the numerical solver is unable to find a step length that decreases the objective function. Figures~\ref{fig:Mora_mod_spline}a-c show the sequence of inverted models at the end of each spline grid. The final recovered model is excellent and manages to accurately reconstruct the velocity values in the shadow zone located between the bottom of the anomaly and the horizontal interface. The sharpness of the anomaly is also well captured, as shown by the vertical (Figure~\ref{fig:Mora_fwime_z1d}) and horizontal (Figure~\ref{fig:Mora_fwime_x1d}) velocity profiles extracted at $x=2$ km and $z=0.6$ km, respectively (the oscillatory behavior of the model is due to the limited frequency range available in the dataset). In addition, the difference between the observed and predicted data computed with the final FWIME model is shown in Figure~\ref{fig:Mora_datDiffFwime} and confirms the quality of the inversion result. In this experiment, the sensitivity of the inverted model with respect to the trade-off parameter $\epsilon$ was very limited. Similar results as the one shown in Figure~\ref{fig:Mora_fwime_spline3} were obtained for $\epsilon$-values ranging within one order of magnitude. 

%%%%%%%%%%%%%%%%%%%%%%%%%%%%%%%% Diving waves %%%%%%%%%%%%%%%%%%%%%%%%%%%%%%%%%%%%%%%%
\subsection{Inversion of diving waves}
\begin{figure}[tbhp]
    \centering
    \subfigure[]{\label{fig:diving_init}\includegraphics[width=0.45\linewidth]{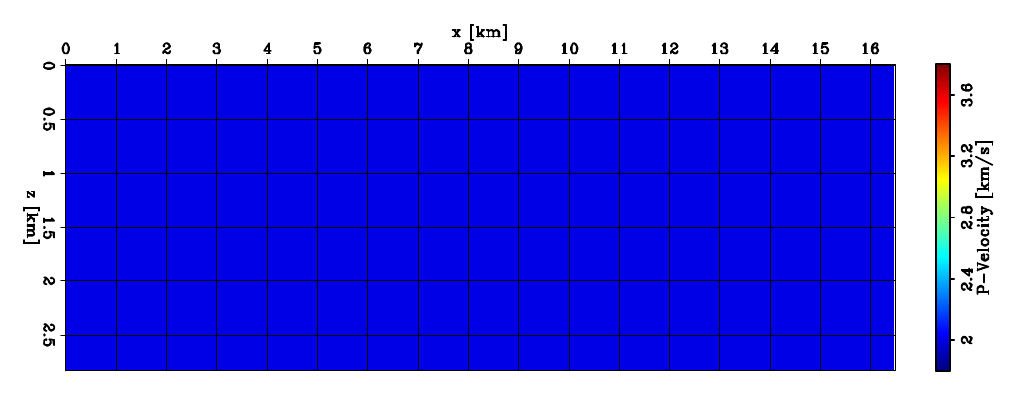}}
    \subfigure[]{\label{fig:diving_fwi}\includegraphics[width=0.45\linewidth]{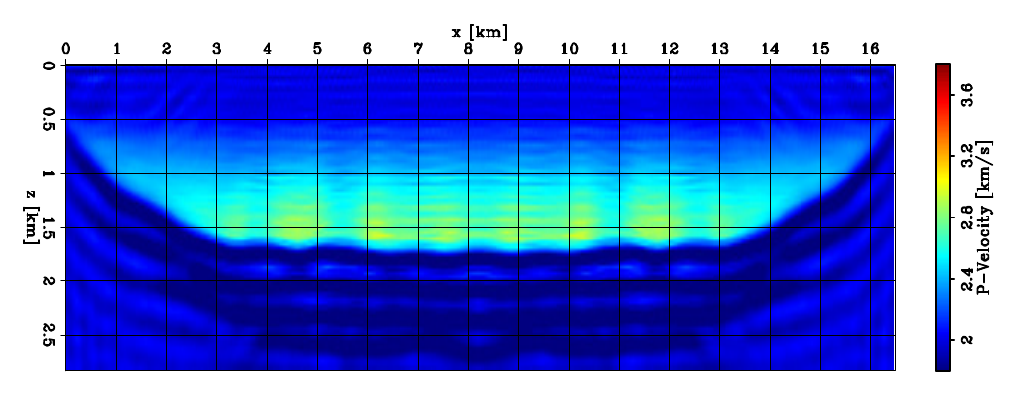}}\\
    \subfigure[]{\label{fig:diving_fwime}\includegraphics[width=0.45\linewidth]{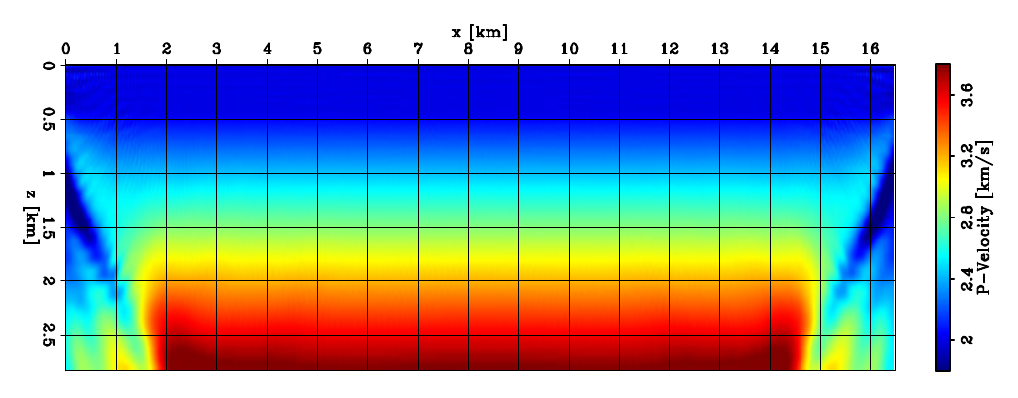}}    
    \subfigure[]{\label{fig:diving_true}\includegraphics[width=0.45\linewidth]{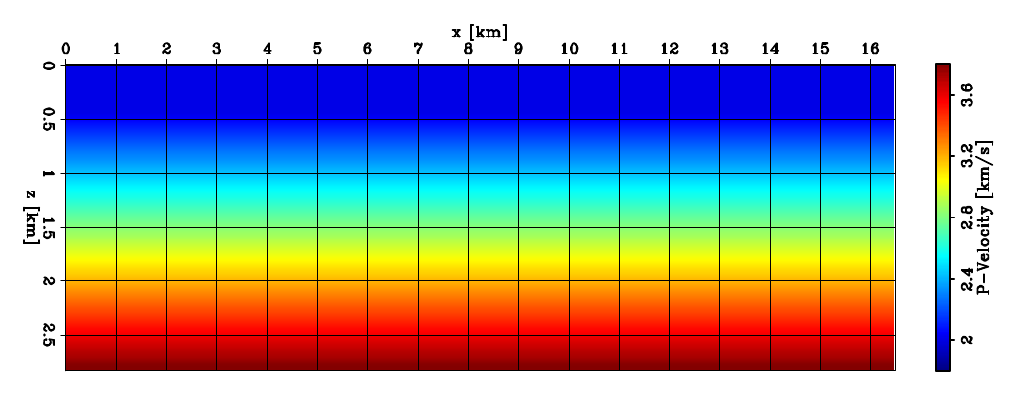}}        
    \caption{2D panels of velocity models. (a) Initial model. (b) Inverted model with a conventional data-space multi-scale FWI (using three frequency bands). (c) Final FWIME inverted model. (d) True model. Panel (c) was obtained using a model-space multi-scale FWIME approach with three different spline grids and a total of 40 L-BFGS iterations.}
    \label{fig:diving_mod}
\end{figure}

\begin{figure}[tbhp]
    \centering
    \subfigure[]{\label{fig:data_true}\includegraphics[width=0.3\linewidth]{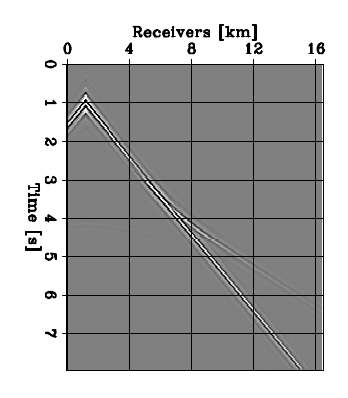}}
    \subfigure[]{\label{fig:data_init}\includegraphics[width=0.3\linewidth]{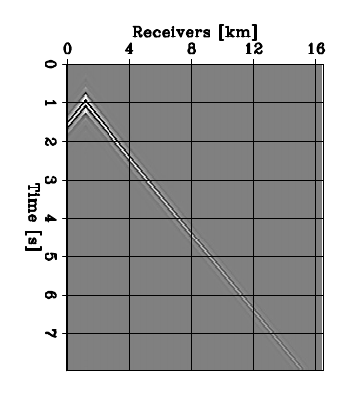}}
    \subfigure[]{\label{fig:data_initDiff}\includegraphics[width=0.3\linewidth]{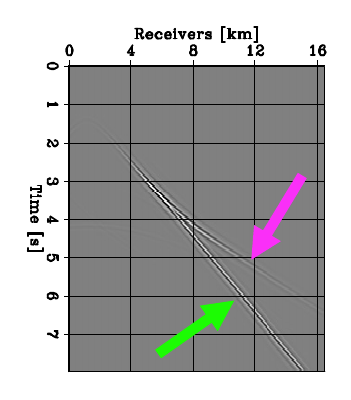}}
    \caption{Representative shot gathers generated with a source located at $x = 1.2$ km. (a) Observed data, $\mathbf{d}^{obs}$. (b) Predicted data with the initial model, $\mathbf{f}(\mathbf{m}_0)$. (c) Initial data difference, $\Delta \mathbf{d}(\mathbf{m}_0) = \mathbf{d}^{obs}-\mathbf{f}(\mathbf{m}_0)$. All panels are displayed with the same grayscale.}
    \label{fig:diving_data_init}
\end{figure}

We invert a dataset solely composed of diving waves where the inaccurate initial velocity model produces very large kinematic errors in the predicted data. The dataset is generated with a source wavelet containing energy strictly limited to the 3-12 Hz frequency range, which prevents conventional FWI from leveraging the low-frequency signal (below 3 Hz) to overcome the cycle-skipping phenomenon. The true model is 16 km wide by 2.8 km deep, and is discretized with a finite-difference grid spacing of 30 m in both directions. It is composed of a 0.4 km-thick homogeneous layer placed on top of a second horizontally-invariant layer whose values linearly increase with depth, as shown in  Figures~\ref{fig:diving_true} and \ref{fig:diving_mod_1d} (black curve). The initial model $\mathbf{m}_0$ is chosen to be unrealistically inaccurate (Figure~\ref{fig:diving_init}). It is homogeneous and set to 2.0 km/s (dark-blue curve in Figure~\ref{fig:diving_mod_1d}). We place 137 sources and 550 receivers at the surface, spaced every 120 m and 30 m, respectively. Figures~\ref{fig:diving_data_init}a-c show a representative shot gather corresponding to the observed data $\mathbf{d}^{obs}$, the initial prediction $\mathbf{f}(\mathbf{m}_0)$, and initial data difference, $\Delta \mathbf{d}(\mathbf{m}_0)=\mathbf{d}^{obs}-\mathbf{f}(\mathbf{m}_0)$ computed for a source placed at $x = 1.2$ km. 

\begin{figure}[tbhp]
    \centering
    \label{fig:diving_mod_1d}\includegraphics[width=0.4\linewidth]{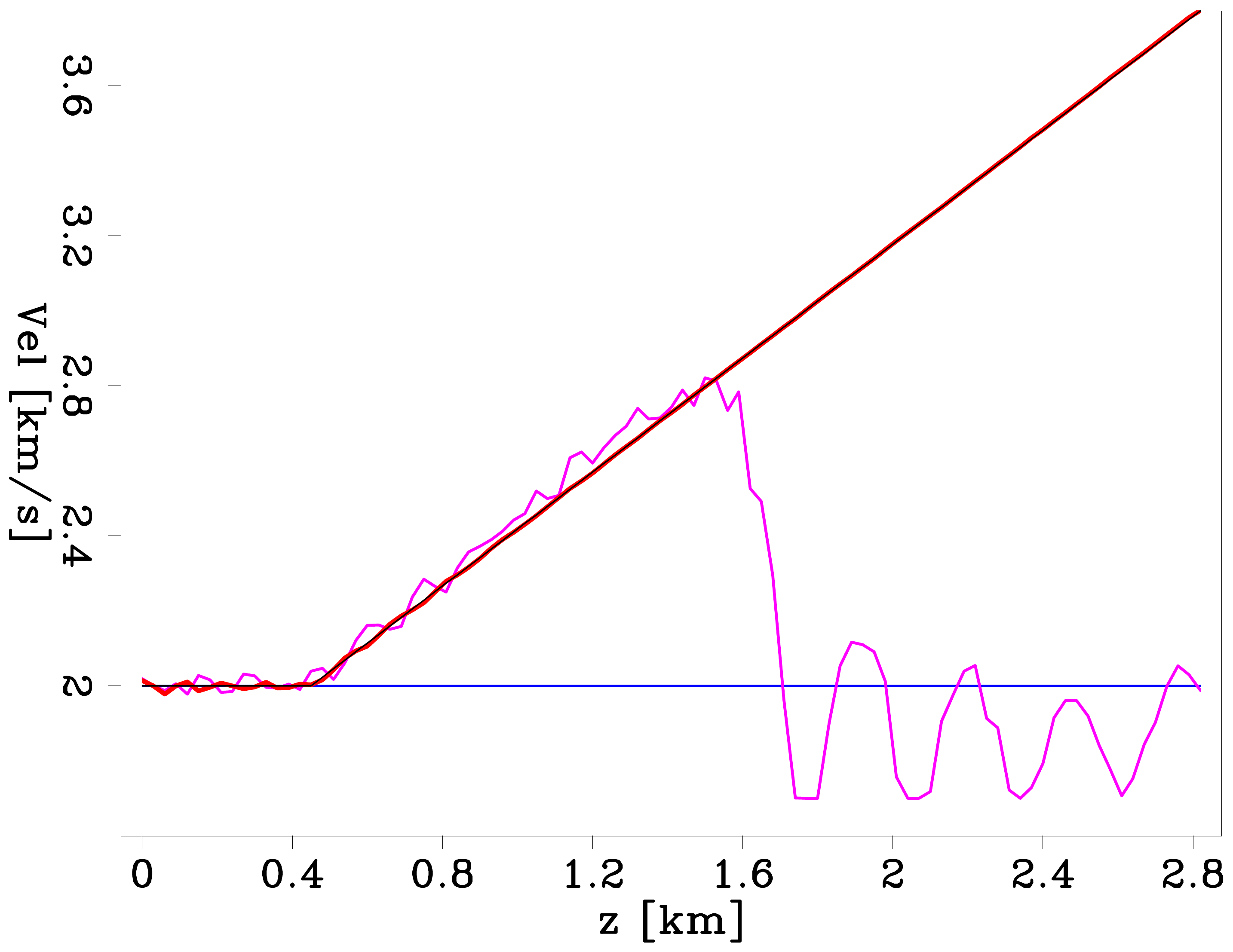}
    \caption{Depth velocity profiles extracted at $x = 8$ km from the initial model (blue curve), the FWI model (pink curve), the FWIME model (red curve), and the true model (black curve). The black and red curves are similar and overlap each other.}
    \label{fig:diving_mod_1d}
\end{figure}

\begin{figure}[tbhp]
    \centering
    \subfigure[]{\label{fig:data_true}\includegraphics[width=0.3\linewidth]{Fig/divingWave2-thesis-3-12-data-true.pdf}}
    \subfigure[]{\label{fig:data_fwi}\includegraphics[width=0.3\linewidth]{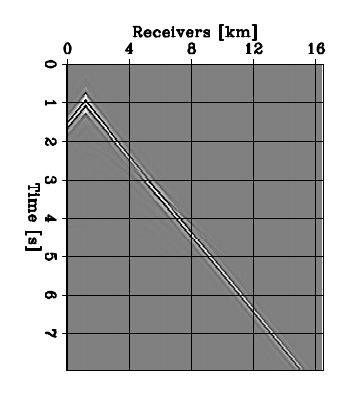}}
    \subfigure[]{\label{fig:data_fwiDiff}\includegraphics[width=0.3\linewidth]{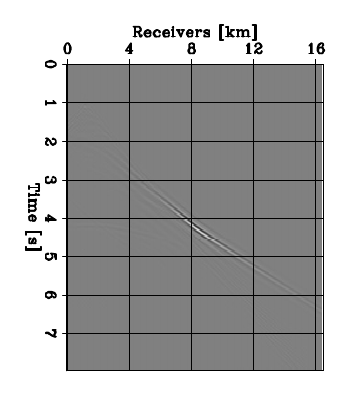}}
    \caption{Representative shot gathers generated with a source located at $x = 1.2$ km. (a) Observed data, $\mathbf{d}^{obs}$. (b) Predicted data with the final FWI inverted model, $\mathbf{f}(\mathbf{m}_{FWI})$. (c) Final data difference, $\Delta \mathbf{d}(\mathbf{m}_{FWI}) = \mathbf{d}^{obs}-\mathbf{f}(\mathbf{m}_{FWI})$. All panels are displayed with the same grayscale.}
    \label{fig:diving_data_fwi}
\end{figure}

We apply a conventional multi-scale FWI workflow using three frequency bands spanning the available 3-12 Hz bandwidth, starting with the uniform model $\mathbf{m}_0$. For each frequency band, we conduct 500 iterations of L-BFGS. FWI fails to recover the correct velocity model for depths greater than 1.6 km, as shown in Figures~\ref{fig:diving_fwi} and \ref{fig:diving_mod_1d} (pink curve). In addition, Figures~\ref{fig:diving_data_fwi}b and \ref{fig:diving_data_fwi}c display the predicted data and data residual computed with the final FWI model and show that the recovered model is unable to accurately predict refracted events (i.e., diving waves) for offsets greater than 7 km. 

For FWIME, we compute $\mathbf{\tilde{p}}_{\epsilon}^{opt}$ with a time-lag extension. The extended axis is composed of 101 points sampled at $\Delta \tau = 32$ ms, which correspond to time lags ranging from $-1.6$ s to $1.6$ s. The full potential of extended modeling (and the ability of the data-correcting term $\tilde{\mathbf{B}}(\mathbf{m}_0) \mathbf{\tilde{p}}_{\epsilon}^{opt}(\mathbf{m}_0)$ to match any data misfit even for inaccurate background velocity models) can be better appreciated by closely examining the first variable projection step in the FWIME workflow. The initial data difference $\Delta \mathbf{d}(\mathbf{m}_0) = \mathbf{d}^{obs} - \mathbf{f}(\mathbf{m}_0)$ contains two events (Figure~\ref{fig:data_initDiff}). The first event possesses a linear moveout (green arrow) and corresponds to the phase mismatch between the direct arrivals from the true and initial models. The second event (pink arrow) is the diving wave present in the observed data that is not modeled by our initial prediction $\mathbf{f}(\mathbf{m}_0)$. On one hand, a data-correcting term $\mathbf{B} (\mathbf{m}_0) \mathbf{p}_{\epsilon}^{opt} (\mathbf{m}_0)$ computed by minimizing equation~\ref{eqn:vp.obj} with a non-extended Born modeling operator, the initial velocity model $\mathbf{m}_0$, and $\epsilon=0$, would have no chance to linearly generate diving waves that would fit the initial data difference. 

\begin{figure}[tbhp]
    \centering
    \subfigure[]{\label{fig:vp_noExt_dataInitDiff}\includegraphics[width=0.3\linewidth]{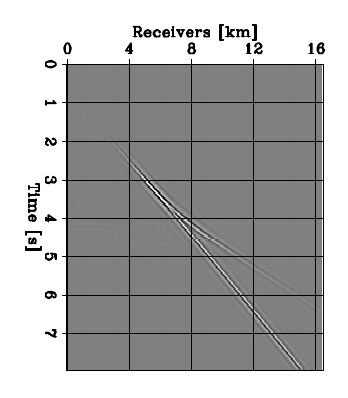}}
    \subfigure[]{\label{fig:vp_noExt_dataPred}\includegraphics[width=0.3\linewidth]{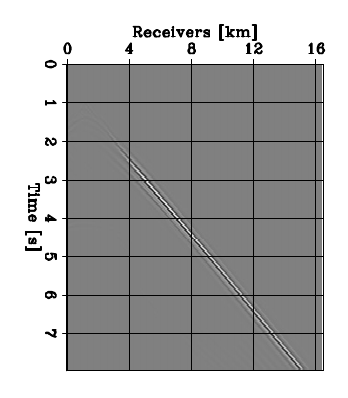}}
    \subfigure[]{\label{fig:vp_noExt_predError}\includegraphics[width=0.3\linewidth]{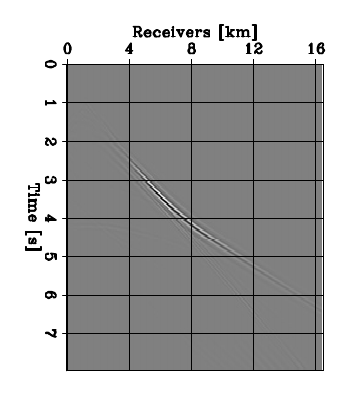}}
    \caption{Representative shot gathers generated with a source located at $x = 1.2$ km. (a) Initial data difference: $\Delta \mathbf{d}(\mathbf{m}_0) = \mathbf{d}^{obs}-\mathbf{f}(\mathbf{m}_0)$. (b) Data-correcting term computed with a non-extended Born operator: $\mathbf{B} (\mathbf{m}_0) \mathbf{p}_{\epsilon}^{opt} (\mathbf{m}_0)$. (c) Prediction error between the data-correcting term and the initial data residual: $\mathbf{r}^{\epsilon}_d (\mathbf{m}_0) = \mathbf{B} (\mathbf{m}_0) \mathbf{p}_{\epsilon}^{opt} (\mathbf{m}_0)-\Delta \mathbf{d}(\mathbf{m}_0)$. Panels (b) and (c) are computed with $\epsilon=0$. All panels are displayed with the same grayscale.}
    \label{fig:vp_data_no_ext}
\end{figure}

\begin{figure}[tbhp]
    \centering
    \subfigure[]{\label{fig:vp_time_dataInitDiff}\includegraphics[width=0.3\linewidth]{Fig/divingWave2-fwime-3-12-time-vp-dataInitDiff.pdf}}
    \subfigure[]{\label{fig:vp_time_dataPred}\includegraphics[width=0.3\linewidth]{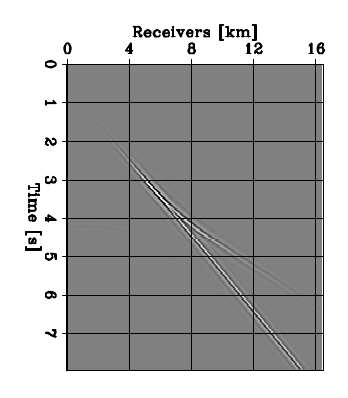}}
    \subfigure[]{\label{fig:vp_time_predError}\includegraphics[width=0.3\linewidth]{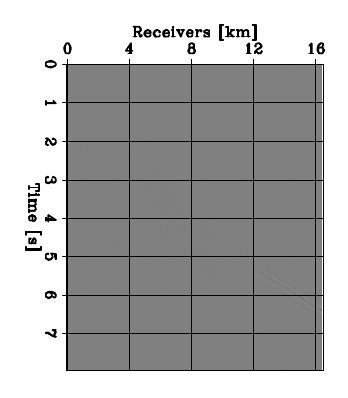}}
    \caption{Representative shot gathers generated with a source located at $x = 1.2$ km. (a) Initial data difference: $\Delta \mathbf{d}(\mathbf{m}_0) = \mathbf{d}^{obs}-\mathbf{f}(\mathbf{m}_0)$. (b) Data correcting term computed with a time-lag extended Born operator: $\tilde{\mathbf{B}} (\mathbf{m}_0) \tilde{\mathbf{p}}_{\epsilon}^{opt} (\mathbf{m}_0)$. (c) Prediction error between the data-correcting term and the initial data residual: $\mathbf{r}^{\epsilon}_d (\mathbf{m}_0) = \tilde{\mathbf{B}} (\mathbf{m}_0) \tilde{\mathbf{p}}_{\epsilon}^{opt} (\mathbf{m}_0)-\Delta \mathbf{d}(\mathbf{m}_0)$. Panels (b) and (c) are computed with $\epsilon=0$. All panels are displayed with the same grayscale.}
    \label{fig:vp_data_time}
\end{figure}

Figure~\ref{fig:vp_noExt_dataPred} shows the data-correcting term computed with a non-extended Born operator and with $\epsilon=0$, $\mathbf{B} (\mathbf{m}_0)\mathbf{p}_{\epsilon}^{opt} (\mathbf{m}_0)$. As expected, it fails to predict the refracted event. This is indeed confirmed by examining the prediction error, $\mathbf{r}^{\epsilon}_d (\mathbf{m}_0)=\mathbf{B} (\mathbf{m}_0)\mathbf{p}_{\epsilon}^{opt} (\mathbf{m}_0)-\Delta \mathbf{d}(\mathbf{m}_0)$ in Figure~\ref{fig:vp_noExt_predError}. On the other hand, the use of an extended Born operator allows the data-correcting term to accurately match the highly nonlinear events (diving waves) present in the initial data misfit with a linear operator (Figure~\ref{fig:vp_data_time}) and a constant velocity model $\mathbf{m}_0$.

\begin{figure}[tbhp]
    \centering
    \subfigure[]{\label{fig:diving_cig-it0}\includegraphics[width=0.23\linewidth]{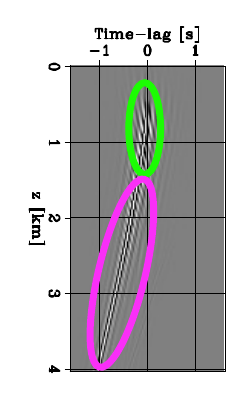}}
    \subfigure[]{\label{fig:diving_cig-it1}\includegraphics[width=0.23\linewidth]{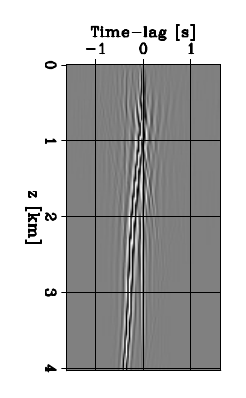}}
    \subfigure[]{\label{fig:diving_cig-it2}\includegraphics[width=0.23\linewidth]{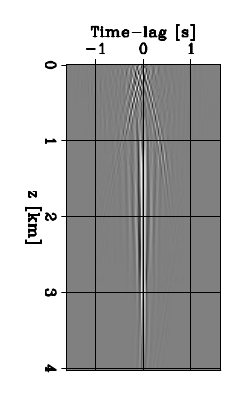}}    
    \subfigure[]{\label{fig:diving_cig-it15}\includegraphics[width=0.23\linewidth]{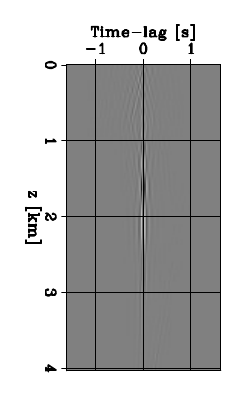}}            
    \caption{TLCIGs extracted at $x = 8$ km from $\mathbf{\tilde{p}}_{\epsilon}^{opt}$ at four stages of FWIME. (a) Initial FWIME step, (b) iteration 3, (c) iteration 10, (d) iteration 25. All panels are displayed with the same grayscale.}
    \label{fig:diving_cig}
\end{figure}

Figure~\ref{fig:diving_cig}a shows a TLCIG extracted at $x=8$ km from $\mathbf{\tilde{p}}_{\epsilon}^{opt}(\mathbf{m}_0)$ computed at the initial step with $\epsilon=7.5 \times 10^{-5}$. In the shallow part of the model where $\mathbf{m}_0$ is accurate (for $z \leq 1.6 $ km), the energy within $\mathbf{\tilde{p}}_{\epsilon}^{opt}(\mathbf{m}_0)$ is focused in the vicinity of the zero-lag axis (region within the green oval). This energy cluster corresponds to the mapping of the direct arrival (green arrow in Figure~\ref{fig:data_initDiff}) into $\mathbf{\tilde{p}}_{\epsilon}^{opt}(\mathbf{m}_0)$, which can be modeled without any extension. In the deeper region, the velocity error increases and the coherent energy is gradually positioned at large-amplitude negative time lags. This energy cluster (pink oval), which corresponds to the mapping of the diving waves (pink arrow in Figure~\ref{fig:data_initDiff}), can only be captured with the use of an extended perturbation and an extended modeling operator. In this particular example, the initial model $\mathbf{m}_0$ was extremely inaccurate and our extended axis $\tau$ had to span an unusually wide range of time lags to fully capture the data misfit. By comparing TLCIGs for diving waves (Figure~\ref{fig:diving_cig}a) with the ones for transmitted waves (Figure~\ref{fig:oned_fwime_cig_init}), we observe that different wave types are mapped into energy clusters that possess various moveout characteristics in the extended space of $\mathbf{\tilde{p}}_{\epsilon}^{opt}$. As we show in this paper, one of the main advantages of FWIME is its capacity to invert any wave type with the same mechanism.

\begin{figure}[tbhp]
    \centering
    \subfigure[]{\label{fig:diving_grad_born}\includegraphics[width=0.45\linewidth]{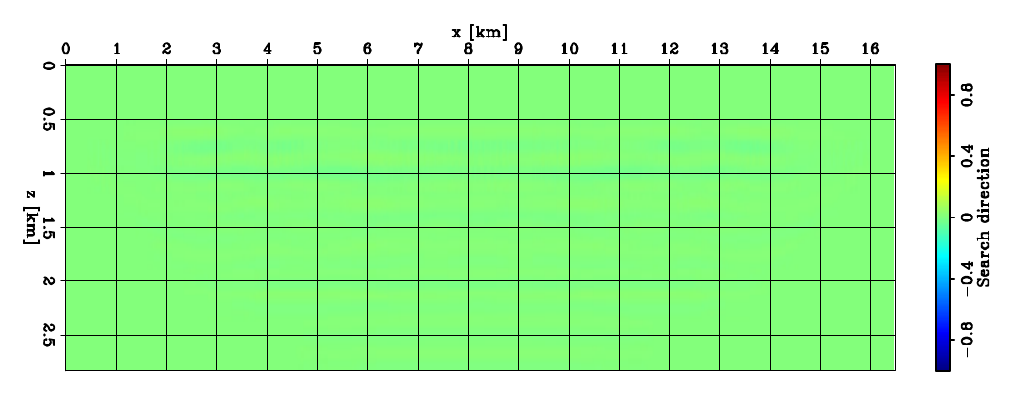}}
    \subfigure[]{\label{fig:diving_grad_tomo}\includegraphics[width=0.45\linewidth]{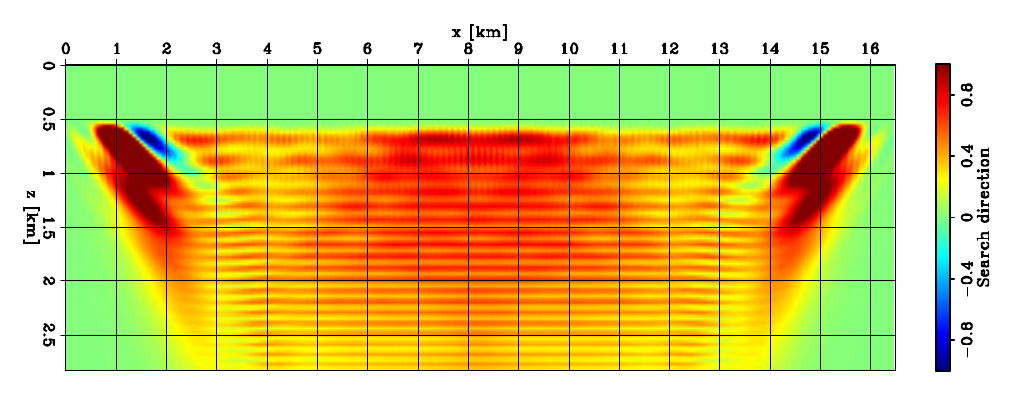}}\\   
    \subfigure[]{\label{fig:diving_grad_total}\includegraphics[width=0.45\linewidth]{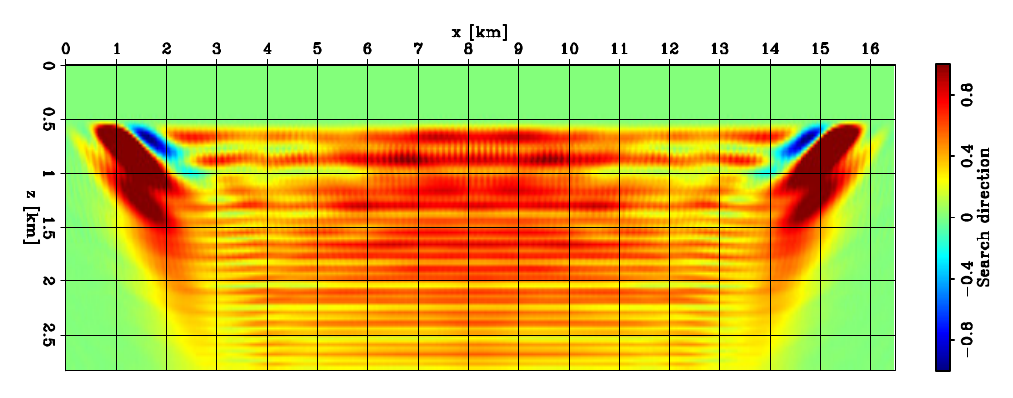}}        
    \subfigure[]{\label{fig:diving_grad_true}\includegraphics[width=0.45\linewidth]{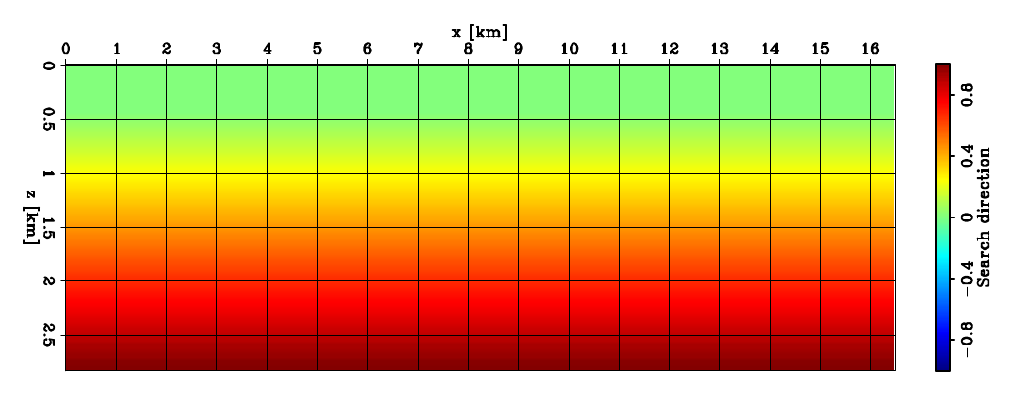}}            
    \caption{2D panels of normalized initial search directions. (a) FWIME Born search direction. (b) FWIME tomographic search direction. (c) FWIME total search direction. (d) Ideal search direction. Panels (a), (b), and (c) are normalized by the same value and displayed on the same color scale.}
    \label{fig:diving_grad}
\end{figure}

\begin{figure}[tbhp]
    \centering
    \subfigure[]{\label{fig:diving_grad_born_spline}\includegraphics[width=0.45\linewidth]{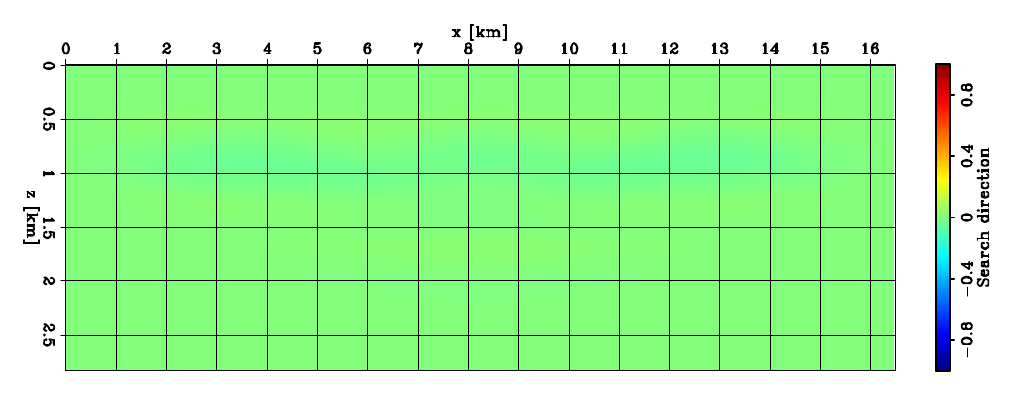}}
    \subfigure[]{\label{fig:diving_grad_tomo_spline}\includegraphics[width=0.45\linewidth]{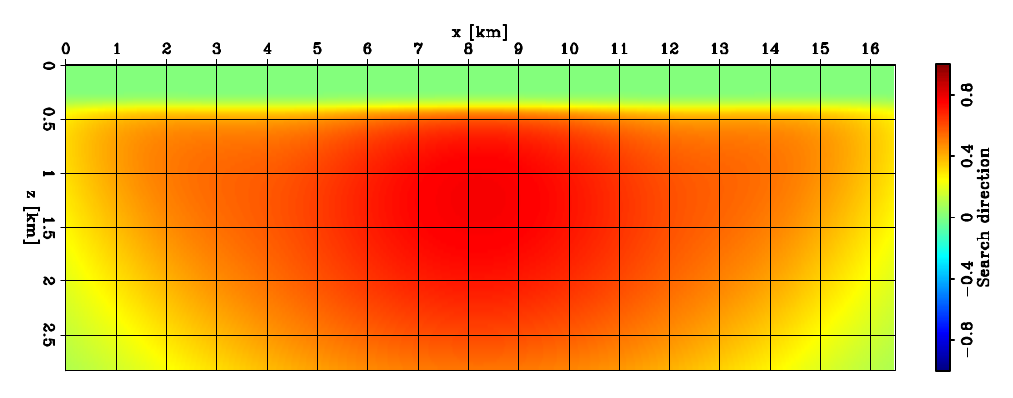}}\\   
    \subfigure[]{\label{fig:diving_grad_total_spline}\includegraphics[width=0.45\linewidth]{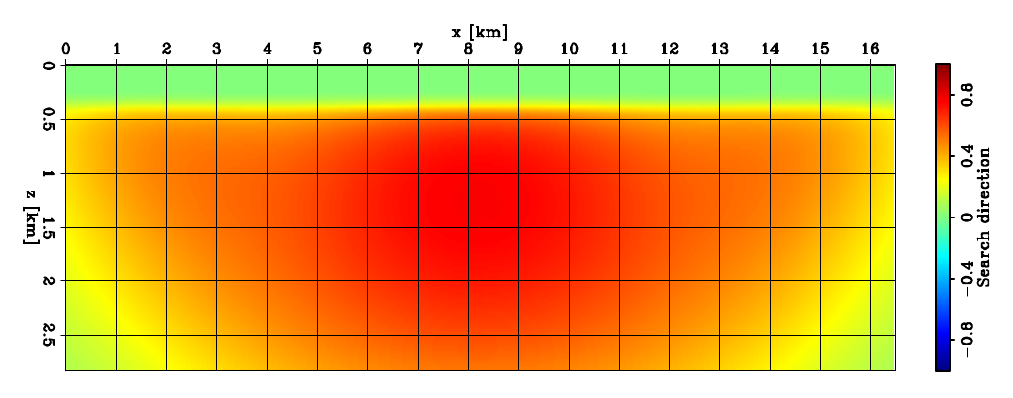}}        
    \subfigure[]{\label{fig:diving_grad_true_spline}\includegraphics[width=0.45\linewidth]{Fig/sep20-true-gradient.pdf}}            
    \caption{2D panels of normalized initial search directions after applying operator $\mathbf{S}\mathbf{S}^*$ (where $\mathbf{S}$ is the spline operator for the first grid). (a) FWIME Born search direction. (b) FWIME tomographic search direction. (c) FWIME total search direction. (d) Ideal search direction. Panels (a), (b), and (c) are normalized by the same value and displayed on the same color scale.}
    \label{fig:diving_grad_spline}
\end{figure}

Figures~\ref{fig:diving_grad}a-c show the Born, tomographic, and total first search directions for our FWIME workflow (computed with $\epsilon = 7.5 \times 10^{-5}$) on the finite-difference grid (no spline mapping is applied yet). The total search direction seems promising and is dominated by the tomographic component. In order to impose lateral smoothness on the inverted model, we begin the multi-scale FWIME scheme by designing a coarse and spatially uniform initial spline grid, where $\Delta z = 1$ km, and $\Delta x = 1.8$ km. The search directions after the application of $\mathbf{S} \mathbf{S}^*$ on the panels in Figures~\ref{fig:diving_grad}a-c are displayed in Figures~\ref{fig:diving_grad_spline}a-c, and are much more geologically realistic. 

We simultaneously invert the full available bandwidth from the data, we fix the $\epsilon$-value to $7.5 \times 10^{-5}$, and we use a sequence of three spline grids (the last grid coincides with the finite-difference grid). The final inverted model obtained after 40 L-BFGS iterations of FWIME is shown in Figure~\ref{fig:diving_fwime}. While it has recovered an excellent solution down to a depth of approximately 2.8 km (red curve in Figure~\ref{fig:diving_mod_1d}), it suffers from small edge artifacts inherent to the limited acquisition aperture. As shown by the evolution of the TLCIG in Figure~\ref{fig:diving_cig}, most of the kinematic errors have been corrected after 10 iterations (the energy within $\tilde{\mathbf{p}}_{\epsilon}^{opt}$ is focused in the vicinity of the zero time-lag axis), at which point the FWIME inverted model could be used as input for the less computationally costly FWI. 

In the FWIME scheme, we embedded prior information (lateral smoothness) by using an extremely coarse initial spline grid. For a fair comparison, we also attempted to improve the results obtained with conventional FWI by using the same spline parametrization sequence, as proposed by \cite{barnier2019waveform}. However (and as expected), this test did not manage to enhance the quality of the FWI inverted model.

%%%%%%%%%%%%%%%%% Discussion %%%%%%%%%%%%%%%%
\section{Discussions}
%%%%%%%%%%%%%%%%%%%%%%%%%%%%%%%%%%%%%%%%%%%%%%%%%%%%%%%%%%%%%%%%%%%%%%%%%%%%%%%%%%%%%%
%%%%%%%%%%%%%%%%%%%%%%%%%%%%%%%%%% Discussion %%%%%%%%%%%%%%%%%%%%%%%%%%%%%%%%%%%%%%%%
%%%%%%%%%%%%%%%%%%%%%%%%%%%%%%%%%%%%%%%%%%%%%%%%%%%%%%%%%%%%%%%%%%%%%%%%%%%%%%%%%%%%%%
% Computational cost
\subsection{The computational bottleneck}
The computation of $\tilde{\mathbf{p}}_{\epsilon}^{opt}(\mathbf{m})$ in the variable projection step (minimization of equation~\ref{eqn:vp.obj} with a linear conjugate gradient algorithm) is equivalent to performing an extended linearized waveform inversion, and accounts for approximately 99$\%$ of the total computational cost of FWIME. In addition, $\tilde{\mathbf{p}}_{\epsilon}^{opt}(\mathbf{m})$ must be re-evaluated each time the velocity model $\mathbf{m}$ is modified. Thus, the objective function estimation is the main computational bottleneck of the workflow (compared to the gradient computation for conventional FWI). For 3D field applications, we observe that approximately 50 iterations of linear conjugate gradient are sufficient to minimize equation~\ref{eqn:vp.obj}. Depending on the length of the extended axis, one iteration of linear conjugate gradient for the variable projection step can be $30 \%$ more costly than one iteration of conventional FWI. Therefore, FWIME is approximately 70 times more computationally demanding than conventional FWI. 

\cite{hou2017alternative} successfully developed a computationally cost-effective method to reduce the number of linear iterations in the variable projection step by one order of magnitude. Their technique, based on an approximate inverse of the space-lag extended Born modeling operator, is designed to minimize quadratic functions of the following form,
\begin{eqnarray}
    \Phi(\mathbf{\mathbf{\tilde{p}}}) &=& \dfrac{1}{2} \left\| \tilde{\mathbf{B}}\mathbf{\tilde{p}}  - \mathbf{d}^{obs} \right\|^2_2.
\end{eqnarray}
However, there are three issues preventing us from directly applying the authors' method to the variable projection step in FWIME. First, their workflow was developed for a space-lag extended Born modeling operator and needs to be adapted to time-lag extension. In addition, the approximate inverse formulation is only valid when the annihilating term in equation~\ref{eqn:vp.obj} is null (i.e., for $\epsilon=0$), and when the data residual $\mathbf{d}^{obs} - \mathbf{f}(\mathbf{m})$ only contains reflected energy. In this paper, we do not propose a solution to this challenge and we leave such investigation for future work. 

\subsection{Using $\tilde{\mathbf{p}}_{\epsilon}^{opt}$ as a quality-control metric}
As we saw from the numerical examples in this paper, $\tilde{\mathbf{p}}_{\epsilon}^{opt}$ is expensive to compute but it can also be used as a quality-control (QC) tool to assess the accuracy of our inverted model. By examining the amount of energy present in the extended space of $\tilde{\mathbf{p}}_{\epsilon}^{opt}$, we can determine when the recovered model is accurate enough to be used as an input for a non-extended inversion strategy such as FWI. 

\subsection{Next steps}
Though not studied here, we carefully analyze how the presence of incoherent and coherent sources of noise impacts the accuracy of our recovered solution in a complementary paper. From a more theoretical standpoint, a formal mathematical proof showing that FWIME produces convex descent paths towards the optimal solution must be investigated.

%%%%%%%%%%%%%%%%% Conclusions %%%%%%%%%%%%%%%
\section{Conclusions}
%%%%%%%%%%%%%%%%%%%%%%%%%%%%%%%%%%%%%%%%%%%%%%%%%%%%%%%%%%%%%%%%%%%%%%%%%%%%%%%%%%%%%%
%%%%%%%%%%%%%%%%%%%%%%%%%%%%%%%%%% Conclusions %%%%%%%%%%%%%%%%%%%%%%%%%%%%%%%%%%%%%%%
%%%%%%%%%%%%%%%%%%%%%%%%%%%%%%%%%%%%%%%%%%%%%%%%%%%%%%%%%%%%%%%%%%%%%%%%%%%%%%%%%%%%%%
We present the theory of our new waveform inversion framework that successfully combines WEMVA with FWI, thereby leveraging the robust convergence properties of the former with the accuracy and high-resolution nature of the latter. We devise a new cost function formulation where we modify the original FWI problem by adding a data-correcting term based on extended modeling to control the level of data fitting. We add an annihilating component to gradually penalize this data-correcting term and guide the inversion towards the global minimum. The use of the variable projection method automatically handles the coupling between the two components of our objective function. From an optimization standpoint, we combine our formulation with a new model-space multi-scale strategy, which is crucial for the success of FWIME. We simultaneously invert the full data bandwidth without the need to manually select specific events, which reduces the number of hyper-parameters to adjust in our technique to only two: the trade-off parameter and the spline grid refinement rate. In this paper, we illustrate the potential of FWIME on simple numerical examples and we show that the same procedure can be used to invert any type of data. In a second paper, we tackle more realistic examples where challenging geological scenarios are considered. 

%%%%%%%%%%%%%%%% Appendices %%%%%%%%%%%%%%%%
\appendix 
\section{FWIME and FWI share the same global minimum}
\label{sameMinimum}
%%%%%%%%%%%%%%%%%%%%%%%%%%%%%%%%%%%%%%%%%%%%%%%%%%%%%%%%%%%%%%%%%%%%%%%%%%%%%%%%%%%%%%%%%%
%%%%%%%%%%%%%%%%%%%%%%%%%%%%%%%%%%%%%% APPENDIX %%%%%%%%%%%%%%%%%%%%%%%%%%%%%%%%%%%%%%%%%%
%%%%%%%%%%%%%%%%%%%%%%%%%% FWIME and FWI have the same minimum %%%%%%%%%%%%%%%%%%%%%%%%%%%
%%%%%%%%%%%%%%%%%%%%%%%%%%%%%%%%%%%%%%%%%%%%%%%%%%%%%%%%%%%%%%%%%%%%%%%%%%%%%%%%%%%%%%%%%%
We show that FWIME and FWI share the same global minimum under the following assumptions,
\begin{enumerate}
    \item There exists a unique global minimum to the FWI objective function, $\mathbf{m}_t$
    \item The observed data are acoustic and noise free, and $\mathbf{f}(\mathbf{m}_t) = \mathbf{d}^{obs}$
    \item $\epsilon > 0$ and $\alpha > 0$, where $\alpha$ is defined in the caption of Figure~\ref{fig:dso_penalty}. 
\end{enumerate}

First, the existence of a global minimizer is straightforward to verify because $\Phi_{\epsilon}(\mathbf{m}_t)=0$. To prove its uniqueness, let us assume there exists another $\mathbf{m}^*$ such that $\Phi_{\epsilon}(\mathbf{m}^*)=0$ and $\mathbf{m}^* \neq \mathbf{m}_t$. Then, both components of the FWIME objective function (equation~\ref{eqn:fwime.obj}) must vanish:

\begin{eqnarray}
    \begin{cases}
    \mathbf{f}(\mathbf{m}^*) + \tilde{\mathbf{B}} (\mathbf{m}^*) \mathbf{\tilde{p}}_{\epsilon}^{opt}(\mathbf{m}^*) - \mathbf{d}^{obs} = \mathbf{0}, \\
    \mathbf{D}{\mathbf{\tilde p}}_{\epsilon}^{opt}(\mathbf{m}^*) = \mathbf{0}.
    \end{cases}
\end{eqnarray}

Moreover, since $\mathbf{D}$ is invertible, $\mathbf{\tilde p}_{\epsilon}^{opt}(\mathbf{m}^*)=\mathbf{0}$. Therefore, $\mathbf{f}(\mathbf{m}^*) - \mathbf{d}^{obs} = \mathbf{0}$, which contradicts the assumption of a unique global minimizer for the FWI objective function.

\section{Convergence of FWIME towards conventional FWI}
\label{fwimeToFwi}
%%%%%%%%%%%%%%%%%%%%%%%%%%%%%%%%%%%%%%%%%%%%%%%%%%%%%%%%%%%%%%%%%%%%%%%%%%%%%%%%%%%%%%%%%%
%%%%%%%%%%%%%%%%%%%%%%%%%%%%%%%%%%%%%% APPENDIX %%%%%%%%%%%%%%%%%%%%%%%%%%%%%%%%%%%%%%%%%%
%%%%%%%%%%%%%%%%%%%%%%%%%%%%%%%% Pointwise convergence %%%%%%%%%%%%%%%%%%%%%%%%%%%%%%%%%%%
%%%%%%%%%%%%%%%%%%%%%%%%%%%%%%%%%%%%%%%%%%%%%%%%%%%%%%%%%%%%%%%%%%%%%%%%%%%%%%%%%%%%%%%%%%
We show that the FWIME objective function converges pointwise in $\mathbf{m} \in \mathbb{R}^{N_m}$ towards the FWI objective function (equation~\ref{eqn:fwi.obj}) when $\epsilon \to +\infty$. To emphasize the fact that $\epsilon$ is now a variable, we re-write $\tilde{\mathbf{p}}^{opt}_{\epsilon}(\mathbf{m})$ and $\Phi_{\epsilon}(\mathbf{m})$ as

\begin{itemize}
    \item $\tilde{\mathbf{p}}^{opt}_{\epsilon}(\mathbf{m}) = \tilde{\mathbf{p}}^{opt}(\mathbf{m},\epsilon)$, and
    \item $\Phi_{\epsilon}(\mathbf{m}) = \Phi(\mathbf{m},\epsilon)$.
\end{itemize}

We conduct the proof in two steps:
\begin{enumerate}
    \item We first prove that $\forall \, \mathbf{m}, \: \underset{\epsilon \to \infty}{\lim} \: \frac{\epsilon^2}{2 }\| \mathbf{D}\tilde{\mathbf{p}}^{opt}({\mathbf{m}},\epsilon) \|^2_2 = 0$, and
    \item We conclude that $\forall \,\mathbf{m}, \: \underset{\epsilon \to \infty}{\lim} \: \Phi(\mathbf{m},\epsilon) = \Phi_{\rm FWI}(\mathbf{m})$,
\end{enumerate}

\noindent where $\Phi$ is the FWIME objective function. The optimal extended perturbation $\mathbf{\tilde{p}}^{opt}$ is given by

\begin{eqnarray}
    \label{eqn:pert.opt.limit}
    \tilde{\mathbf{p}}^{opt}(\mathbf{m},\epsilon) &=& \big [ \tilde{\mathbf{B}}^*(\mathbf{m}) \tilde{\mathbf{B}}(\mathbf{m}) + \epsilon^2 {\mathbf{D}^*} {\mathbf{D}} \big ]^{-1} \tilde{\mathbf{B}}^*(\mathbf{m}) \left ( \mathbf{d}^{obs} - \mathbf{f}(\mathbf{m}) \right ) \\
    &=& \dfrac{1}{\epsilon^2} \big [ \frac{1}{\epsilon^2}\tilde{\mathbf{B}}^*(\mathbf{m}) \tilde{\mathbf{B}}(\mathbf{m})+ {\mathbf{D}^*} {\mathbf{D}} \big ]^{-1} \tilde{\mathbf{B}}^*(\mathbf{m}) \left ( \mathbf{d}^{obs} - \mathbf{f}(\mathbf{m}) \right ).  \nonumber
\end{eqnarray}

Let us assumed that the application of $\tilde{\mathbf{B}}$ and $\mathbf{D}$ (and their respective adjoints) on any bounded vector is bounded. When $\epsilon \to +\infty$, we the following approximation holds,
\begin{eqnarray}
\label{eqn:opt.limit.approx}
\dfrac{1}{\epsilon^2}\tilde{\mathbf{B}}^*(\mathbf{m}) \tilde{\mathbf{B}}(\mathbf{m})+ {\mathbf{D}^*} {\mathbf{D}} \approx {\mathbf{D}^*} {\mathbf{D}}.
\end{eqnarray}
Therefore,
\begin{eqnarray}
\label{eqn:pert.opt.limit.approx}
\tilde{\mathbf{p}}^{opt}(\mathbf{m},\epsilon) &\approx&  \dfrac{1}{\epsilon^2} \big [ {\mathbf{D}^*} {\mathbf{D}} \big ]^{-1} \tilde{\mathbf{B}}^*(\mathbf{m}) \left ( \mathbf{d}^{obs} - \mathbf{f}(\mathbf{m}) \right ) \\
&\approx& \dfrac{1}{\epsilon^2} \tilde{\mathbf{q}}(\mathbf{m}), \nonumber
\end{eqnarray}
where $\tilde{\mathbf{q}}(\mathbf{m}) = \big [ {\mathbf{D}^*} {\mathbf{D}} \big ]^{-1} \tilde{\mathbf{B}}^*(\mathbf{m}) \left ( \mathbf{d}^{obs} - \mathbf{f}(\mathbf{m}) \right )$. Assuming that $\tilde{\mathbf{q}}(\mathbf{m})$ is bounded, we can deduce that for any $\mathbf{m}$,
\begin{itemize}
\item $\| \tilde{\mathbf{p}}^{opt}(\mathbf{m},\epsilon) \|^2_2 \underset{\epsilon \to \infty}{\to} 0$, and that
\item $\dfrac{\epsilon^2}{2}\| \mathbf{D}\tilde{\mathbf{p}}^{opt}(\mathbf{m},\epsilon) \|^2_2 \underset{\epsilon \to \infty}{\to} 0$.
\end{itemize}
\noindent From equations~\ref{eqn:fwi.obj} and \ref{eqn:fwime.obj}, we can write
\begin{eqnarray}
\label{eqn:phi.diff}
\lefteqn{ \Phi(\mathbf{m},\epsilon) - \Phi_{\rm FWI}(\mathbf{m}) = } \\
& & \dfrac{1}{2} \| \mathbf{f}({\mathbf{m}}) + \tilde{\mathbf{B}} (\mathbf{m}) \mathbf{\tilde{p}}^{opt}(\mathbf{m},\epsilon) - \mathbf{d}^{obs} \|^2_2 + \dfrac{\epsilon^2}{2} \| \mathbf{D}\tilde{\mathbf{p}}^{opt}(\mathbf{m},\epsilon) \|^2_2 - \dfrac{1}{2} \| \mathbf{f}(\mathbf{m}) - \mathbf{d}^{obs} \|_2^2. \nonumber
\end{eqnarray}
\noindent Moreover,
\begin{eqnarray}
\label{eqn:triangle2}
\lefteqn{\| \mathbf{f}(\mathbf{m}) + \tilde{\mathbf{B}} (\mathbf{m}) \mathbf{\tilde{p}}_{opt}(\mathbf{m}) - \mathbf{d}^{obs} \|^2_2= } \\
& & \| \mathbf{f}(\mathbf{m}) - \mathbf{d}^{obs} \|^2_2 + \| \tilde{\mathbf{B}} (\mathbf{m}) \mathbf{\tilde{p}}_{opt}(\mathbf{m}) \|^2_2 + 2 \: \left (\mathbf{f}(\mathbf{m}) - \mathbf{d}^{obs} \right )^{*} \tilde{\mathbf{B}} (\mathbf{m}) \mathbf{\tilde{p}}^{opt}(\mathbf{m},\epsilon). \nonumber
\end{eqnarray}
\noindent $\| \tilde{\mathbf{p}}^{opt}(\mathbf{m},\epsilon) \|^2_2 \underset{\epsilon \to \infty}{\to} 0$ implies that $\| \tilde{\mathbf{B}} (\mathbf{m}) \mathbf{\tilde{p}}^{opt}(\mathbf{m},\epsilon) \|^2_2 \underset{\epsilon \to \infty}{\to} 0$, and by the Cauchy-Schwarz inequality,
\begin{eqnarray}
\label{eqn:triangle2}
\Big | \left (\mathbf{f}(\mathbf{m}) - \mathbf{d}^{obs} \right )^{*} \tilde{\mathbf{B}} (\mathbf{m}) \mathbf{\tilde{p}}^{opt}(\mathbf{m},\epsilon) \Big | \le \| \mathbf{f}(\mathbf{m}) - \mathbf{d}^{obs} \|_2  \: \| \tilde{\mathbf{B}} (\mathbf{m}) \mathbf{\tilde{p}}^{opt}(\mathbf{m},\epsilon) \|_2.
\end{eqnarray}
\noindent Therefore,
\begin{itemize}
\item $ \left (\mathbf{f}(\mathbf{m}) - \mathbf{d}^{obs} \right )^{*} \tilde{\mathbf{B}} (\mathbf{m}) \mathbf{\tilde{p}}^{opt}(\mathbf{m},\epsilon) \underset{\epsilon \to \infty}{\to} 0$, and
\item $\| \mathbf{f}(\mathbf{m}) + \tilde{\mathbf{B}} (\mathbf{m}) \mathbf{\tilde{p}}^{opt}(\mathbf{m},\epsilon) - \mathbf{d}^{obs} \|^2_2 - \| \mathbf{f}(\mathbf{m}) - \mathbf{d}^{obs} \|_2^2 \underset{\epsilon \to \infty}{\to} 0$.
\end{itemize}
\noindent Finally, we deduce that
\begin{eqnarray}
\label{eqn:convergence}
\forall \,\mathbf{m}, \: \underset{\epsilon \to \infty}{\lim} \: \Phi(\mathbf{m},\epsilon) = \Phi_{FWI}(\mathbf{m}).
\end{eqnarray}

\section{Derivation of the FWIME gradient}
\label{fwimeGradient}
%%%%%%%%%%%%%%%%%%%%%%%%%%%%%%%%%%%%%%%%%%%%%%%%%%%%%%%%%%%%%%%%%%%%%%%%%%%%%%%%%%%%%%%%%%
%%%%%%%%%%%%%%%%%%%%%%%%%%%%%%%%%%%% APPENDIX %%%%%%%%%%%%%%%%%%%%%%%%%%%%%%%%%%%%%%%%%%%%
%%%%%%%%%%%%%%%%%%%%%%%%%%%%%%%% GRADIENT DERIVATION %%%%%%%%%%%%%%%%%%%%%%%%%%%%%%%%%%%%%
%%%%%%%%%%%%%%%%%%%%%%%%%%%%%%%%%%%%%%%%%%%%%%%%%%%%%%%%%%%%%%%%%%%%%%%%%%%%%%%%%%%%%%%%%%
We derive the gradient of the objective function expressed in equation~\ref{eqn:fwime.gradient}. First, we define
\begin{eqnarray}
\label{eqn:obj.rd}
\mathbf{r}_d (\mathbf{m}) &=& \mathbf{f}(\mathbf{m}) + \tilde{\mathbf{B}} (\mathbf{m}) \tilde{\mathbf{p}}^{opt}_{\epsilon}(\mathbf{m}) - \mathbf{d}^{obs} \\
\mathbf{r}_{\tilde p} (\mathbf{m}) &=& \mathbf{D} \tilde{\mathbf{p}}^{opt}_{\epsilon}(\mathbf{m}),
\end{eqnarray}
where $\mathbf{r}_d \in \mathbb{R}^{N_d}$ is the FWIME data residual, and $\mathbf{r}_{\tilde p} \in \mathbb{R}^{N_{\tilde{p}}}$ is the argument of the annihilating component. Therefore, the FWIME objective function can be written as 
\begin{eqnarray}
\label{eqn:fwime.obj.compact}
\Phi_{\epsilon}(\mathbf{m}) &=& \frac{1}{2} \| \mathbf{r}_d (\mathbf{m})  \|_2^{2} + \frac{\epsilon^2}{2} \| \mathbf{r}_{\tilde p} (\mathbf{m})  \|_2^{2}.
\end{eqnarray}
The gradient of $\Phi_{\epsilon}$ is given by
\begin{eqnarray}
    \label{eqn:fwime.gradient.tmp1}
    \nabla \Phi_{\epsilon}(\mathbf{m}) &=& \left ( \dfrac{\partial \mathbf{r}_d (\mathbf{m})}{\partial \mathbf{m}} \right )^* \mathbf{r}_d (\mathbf{m}) + \epsilon^2 \left ( \dfrac{\partial \mathbf{r}_{\tilde p} (\mathbf{m})}{\partial\mathbf{m}} \right )^* \mathbf{r}_{\tilde p} (\mathbf{m}).
\end{eqnarray}
where $\nabla \Phi_{\epsilon}(\mathbf{m})\in \mathbb{R}^{N_m}$. The Jacobian of $\mathbf{r}_d(\mathbf{m})$ is an operator mapping velocity perturbations into data perturbations,
\begin{eqnarray}
    \dfrac{\partial \mathbf{r}_d (\mathbf{m})}{\partial \mathbf{m}}: \mathbb{R}^{N_m} \mapsto \mathbb{R}^{N_d},
\end{eqnarray}
and its expression is given by
\begin{eqnarray}
    \label{eqn:data.jacobian1}
    \dfrac{\partial \mathbf{r}_d (\mathbf{m})}{\partial \mathbf{m}} &=& \dfrac{\partial \mathbf{f} (\mathbf{m})}{\partial \mathbf{m}} + \dfrac{\partial \left (\tilde{\mathbf{B}} (\mathbf{m}) \tilde{\mathbf{p}}^{opt}_{\epsilon}(\mathbf{m}) \right ) }{\partial \mathbf{m}} \nonumber \\
    &=& \mathbf{B}(\mathbf{m})+ \dfrac{\partial \left ( \tilde{\mathbf{B}} (\mathbf{m}) \tilde{\mathbf{p}}^{opt}_{\epsilon}(\mathbf{m})\right ) }{\partial \mathbf{m}} \Big |_{\tilde{\mathbf{p}}^{opt}_{\epsilon}} + \mathbf{\tilde{B}} (\mathbf{m}) \dfrac{\partial \tilde{\mathbf{p}}^{opt}_{\epsilon}(\mathbf{m})}{\partial \mathbf{m}}.
\end{eqnarray}
The first term of the right side of equation~\ref{eqn:data.jacobian1}, is the conventional non-extended Born modeling operator, $\mathbf{B}(\mathbf{m}): \mathbb{R}^{N_m} \mapsto \mathbb{R}^{N_d}$. The second term of the right side of equation~\ref{eqn:data.jacobian1} characterizes linear variations of the Born-modeled data (using an extended Born modeling operator) with variations in the velocity model $\mathbf{m}$, given a fixed extended perturbation $\mathbf{p}^{opt}_{\epsilon}$. Therefore, 
\begin{eqnarray}
    \label{eqn:tomo.connection}
    \mathbf{T}(\mathbf{m}, \mathbf{\tilde{p}}_{\epsilon}^{opt}) &=& \dfrac{\partial \left (\tilde{\mathbf{B}} (\mathbf{m}) \tilde{\mathbf{p}}^{opt}_{\epsilon}(\mathbf{m}) \right ) }{\partial \mathbf{m}} \Big |_{\tilde{\mathbf{p}}^{opt}_{\epsilon}}
\end{eqnarray}
where $\mathbf{T}(\mathbf{m}, \mathbf{\tilde{p}}_{\epsilon}^{opt}): \mathbb{R}^{N_m} \mapsto \mathbb{R}^{N_d}$ is the data-space tomographic operator (equation~\ref{eqn:tomo_operator_definition}). In the following, to simplify notations, we do not explicitly write the dependency of $\mathbf{T}$ with respect to $\mathbf{\tilde{p}}_{\epsilon}^{opt}$. The last term of right side of equation~\ref{eqn:data.jacobian1} is the composition of the following two operators:
\begin{eqnarray}
    \dfrac{\partial \tilde{\mathbf{p}}^{opt}_{\epsilon}(\mathbf{m})}{\partial \mathbf{m}}&:& \mathbb{R}^{N_m} \mapsto \mathbb{R}^{N_{\tilde{p}}} \\
    \tilde{\mathbf{B}} (\mathbf{m})&:& \mathbb{R}^{N_{\tilde{p}}} \mapsto \mathbb{R}^{N_d}.    
\end{eqnarray}
Therefore, the Jacobian of the FWIME data residual is given by
\begin{eqnarray}
    \label{eqn:data.jacobian2}
    \dfrac{\partial \mathbf{r}_d (\mathbf{m})}{\partial \mathbf{m}} &=& \mathbf{B}(\mathbf{m}) + \mathbf{T}(\mathbf{m}) + \mathbf{\tilde{B}} (\mathbf{m}) \dfrac{\partial \tilde{\mathbf{p}}^{opt}_{\epsilon}(\mathbf{m})}{\partial \mathbf{m}}, 
\end{eqnarray}
and its adjoint is expressed by
\begin{eqnarray}
    \label{eqn:data.gradient}
    \left ( \dfrac{\partial \mathbf{r}_d (\mathbf{m})}{\partial \mathbf{m}} \right )^* &=& \mathbf{B}^*(\mathbf{m}) + \mathbf{T}^*(\mathbf{m}) + \left ( \dfrac{\partial \tilde{\mathbf{p}}^{opt}_{\epsilon}(\mathbf{m})}{\partial \mathbf{m}} \right )^* \tilde{\mathbf{B}} (\mathbf{m})^*, 
\end{eqnarray}
where $\left ( \dfrac{\partial \mathbf{r}_d (\mathbf{m})}{\partial \mathbf{m}} \right )^*: \mathbb{R}^{N_d} \mapsto \mathbb{R}^{N_m}$. For the second term of right side of equation~\ref{eqn:fwime.gradient.tmp1}, we have
\begin{eqnarray}
\label{eqn:model.gradient}
\left ( \dfrac{\partial \mathbf{r}_{\tilde p} (\mathbf{m})}{\partial \mathbf{m}} \right )^* &=& \left ( \dfrac{\partial \tilde{\mathbf{p}}^{opt}_{\epsilon}(\mathbf{m})}{\partial \mathbf{m}} \right )^* \mathbf{D}^*,
\end{eqnarray}
where $\left ( \dfrac{\partial \mathbf{r}_{\tilde p} (\mathbf{m})}{\partial \mathbf{m}} \right )^*: \mathbb{R}^{N_d} \mapsto \mathbb{R}^{N_m}$. Therefore, equation~\ref{eqn:fwime.gradient.tmp1} becomes
\begin{eqnarray}
\label{eqn:fwime.gradient.vp}
\nabla \Phi_{\epsilon}(\mathbf{m}) &=& \Big [ \mathbf{B}^*(\mathbf{m}) + \mathbf{T}^*(\mathbf{m}) + \left ( \dfrac{\partial \tilde{\mathbf{p}}^{opt}_{\epsilon}(\mathbf{m})}{\partial \mathbf{m}} \right )^* \tilde{\mathbf{B}} (\mathbf{m})^* \Big ] \mathbf{r}_d (\mathbf{m}) + \epsilon^2 \left ( \dfrac{\partial \tilde{\mathbf{p}}^{opt}_{\epsilon}(\mathbf{m})}{\partial \mathbf{m}} \right )^* \mathbf{D}^* \mathbf{r}_{\tilde p} (\mathbf{m}) \nonumber \\
&=&  \Big [ \mathbf{B}^*(\mathbf{m}) + \mathbf{T}^*(\mathbf{m}) \Big ] \mathbf{r}_d (\mathbf{m}) \nonumber \\
&+& \left ( \dfrac{\partial \tilde{\mathbf{p}}^{opt}_{\epsilon}(\mathbf{m})}{\partial \mathbf{m}} \right )^* \Big [ \tilde{\mathbf{B}}^* (\mathbf{m}) \mathbf{r}_d (\mathbf{m}) + \epsilon^2 \mathbf{D}^* \mathbf{r}_{\tilde p} (\mathbf{m}) \Big ].
\end{eqnarray}
Now, we are going to show that 
\begin{eqnarray}
\label{eqn:res.vp.null}
\tilde{\mathbf{B}}^* (\mathbf{m}) \mathbf{r}_d (\mathbf{m}) + \epsilon^2 \mathbf{D}^* \mathbf{r}_{\tilde p} (\mathbf{m}) = \mathbf{0}.
\end{eqnarray}
Since $\tilde{\mathbf{p}}^{opt}_{\epsilon}$ satisfies equation~\ref{eqn:pert.opt} (variable projection step), we can write
\begin{eqnarray}
\label{eqn:pert.opt.recast}
\Big [ \tilde{\mathbf{B}}^*(\mathbf{m}) \tilde{\mathbf{B}}(\mathbf{m}) + \epsilon^2 {\mathbf{D}^*} {\mathbf{D}} \Big ] \tilde{\mathbf{p}}^{opt}_{\epsilon}(\mathbf{m}) = \tilde{\mathbf{B}}^*(\mathbf{m}) \left ( \mathbf{d^{obs}} - \mathbf{f}(\mathbf{m}) \right ).
\end{eqnarray}
Therefore,
\begin{eqnarray}
\label{eqn:vp.orthogonality}
\tilde{\mathbf{B}}^* (\mathbf{m}) \mathbf{r}_d (\mathbf{m}) + \epsilon^2 \mathbf{D}^* \mathbf{r}_{\tilde p} (\mathbf{m}) &=& \tilde{\mathbf{B}}^* ( \mathbf{m}) \left ( \mathbf{f}(\mathbf{m}) + \tilde{\mathbf{B}} (\mathbf{m}) \tilde{\mathbf{p}}^{opt}_{\epsilon}(\mathbf{m}) - \mathbf{d^{obs}} \right ) + \epsilon^2 \mathbf{D}^* \mathbf{D}  \tilde{\mathbf{p}}^{opt}_{\epsilon}(\mathbf{m}) \nonumber \\
&=& \Big [ \tilde{\mathbf{B}}^* (\mathbf{m}) \tilde{\mathbf{B}} (\mathbf{m}) + \epsilon^2 \mathbf{D}^* \mathbf{D} \Big ] \tilde{\mathbf{p}}^{opt}_{\epsilon}(\mathbf{m}) - \tilde{\mathbf{B}}^*(\mathbf{m}) (\mathbf{d^{obs}} - \mathbf{f}(\mathbf{m})) \nonumber \\
&=& \mathbf{0}.
\end{eqnarray}
The result from Equation~\ref{eqn:res.vp.null} is very useful because it allows me to avoid computing the following term:
\begin{eqnarray}
 \left ( \dfrac{\partial \tilde{\mathbf{p}}^{opt}_{\epsilon}(\mathbf{m})}{\partial \mathbf{m}} \right )^* \Big [ \tilde{\mathbf{B}}^* (\mathbf{m}) \mathbf{r}_d (\mathbf{m}) + \epsilon^2 \mathbf{D}^* \mathbf{r}_{\tilde p} (\mathbf{m}) \Big ].
\end{eqnarray}
Notice that Equation~\ref{eqn:res.vp.null} is true provided that equation~\ref{eqn:pert.opt} is satisfied, which highlights the importance to conduct enough linear conjugate gradient iterations during the FWIME variable projection step. Finally, equation~\ref{eqn:fwime.gradient.vp} reduces to
\begin{eqnarray}
\label{eqn:vptfwi.gradient.vp}
\nabla \Phi(\mathbf{m}) &=& \Big [ \mathbf{B}^*(\mathbf{m}) + \mathbf{T}^*(\mathbf{m}) \Big ] \left ( \mathbf{f}(\mathbf{m}) + \tilde{\mathbf{B}} (\mathbf{m}) \tilde{\mathbf{p}}^{opt}_{\epsilon}(\mathbf{m}) - \mathbf{d^{obs}} \right ) \nonumber \\
&=& \Big [ \mathbf{B}^*(\mathbf{m}) + \mathbf{T}^*(\mathbf{m}) \Big ] \mathbf{r}_d(\mathbf{m}).
\end{eqnarray}

\bibliographystyle{unsrt}  
\bibliography{references}  

\end{document}